\documentclass[twocolumn,a4paper,accepted=2022-03-08]{quantumarticle}
\pdfoutput=1

\usepackage[numbers]{natbib}
\usepackage{shorttoc}

\usepackage[toc,title]{appendix}

\usepackage{amsmath}
\usepackage{amssymb}
\usepackage{amsthm}
\usepackage{bbold}
\usepackage{enumitem}
\usepackage[hidelinks,breaklinks]{hyperref}
\hypersetup{pdfauthor={V. Gitton, M. P. Woods},pdftitle={Solvable Criterion for the Contextuality of any Prepare-and-Measure Scenario}}
\usepackage{bm}
\usepackage{comment}
\usepackage[normalem]{ulem}
\usepackage{thm-restate}
\usepackage{thmtools,nameref,cleveref}
\usepackage{chngcntr}
\usepackage{apptools}
\usepackage{graphicx}

\declaretheorem[name=Theorem,
	refname={theorem,theorems},
	Refname={Theorem,Theorems}]{theorem}
\declaretheorem[name=Proposition,
	refname={proposition,propositions},
	Refname={Proposition,Propositions}]{prop}
\declaretheorem[name=Lemma,
	refname={lemma,lemmas},
	Refname={Lemma,Lemmas},
	numberlike=prop]{lemma}
\declaretheorem[name=Corollary,
	refname={corollary,corollaries},
	Refname={Corollary,Corollaries},
	numberlike=prop]{corollary}
\declaretheorem[name=Definition,
	refname={definition,definitions},
	Refname={Definition,Definitions},
	numberlike=prop]{definition}

\theoremstyle{remark}

\newtheorem*{remark*}{Remark}

\numberwithin{equation}{section}

\AtAppendix{\numberwithin{prop}{section}}
\AtAppendix{\numberwithin{lemma}{section}}
\AtAppendix{\numberwithin{theorem}{section}}
\AtAppendix{\numberwithin{definition}{section}}
\AtAppendix{\numberwithin{corollary}{section}}

\newenvironment{myproof}
{
\vspace{0.1cm}
\begin{proof}
}
{
\end{proof}
\vspace{0.9cm}
}

\newcommand{\vgr}[1]{\textcolor{blue}{#1}}
\usepackage{color}

\newcommand\mpwS[1]{{\let\helpcmd\sout\parhelp#1\par\relax\relax} }
\long\def\parhelp#1\par#2\relax{%
	\helpcmd{#1}\ifx\relax#2\else\par\parhelp#2\relax\fi%
}

\newcommand{\vgrSC}[1]{\textcolor{red}{\mpwS{#1}}}

\newcommand{\commentconditional}[1]{#1}

\def\nocomments{1}

\ifx\nocomments\undefined%
\else%
\renewcommand{\vgrSC}[1]{}%
\renewcommand{\vgr}[1]{#1}%
\renewcommand{\commentconditional}[1]{}%
\fi

\newcommand{\app}{\text{appendix}}

\newcommand{\doc}{\text{manuscript}}  

\newcommand{\hil}{\mathcal H}
\newcommand{\lin}[1]{L(#1)}
\newcommand{\linherm}[1]{\mathcal L(#1)}
\newcommand{\lh}{\linherm\hil}
\newcommand{\lhplus}{\mathcal P(\hil)}
\newcommand{\states}{\mathcal S(\mathcal H)}
\newcommand{\effects}{\mathcal E(\mathcal H)}

\newcommand{\sete}{\mathtt e}
\newcommand{\sets}{\mathtt s}
\newcommand{\mainr}{\mathcal R}
\newcommand{\altr}[1]{#1_{\textup{alt}}}

\newcommand{\LargeNegBlank}{\hspace{-0.08em}}
\newcommand{\SmallNegBlank}{\hspace{-0.00cm}}
\newcommand{\proj}[2]{P_{#1}\LargeNegBlank\left(\SmallNegBlank#2\SmallNegBlank\right)}
\newcommand{\rede}{\proj{\mainr}{\sete}}
\newcommand{\reds}{\proj{\mainr}{\sets}}

\newcommand{\scal}[3]{\left<#1,#2\right>_{#3}}
\newcommand{\norm}[2]{\left\|#1\right\|_{#2}}

\newcommand{\id}[1]{\mathbb{1}_{#1}}
\newcommand{\tr}[2]{\textup{Tr}_{#1}\!\left[#2\right]}

\newcommand{\myspan}[1]{\textup{span}\LargeNegBlank\left(\SmallNegBlank#1\SmallNegBlank\right)}
\newcommand{\spane}{\myspan{\sete}}
\newcommand{\spans}{\myspan{\sets}}

\newcommand{\st}{:\ }
\newcommand{\red}[1]{\bar #1}

\newcommand{\ospace}{\Lambda}
\newcommand{\hv}{\lambda}
\newcommand{\ostate}[2]{\mu(#1,#2)}
\newcommand{\orep}[2]{\xi(#1,#2)}

\newcommand{\otrace}[1]{\scal{#1}{\proj{\mainr}{\id\hil}}{\mainr}}

\newcommand{\myint}[3]{\int_{#1}\LargeNegBlank\textup{d}#2\,#3}
\newcommand{\inthv}[1]{\myint{\ospace}{\hv}{#1}}

\newcommand{\prepproc}{\bm{P}}
\newcommand{\measproc}{\bm{M}}

\newcommand{\realpos}{\mathbb R_{\geq 0}}
\newcommand{\funcrange}[2]{:#1\rightarrow#2}

\newcommand{\polar}[2]{#1^{+_{#2}}}
\newcommand{\bigpolar}[2]{#1^{+\scalebox{0.6}{$#2$}}}

\newcommand{\prodes}{\mathtt{Prod}(\sets,\sete)}

\newcommand{\myconv}[1]{\textup{conv}\LargeNegBlank\left(#1\right)}
\newcommand{\mycone}[1]{\textup{coni}\LargeNegBlank\left(#1\right)}

\newcommand{\sepes}{\mathtt{Sep}(\sets,\sete)}

\newcommand{\choi}[1]{\mathbb J(#1)}
\newcommand{\invchoi}[1]{\mathbb J^{-1}[#1]}

\newcommand{\card}[1]{\big|#1\big|}
\newcommand{\scard}[1]{|#1|}

\newcommand{\rank}[1]{\textup{rank}\LargeNegBlank\left(#1\right)}

\newcommand{\closure}[1]{\overline{#1}}

\newcommand{\sgn}[1]{\textup{sgn}(#1)}

\newcommand{\probcond}[2]{\textup{Pr}[#1|#2]}

\newcommand{\ket}[1]{\left|#1\right>}
\newcommand{\ketbraa}[2]{\left|#1\right>\!\left<#2\right|}
\newcommand{\ketbra}[1]{\ketbraa{#1}{#1}}
\newcommand{\zerostate}{\ketbra{0}}
\newcommand{\onestate}{\ketbra{1}}
\newcommand{\plusstate}{\ketbra{+}}
\newcommand{\minusstate}{\ketbra{-}}

\newcommand{\restrict}[2]{\left.#1\right|_{\scalebox{0.9}{$#2$}}}
\newcommand{\hvdis}{\hv^{\textup{(dis)}}}

\newcommand{\im}[1]{\textup{Im}(#1)}

\newcommand{\convextr}[1]{\textup{ep}(#1)}
\newcommand{\coneextr}[1]{\textup{extr}(#1)}
\newcommand{\venum}[2]{\mathtt{V}.\mathtt{E}_{#1}\big[#2\big]}

\newcommand{\scala}[2]{\mathtt{S}_1(#1,#2)}
\newcommand{\scalb}[2]{\mathtt{S}_2(#1,#2)}

\newcommand{\couts}{\mathcal C_\sets^{\textup{(out)}}}
\newcommand{\coute}{\mathcal C_\sete^{\textup{(out)}}}
\newcommand{\cins}{\mathcal C_\sets^{\textup{(in)}}}
\newcommand{\cine}{\mathcal C_\sete^{\textup{(in)}}}
\newcommand{\halfl}{\mathfrak{l}}

\newcommand{\oprod}{\otimes_{\textup{set}}}

\newcommand{\resources}{$(\sets,\sete)$}
\newcommand{\polars}{\polar{\reds}{\mainr}}
\newcommand{\polare}{\polar{\rede}{\mainr}}
\newcommand{\sepout}{\mathtt{Sep}^{\textup{(out)}}}
\newcommand{\sepin}{\mathtt{Sep}^{\textup{(in)}}}
\newcommand{\choijamiolkowsky}{Choi-Jamio\l kowsky}
\newcommand{\qsep}{\mathtt{Q.Sep}}
\newcommand{\unitsep}{unit separability}
\newcommand{\Unitsep}{Unit separability}

\newenvironment{myitem}
{\begin{enumerate}[label=\textup{(\roman*)}]
\setlength\itemsep{1pt}
\setlength\parsep{-5pt}
}
{
\end{enumerate}
}

\author{Victor Gitton}
\author{Mischa P. Woods}
\affiliation{Institute for Theoretical Physics, ETH Zürich, Switzerland}

\begin{document}

\title{Solvable Criterion for the Contextuality of any Prepare-and-Measure Scenario}

\begin{abstract}

Starting from arbitrary sets of quantum states and measurements,
referred to as the prepare-and-measure scenario,
\vgrSC{a non-contextual}\vgr{an operationally noncontextual} ontological model
of the quantum statistics associated with the prepare-and-measure scenario is constructed.
The \vgrSC{non-contextual}\vgr{operationally noncontextual} ontological model coincides with standard Spekkens noncontextual ontological models for tomographically complete scenarios,
while covering the non-tomographically complete case with a new notion of a reduced space, which we motivate following the guiding principles of noncontextuality.
A mathematical criterion, called \textit{\unitsep}, is formulated as the relevant classicality criterion ---
the name is inspired by the usual notion of quantum state separability.
Using this criterion, we derive a new upper bound on the cardinality of the ontic space.
Then, we recast the \unitsep{} criterion as a (possibly infinite) set of linear constraints, from which \vgr{we obtain} two separate hierarchies of algorithmic tests to witness \vgr{the} non-classicality or certify \vgr{the} classicality \vgr{of a scenario}.
Finally, we reformulate our results in the framework of generalized probabilistic theories and discuss the implications
for simplex-embeddability in such theories.
\end{abstract}

\maketitle

\makeatletter
\def\p@paragraph{\thesection\,}
\makeatother


\setcounter{tocdepth}{1}
{\expandafter\def\csname @starttoc\endcsname#1{\InputIfFileExists{\jobname.#1}{}{}}%
\tableofcontents}


\section{Introduction}

\subsection{Background: previous notions of classicality}

Studying the non-classicality of quantum mechanics is a field that originated from the collective effort of the scientific community to obtain meaningful interpretations of the ontologically opaque yet undoubtedly successful theory of quantum mechanics.
One of the early influential works highlighting how quantum mechanics departs significantly from classical mechanics was that of Einstein, Podolsky and Rosen~\cite{EPR}: there, it was brought to light that local realism, a natural notion of classicality, is in conflict with the quantum description of nature.
Realism means that one posits the existence of a hidden state that should describe the actual physics behind the scenes, the \emph{ontic} (actual) state of the system.
Local realism means that the ontic state cannot be updated from a spacelike-separated spacetime region.
This notion was further studied and turned into an experimentally verifiable \emph{no-go theorem} by Bell~\cite{Bell}: the no-go theorem states that quantum mechanics cannot be described by a local hidden variable model.
For the perspective of this \doc{}, it is important to notice that this notion of classicality only applies to spacelike-separated systems, whereas a single quantum system is not eligible to be tested via the prism of local realism.

A natural generalization of local realism is that of Kochen-Specker noncontextuality, where the associated classical model is called noncontextual hidden variable model.
This notion of classicality assumes that at the ontic state level, the outcome statistics of one measurement are 1) statistically independent from the outcome statistics of any other commuting
measurement and 2) non-varying with respect to changing the jointly-measured commuting measurement. This notion was formalized and shown to be inconsistent with quantum mechanics by Kochen and Specker~\cite{KS}.
Only commuting measurements may be tested through the prism of Kochen-Specker noncontextuality but possibly on a single quantum \vgrSC{systems}\vgr{system}, which was not the case with local realism.

The work of Spekkens~\cite{Spekkens05} lead to a new notion of noncontextuality that subsumes Bell's local hidden variables and Kochen-Specker noncontextuality.
 The assumption of realism is similar to that of the previously mentioned notions of classicality, but the scope of noncontextuality is more universal.
The first step towards formulating an assumption of noncontextuality is to formulate a notion of operational equivalence, such as e.g.\ the operational equivalence of an electron spin-$\frac{1}{2}$ degree of freedom and a photon polarization degree of freedom as two implementations of a qubit.
The corresponding assumption of noncontextuality is to posit that operationally equivalent procedures have an identical representation at the level of the ontic model.
In~\cite{Spekkens05}, several no-go theorems are presented to show the incompatibility of quantum mechanics with respect to Spekkens noncontextuality.
Quantum procedures may be eligible for testing their classicality with respect to Spekkens noncontextuality irrespective of the existence of commuting measurements.
Furthermore, the incompatibility of quantum mechanics and Spekkens noncontextuality has known links with computational efficiency of quantum protocols, see e.g.~\cite{SHVStateDiscrimination,SHVPostSelectedMetrology}, which supplement the existing links between computational efficiency and violations of Kochen-Specker noncontextuality (a special case of Spekkens noncontextuality), see e.g.~\cite{NCHVAdv1,NCHVAdv2}.

\subsection{The objective notion of classicality}
\label{par:motivation se}

The present work aims at obtaining a notion of classicality that is applicable to an arbitrary prepare-and-measure scenario and that provides an answer to the question of whether the scenario is classical or not with respect to that notion of classicality.
The prepare-and-measure scenario may consist of all states and measurements allowed by quantum mechanics within a given Hilbert space, but it can also consist of strict subsets of these:
this would be interesting if for instance one has an apparatus that only allows to produce certain types of states or perform certain types of measurements.
Then, one could answer the question of whether this specific apparatus has a classical description or not.
Alternatively, one can associate to a given quantum protocol a corresponding prepare-and-measure scenario that only features the states and measurements relevant for the protocol.
For instance, the set of states of the scenario could be special types of multi-qubit states of a quantum computer that are relevant for a given algorithm.
Then, assessing the classicality of the prepare-and-measure scenario associated to the protocol is an indirect way of assessing the classicality of the protocol itself.
This assessment may help identify resources that are most useful for efficient protocols.

Local realism and Kochen-Specker noncontextuality are well-motivated and widely useful notions of classicality, but they 
do not quite fulfill the \vgr{above} requirement \vgrSC{that any set of states and set of measurements are eligible for a test of classicality}\vgr{of being applicable to arbitrary prepare-and-measure scenarios.}
Indeed, local realism specializes to local measurements on spacelike separated systems, and Kochen-Specker noncontextuality specializes to commuting measurements.
On the other hand, the universality of Spekkens' notion of noncontextuality makes it a promising basis for the formulation of our objective notion of classicality.

\subsection{Content overview}

\Cref{sec:PresentationClassicalModel} will formalize the quantum prepare-and-measure scenarios under consideration, motivate 
Spekkens noncontextual ontological models, and define the adjustments thereof that we posit in the case of non-tomographically complete scenarios --- the \vgrSC{classical model}\vgr{resulting \emph{operationally noncontextual ontological model}} that we will \vgrSC{use}\vgr{consider} is specified in \cref{def:classicalmodel}.
In particular, \cref{subsection:discussion of gshv} will discuss the motivation for \vgrSC{the reduced space prescription that appears in the classical model}\vgr{our operationally noncontextual ontological models}.
Then, in \cref{sec:ExistenceClassicalModel}, the \vgrSC{classicality of}\vgr{existence of an operationally noncontextual ontological model for} a given prepare-and-measure scenario is turned into the \unitsep{} criterion in \cref{th:maincriterion}.
An explicit example of the application of this criterion is given in \cref{subsec:example}.
This criterion allows one to extract theoretical properties of the \vgrSC{classical}\vgr{operationally noncontextual ontological} model, such as the ontic space cardinality bounds of \cref{th:CardinalityBounds}.
Furthermore, an algorithmic formulation that evaluates the criterion for a given scenario is presented in  \cref{sec:algo}. 
In \cref{sec:Connection}, parallel independent work treating generalized probabilistic theories is discussed and connected to the content of this \doc.

%
%

\section{\vgr{Operationally noncontextual ontological model}}
\label{sec:PresentationClassicalModel}

\subsection{Prepare-and-measure scenarios}

Let $\hil$ be a finite dimensional Hilbert space corresponding to the quantum system. The set of Hermitian matrices acting on $\hil$ is denoted $\lh$. $\lh$ has the structure of a real inner product space of dimension $\dim(\lh) = \dim(\hil)^2$: its inner product, often referred to as the Hilbert-Schmidt inner product, is defined by $\scal{a}{b}{\lh} := \tr{\hil}{ab}$ for all $a,b\in\lh$.
The set of density matrices, i.e., positive semi-definite, trace-one hermitian matrices acting on $\hil$, is denoted $\states$.
The set of quantum effects, i.e., positive semi-definite matrices $E$ acting on $\hil$ such that $\id\hil-E$ is also positive semi-definite, is denoted $\effects$.\footnote{We will use throughout the notation where $\id{\mathcal V}$ (resp.\ $0_{\mathcal V}$) is the identity (resp.\ zero operator) on any vector space $\mathcal V$.}

\begin{definition}[State space]
Let $\sets\subseteq\states$ be a nonempty subset of states that is convex.\label{def:available states}
\end{definition}
Physically, any non-convex set $S_1$ of density matrices together with the possibility of taking classical probabilistic mixtures leads to a set of states $S_2$ that is the convex hull of $S_1$, and hence $S_2$ is convex.

\begin{definition}[Effect space]
\label{def:available effects}
Let $\sete\subseteq\effects$ be a subset of effects such that
\begin{myitem}
\item $\sete$ is convex;
\item $0_\hil,\id{\hil}\in\sete$;
\item if $E\in\sete$, then there exists a completion $\{E_k\in\sete\}_k$ such that $E + \sum_k E_k = \id\hil$.
\end{myitem}
\end{definition}   
The convexity requirement (i) for $\sete$ is motivated by allowing classical probabilistic mixtures of different measurements, see \app~\ref{app:qexp} for an explicit example.
The requirement (ii) comes from the fact that the trivial effects should always be allowed and faithfully be represented in the ontological model.
The requirement (iii) reflects the fact that in any practical application, the effects in $\sete$ will come from complete POVM sets.
\Cref{prop:coarsegrainings} will formalize the fact that including or not incoherent coarse-grainings of measurements in $\sete$ does not make a difference for our purposes.
Furthermore, \cref{prop:ancilla} will formalize the fact that distinct but operationally equivalent quantum descriptions of a prepare-and-measure scenario are equivalent as far as the \vgrSC{classical}\vgr{operationally noncontextual ontological} model is concerned.

The pair \resources{} is referred to as being an instance of a quantum prepare-and-measure scenario, or just a scenario for brevity.
Recall the motivation for taking \resources{} to be strict subsets of all states and effects on the Hilbert space in \cref{par:motivation se}.

\subsection{Reduced space}

Since we are primarily concerned with quantum protocols that involve preparing a given state $\rho\in\sets$ and measuring it once with a complete set of effects where each effect $E$ belongs to $\sete$,
the experimental predictions of quantum mechanics for such protocols are entirely encoded in the probabilities $\scal \rho E \lh$ for all $\rho\in\sets$ and $E\in\sete$.\footnote{One can go beyond this setting by including post-measurement states in the set $\sets$ which is a minimalistic way of accounting for multiple consecutive measurements.}
Let us introduce the \emph{reduced space}, which is meant to capture the minimal amount of information needed to reconstruct these operational statistics.
For any set $X\subseteq\lh$, we denote the linear span of its elements as $\myspan X\subseteq\lh$, which is the minimal vector subspace that contains $X$.
For any $a\in\lh$, the projection of $a$ over any vector subspace $\mathcal V \subseteq \lh$ equipped with an orthonormal basis $\{v_i\in\mathcal V\}_i$ is denoted $\proj{\mathcal V}{a} := \sum_i \scal{v_i}{a}{\lh}v_i$.
The projection of a set $X\subseteq\lh$ over $\mathcal V$ is denoted $\proj{\mathcal V}{X} := \{\proj{\mathcal V}{x}\st x\in X\}.$

\begin{definition}[Reduced space]
\label{def:mainr}
Let
\begin{equation}
\mainr := \proj{\spane}{\spans}
\end{equation}
be the \emph{reduced space} associated with the scenario \resources. $\mainr\subseteq\lh$ is a vector space that we equip with the inner product inherited from $\lh$.
\end{definition}
Note that $\dim(\mainr)\leq\dim(\lh) = \dim(\hil)^2$. The main property of the reduced space is the following. See \app~\ref{app:mainr} for a proof.
\begin{prop}
\label{prop:reducedscalar}
For all $\rho\in\sets$, for all $E\in\sete$,
\begin{equation}
\scal{\rho}{E}{\lh} = \scal{\proj{\mainr}{\rho}}{\proj{\mainr}{E}}{\mainr}.
\end{equation}
\end{prop}
\Cref{prop:reducedscalar} shows that we can in fact restrict the analysis of the probabilities associated to \resources{} to the analysis of all probabilities $\scal{\red\rho}{\red E}{\mainr}$ for all $\red\rho\in\reds$ and for all $\red E\in\rede$.

Since the reduced space will play an important role in the definition of the \vgrSC{classical}\vgr{operationally noncontextual ontological} model (\cref{subsection:definition of classical model}), it is useful to distinguish the case where the reduced space projection is irrelevant.
The following definition formalizes this case --- note that the wording ``tomographically complete" could be used differently in other contexts.
\begin{definition}[Tomographically complete scenarios]
\label{def:tomo}
A tomographically complete scenario \resources{} is one where it holds that
\begin{equation}
\spans = \spane \subseteq \lh.
\end{equation}
\end{definition}
In such scenarios, 
the projection of $\sets$ and $\sete$ over the reduced space is trivial, so that for our purposes we may replace $P_\mainr$ with $\id\lh$ --- specifically, it holds that $P_\mainr|_{\spans} = \id\spans$ and $P_\mainr|_{\spane} = \id\spane$.
Furthermore, \cref{def:tomo} puts states and effects on an equal footing: for instance, a scenario in which $\spans \subset \spane$ could be turned into a tomographically complete scenario by either adding some states to $\spans$ or removing some effects from $\spane$.
Also, it is readily apparent that tomographic completeness is a weaker requirement than having \resources{} consist of all states and effects on the Hilbert space.

\subsection{Definition of the \vgr{operationally noncontextual ontological} model}
\label{subsection:definition of classical model}

We now motivate the construction of the operationally noncontextual ontological model that we are considering for the scenario \resources. This model is largely based on the mathematical description and notation introduced in~\cite{Spekkens05}. 
We will motivate the conditions 
that we impose in the non-tomographically complete case as we go along, and we will further argue in favor of these in \cref{subsection:discussion of gshv}.
While this work was in development, a similar notion of noncontextuality was considered in~\cite{Spekkens19} --- see \cref{sec:SimplexEmbeddability} for differences and similarities in the results.

\subsubsection{Ontological model}

We introduce the notion of an ontic state space, or ontic space for short, denoted $\ospace$. An ontic state $\hv\in\ospace$ is meant to describe a classical state of the system, so that $\ospace$ can be thought of as a classical phase space that will be assigned to the quantum setup.

Let $\prepproc$ denote a preparation procedure, i.e., a set of operational instructions that fully specify the steps one needs to take to obtain the same preparation. The first idea of the ontological model is to associate to each preparation procedure $\prepproc$ a classical probability distribution, i.e., normalizable and nonnegative, over $\ospace$. We refer to these probability distributions as the ontic state distributions.
The ontic state distribution gives the probability $\probcond{\hv}{\prepproc}$ that the system is in the ontic state $\hv$ after having been prepared by the preparation procedure $\prepproc$.

Let $\measproc$ be a measurement procedure with outcomes labeled by $k$. We denote by $\measproc_k$ the event that the outcome $k$ occurred when the measurement procedure $\measproc$ was carried out. Any operational detail should be included in the specification of $\measproc$. In the ontological model, the measurements will be represented as classical probability distributions over the outcomes $k$; but these probability distributions, referred to as the response functions, will not depend on the quantum states directly.
Instead, the response functions will ``read off" the value of a given ontic state $\hv$ to produce the outcome statistics.
The response function is thus  represented by the conditional probabilities $\probcond{\measproc_k}{\hv}$.
The actual outcome statistics, given a preparation $\prepproc$ and an event $\measproc_k$, will be the outcome statistics $\probcond{\measproc_k}{\hv}$ averaged over the probability that the system was in the ontic state $\hv$, which is specified by the ontic state distribution $\probcond{\hv}{\prepproc}$:
\begin{equation}
\label{eq:OnticProbRule}
\probcond{\measproc_k}{\prepproc} = \inthv{\probcond{\hv}{\prepproc}\probcond{\measproc_k}{\hv}}.
\end{equation}

\subsubsection{Noncontextual state representation}

In complete generality, the probability $\probcond{\hv}{\prepproc}$ could depend on any detail of the preparation procedure $\prepproc$.
This is not very satisfactory: we know from quantum mechanics that all possible measurement statistics are uniquely determined from the density matrix $\rho(\prepproc)$ associated to the preparation procedure $\prepproc$.

The standard assumption of noncontextuality that would prevail here was introduced by Spekkens in~\cite{Spekkens05}.
There, it is justified that any detail of the preparation procedure $\prepproc$ which is not reflected in the density matrix $\rho(\prepproc)$ is part of the context.
The corresponding assumption of noncontextuality is that the noncontextual ontic state distribution only depends on the density matrix $\rho(\prepproc)$: thus, we make the replacement
\begin{equation}
\probcond{\hv}{\prepproc} \rightarrow \probcond{\hv}{\rho(\prepproc)}.
\end{equation}
For example, in the case of a mixed quantum state, the ontic state distribution associated with that quantum state does not depend on which ensemble decomposition the mixed state may have originated from.
Another example is the case where a mixed state originated from the partial trace of a pure entangled state on a larger Hilbert space: the ontic state distribution does not distinguish among the different purifications.

In the setup considered here, the only states available are in the set $\sets$, so that it would be reasonable to require that there exists a valid ontic state distribution $\probcond{\hv}{\rho}$ for any $\rho\in\sets$, without requiring anything else for the other quantum states in $\states \setminus \sets$.
However, in the non-tomographically complete case (\cref{def:tomo}), we argue that this is still too permissive given that the only measurements available are those taken out of the set $\sete$.
Indeed, it is clear from \cref{prop:reducedscalar} that any detail of the preparation procedure that is reflected in $\rho\in\sets$ but that is not reflected in the reduced density matrix $\proj{\mainr}{\rho}$ will not be resolved by the available measurement resource $\sete$ and is thus part of a context.
Our notion of noncontextuality\vgrSC{, following the guiding principles of~\cite{Spekkens05},} is that the ontic state distribution only depends on $\proj{\mainr}{\rho}$, i.e., we make the further replacement
\begin{equation}
\probcond{\hv}{\rho} \rightarrow \probcond{\hv}{\proj\mainr\rho}.
\end{equation}
The motivation for this requirement will be discussed in \cref{subsection:discussion of gshv}.
The conventional label for the ontic state distribution is $\mu$~\cite{Spekkens05}: for all $\hv\in\ospace$, for all $\rho\in\sets$,
\begin{equation}
\ostate{\proj{\mainr}{\rho}}{\hv} := \probcond{\hv}{\proj{\mainr}{\rho}}.
\end{equation}
This means that $\mu$ has the following domain:
\begin{subequations}
\begin{equation}
\label{eq:mumapping}
\mu\funcrange{\reds\times \ospace}{\mathbb R}.
\end{equation}
The normalization and nonnegativity of the probability distributions read
\begin{align}
\forall \red\rho\in\reds\st&& \inthv{\ostate{\red\rho}{\hv}} &= 1, \label{eq:NormMu}\\
\forall\hv\in\ospace,\forall \red\rho\in\reds\st&& \ostate{\red\rho}{\hv} &\geq 0. \label{eq:basicmupos}
\end{align}
It is also reasonable to require that the ontic state distribution mapping represents classical probabilistic mixtures of quantum states by classical probabilistic mixtures of ontic states. This is formulated as a convex-linearity requirement of the form:
\begin{multline*}
\forall\hv\in\ospace,\forall p\in[0,1],\forall \red\rho_1,\red\rho_2\in\reds\st \\
\end{multline*}
\vspace{-1.4cm}
\begin{multline}
\ostate{p\red\rho_1 + (1-p)\red\rho_2}{\hv} \\
= p \ostate{\red\rho_1}{\hv}
+ (1-p)\ostate{\red\rho_2}{\hv}. \label{eq:convexlinearityrho}
\end{multline}
\end{subequations}

\subsubsection{Noncontextual measurement representation}

As previously stated, the response function distribution $\probcond{\measproc_k}{\hv}$ could in principle depend on all operational details of $\measproc$.
The notion of noncontextuality that would prevail here~\cite{Spekkens05} would be that the response function distribution does not depend on more than the POVM $\{E_k(\measproc_k)\}_k$ associated to the measurement procedure $\measproc$. 
Thus, we make the replacement
\begin{equation}
\label{eq:ResponseFunctionPOVM}
\probcond{\measproc_k}{\hv} \rightarrow \probcond{E_k(\measproc_k)}{\hv}.
\end{equation}
This is motivated by the fact that in quantum mechanics, two distinct measurement procedures which lead to the same POVM  are equivalent with respect to the statistics that are produced upon measuring any state. Equation \eqref{eq:ResponseFunctionPOVM} implies that the response function does not resolve whether a POVM originated from a coarse-graining of a finer POVM; nor does it resolve whether the POVM originated from tracing out the result of a projective measurement on a larger Hilbert space.

In our setup where all available POVM elements belong to $\sete$, it is reasonable to require that there exists a valid response function $\probcond{E}{\hv}$ for all $E\in\sete$,
irrespective of what is predicted for other quantum effects in $\effects\setminus\sete$.
This is however too permissive: consider distinct quantum effects $E_1,E_2\in\sete$. Given the set $\sets$ and \cref{prop:reducedscalar}, it could be that $\proj{\mainr}{E_1} = \proj{\mainr}{E_2}$ so that the effects are indistinguishable in this setup.
Thus, the part of a quantum effect $E\in\sete$ which is not reflected in $\proj{\mainr}{E}$ is part of a new kind of context, and we make the further replacement
\begin{equation}
\probcond{E}{\hv} \rightarrow \probcond{\proj{\mainr}{E}}{\hv},
\end{equation}
which is, again, nontrivial for non-tomographically complete scenarios (\cref{def:tomo}), and will be further commented in \cref{subsection:discussion of gshv}.
The mapping that associates a response function to each quantum effect is denoted $\xi$, following the notation of~\cite{Spekkens05}: for all $\hv\in\ospace$, for all $E\in\sete$,
\begin{equation}
\orep{\proj{\mainr}{E}}{\hv} :=
\probcond{\hv}{\proj{\mainr}{E}}.
\end{equation}
The domain of $\xi$ is then:
\begin{subequations}
\begin{equation}
\label{eq:ximapping}
\xi\funcrange{\rede\times\ospace}{\mathbb R}.
\end{equation}
The explicit normalization and nonnegativity are imposed as follows:
\begin{multline}
\forall \hv\in\ospace,\forall K\in\mathbb N\cup\{+\infty\}, \\
\forall \left\{ E_k\in\sete\st\textstyle{\sum_{k=1}^K E_k} = \id\hil\right\}_{k=1}^K: \\
\textstyle \sum_{k=1}^K \orep{\proj{\mainr}{E_k}}{\hv} = 1, \label{eq:explicitnormalizationxi}
\end{multline}
\vspace{-0.7cm}
\begin{align}
\label{eq:basicxipos}
&\forall\hv\in\ospace,\forall \red E\in\rede\st&&&&&&&& \orep{\red E}{\hv} \geq 0.
\end{align}
In addition to the properties already specified, the response function mapping should represent classical probabilistic mixtures of quantum effects as classical probabilistic mixtures of response functions:
\begin{multline*}
\forall\hv\in\ospace,\forall p\in[0,1],\forall \red E_1,\red E_2\in\rede\st \\
\end{multline*}
\vspace{-1.3cm}
\begin{multline}
\label{eq:convexlinearityxi}
\orep{p\red E_1 + (1-p)\red E_2}{\hv} \\
= p \orep{\red E_1}{\hv}
+ (1-p)\orep{\red E_2}{\hv}.
\end{multline}
\end{subequations}

We are now able to formulate our definition of \vgrSC{non-contextual}\vgr{operationally noncontextual} ontological model.

\begin{definition}[\vgr{Operationally noncontextual} ontological model]
\label{def:classicalmodel}
The \vgrSC{classical}\vgr{operationally noncontextual ontological} model for \resources{} is specified as follows. 
Let $\mu$ be the ontic state mapping that has domain \eqref{eq:mumapping} and that satisfies  \eqref{eq:NormMu}, \eqref{eq:basicmupos}, and \eqref{eq:convexlinearityrho}. Let $\xi$ be the response function mapping that has domain \eqref{eq:ximapping} and that satisfies \eqref{eq:explicitnormalizationxi}, \eqref{eq:basicxipos}, and \eqref{eq:convexlinearityxi}.
The \vgrSC{classical}\vgr{operationally noncontextual ontological} model is required to reproduce the statistics that quantum mechanics predicts for the available states and measurements. Using \cref{prop:reducedscalar} to write down the quantum probabilities
and equation \eqref{eq:OnticProbRule} to write down the \vgrSC{probability in the classical model}\vgr{ontological probabilities},
this requirement may be formulated in the reduced space $\mainr$ as follows:
\begin{multline}
\label{eq:consistmuxi}
\forall\red \rho\in\reds,\forall \red E\in\rede\st
\\
\scal{\red\rho}{\red E}{\mainr}
= \inthv{\ostate{\red \rho}{\hv}\orep{\red E}{\hv}}.
\end{multline}
\end{definition}

\subsection{Motivation for the reduced space prescription}
\label{subsection:discussion of gshv}

In the tomographically complete case (\cref{def:tomo}), the definition of the \vgrSC{classical}\vgr{operationally noncontextual ontological} model, \cref{def:classicalmodel}, is standard in the literature~\cite{Spekkens05,Spekkens19,Spekkens08}.
In the non-tomographically complete case, however, the situation is different. In order to construct a noncontextual ontological model, it is crucial to identify the operational equivalences that the model should implement as ontological equivalences.
In the tomographically complete case, compelled by the success and apparent operational completeness of non-relativistic quantum mechanics in its scope of application, the standard prescription is to say that any two systems with the same quantum description should be ontologically equivalent~\cite{Spekkens05}.

An interesting question is to inquire about the case of a system that may be entangled in two distinct joint states with an ancilla, yet whose reduced density matrix is identical in both instances of the joint state.
These two states are clearly not ``globally" operationally equivalent, but the two states restricted to the system of interest are.
Hence, noncontextuality asks for the ontological description, restricted to the system of interest, to be identical in these two cases. The operation to restrict to the system of interest is the partial trace. 
We argue that projecting states and effects over the reduced space (as in \cref{def:classicalmodel}) is equally well motivated as taking the partial trace of states or effects for the purpose of restricting the focus to the system of interest.
The main point of tension about this claim lies in the ontology that one attributes to the notion of a ``system", whether described by a reduced space $\mainr$ or a Hilbert space $\hil$: indeed, it seems intuitive that the Hilbert space $\hil$ describes a ``real" system, at least in some cases, while the reduced space is just a description of an effective system. For instance, an effective qubit Hilbert space that corresponds to some 2-dimensional subspace of a larger Hilbert space, say, that of a large quantum computer, is perhaps not the best example of a ``real qubit system". Instead, one has in mind, as an example of a ``real qubit", a single spin-1/2, for instance that of a free electron.\footnote{For argument's sake, we forget about the electron's position and momentum.}
However, we argue that there is in fact no consistent operational way to distinguish a ``real system" from an ``effective" one: operationally, a system can only be defined through the specification of the set of operations that an agent is considering.
Of course, certain systems have been found to be more typical or easier to isolate than others, but typicality is arguably subjective and isolation is not fundamental --- one can simply think of vacuum fluctuations continuously happening around a tentatively isolated system. Ultimately, the cut between ``the system" and ``the rest" is nothing more but a specification of the relevant operational procedures that the agents restrict themselves to consider.
In fact, we give in \app~\ref{subsec:example partial trace} an example construction where the reduced space projection and a partial trace in a suitably chosen tensor product structure are equivalent.

Thus, using this premise that all systems are defined only effectively through the specification of the relevant agent operations, the motivation for not allowing an ontological model to depend on the system's surroundings can be equally well applied upon defining the system through some reduced space projection or some partial trace, depending on the type of prepare-and-measure scenario under consideration.
This motivation can be justified, for instance, from desiring a minimal ontological description.
Alternatively, if one embeds the system (whether a reduced space or the special case of a full Hilbert space system) within a larger system, for instance by repeating the experiment in a different lab without affecting the measurement statistics,
it is natural to expect the ontological description to be invariant under these ``irrelevant" modifications.
Of course, one may reject this motivation both for tomographically complete and non-tomographically complete scenarios,
at the expense of having to have a prescription for the extent of the information upon which a \emph{contextual} ontological model would depend.
One could be worried that the mathematical definition of the reduced space $\mainr$ (\cref{def:mainr}) bears some arbitrariness with respect to the above motivation: in \cref{sec:altr}, we describe more general reduced space constructions and the equivalence between any such constructions.

Independently of the above arguments, a minimal motivation for the reduced space prescription is the following:
in the general framework of~\cite{Spekkens18}, the user of the framework may incorporate arbitrary sets of operational equivalences in the noncontextual ontological model, as long as they are compatible with the statistics of the prepare-and-measure scenario. The reduced space prescription essentially amounts to incorporating the maximal set of operational equivalences. 
When viewed from this perspective, the reduced space prescription induces a strengthening of the classicality requirements compared with a certain class of ``minimal" Spekkens noncontextual ontological models when the prepare-and-measure scenario is not tomographically complete, as is formalized in the following proposition. 
A proof is given in \app~\ref{app:mainr}.

\begin{restatable}{prop}{PropBadClassicalModel}
\label{prop:badclassicalmodel}
We let a \emph{minimal Spekkens noncontextual ontological model} for \resources{} be defined as in \cref{def:classicalmodel} but under the replacement of
\begin{subequations}
\label{eq:standard NCHV}
\begin{equation}
\hspace{1cm}P_\mainr\funcrange{\lh}{\mainr\ (\subseteq\lh)}
\end{equation}
with
\begin{equation}
\id{\lh}\funcrange{\lh}{\lh}.
\end{equation}
\end{subequations}
In particular, the ontic state mapping $\mu$ and response function mapping $\xi$, in this minimal Spekkens noncontextual ontological model, have as quantum arguments $\rho\in\sets$ and $E\in\sete$, respectively, instead of $\proj{\mainr}{\rho}\in\proj{\mainr}{\sets}$ and $\proj{\mainr}{E}\in\proj{\mainr}{\sete}$.
Then, it holds that
\begin{myitem}
\item any \resources{} that admits \vgrSC{a classical}\vgr{an operationally noncontextual ontological} model (\cref{def:classicalmodel})
also admits a minimal Spekkens noncontextual ontological model;
\item there exists \resources{} that admits a minimal Spekkens noncontextual ontological model but no \vgrSC{classical}\vgr{operationally noncontextual ontological} model (\cref{def:classicalmodel}).
\end{myitem}
\end{restatable}

Furthermore, it is also possible to prove that
two scenarios \resources{} and $(\tilde\sets,\tilde\sete)$ that produce the same operational statistics are equivalent with respect to operationally noncontextual ontological models.
This proves that it does not matter whether e.g.\ one chooses to use a larger Hilbert space including an ancilla so that the states are purified and/or the measurements are implemented as projective measurements in accordance with Naimark's theorem. The proof is given in \app~\ref{app:BasicCriterion}. 
Note that such a statement does not necessarily hold in standard Spekkens noncontextual ontological models~\cite{spekkens_status_2014}.

\begin{restatable}{prop}{PropAncilla}
\label{prop:ancilla}
Let $\hil$ and $\tilde\hil$ denote two \vgrSC{(potentially different)}
finite dimensional Hilbert spaces. Let $I$ be a discrete or continuous range of indices. Consider two scenarios $(\sets,\sete)$ and $(\tilde\sets,\tilde\sete)$ that take the form
\begin{subequations}
\begin{align}
\sets &= \myconv{\{\rho_k\in\lh\}_{k\in I}}, \\
\sete &= \myconv{\{E_k\in\lh\}_{k\in I}},
\end{align}
\end{subequations}
and
\begin{subequations}
\begin{align}
\tilde\sets &= \textup{conv}\big(\{\tilde\rho_k\in\linherm{\tilde\hil}\}_{k\in I}\big), \\
\tilde\sete &= \textup{conv}\big(\{\tilde E_k\in\linherm{\tilde\hil}\}_{k\in I}\big),
\end{align}
\end{subequations}
both assumed to satisfy \cref{def:available states,def:available effects}, and that satisfy, for all $k,l\in I$,
\begin{equation}
\tr{\hil}{\rho_k E_l} = \textup{Tr}_{\tilde\hil}\big[\tilde \rho_k \tilde E_l\big].
\label{eq:ancillastats}
\end{equation}
Then, these two scenarios define reduced spaces $\mainr = \proj{\spane}{\spane}$ and $\tilde\mainr = P_{\textup{span}(\tilde\sete)}\big(\textup{span}(\tilde\sets)\big)$ that have the same dimension and the scenario \resources{} admits \vgrSC{a classical}\vgr{an operationally noncontextual ontological} model if and only if the scenario $(\tilde\sets,\tilde\sete)$ does.
In fact, if one of the two scenarios admits \vgrSC{a classical}\vgr{an operationally noncontextual ontological} model with ontic space $\ospace$, so does the other with the same ontic space $\ospace$.
\end{restatable}

We will return to the computational equivalence of working with either version of the prepare-and-measure scenario in \cref{prop:ancillacomp}.

\subsection{Structure of the \vgr{operationally noncontextual ontological} model}

Let us now use the properties of the \vgrSC{classical}\vgr{operationally noncontextual ontological} model to derive basic results related to its structure which will be useful for our later endeavors. The following proposition is proven in \app~\ref{app:LinExt}, and is motivated by the analysis of the no-go theorem developed in~\cite{Spekkens08}.

\begin{restatable}{prop}{PropExtensions}
\label{prop:extensions}
Let $\hv\in\ospace$ be arbitrary. Starting from the convex-linear mappings
\begin{subequations}
\begin{align}
\ostate{\cdot}{\hv}\funcrange{\reds}{\realpos}, \\
\orep{\cdot}{\hv}\funcrange{\rede}{\realpos},
\end{align}
\end{subequations}
there exist unique linear extensions
\begin{subequations}
\begin{align}
\ostate{\cdot}{\hv}\funcrange{\mainr}{\mathbb R},\\
\orep{\cdot}{\hv}\funcrange{\mainr}{\mathbb R}.
\end{align}
\end{subequations}
\end{restatable}

Following~\cite{Spekkens08}, we may apply Riesz' representation theorem, stated in \app~\ref{app:LinExt}, for any fixed $\hv\in\ospace$ to obtain that there exist unique $F(\hv)\in\mainr$, $\sigma(\hv)\in\mainr$ such that for all $\hv\in\ospace$, $r\in\mainr$:
\begin{subequations}
\label{eq:OnticScalarReprs}
\begin{align}
\ostate{r}{\hv} &= \scal{r}{F(\hv)}{\mainr} \\
\orep{r}{\hv} &= \scal{\sigma(\hv)}{r}{\mainr}.
\end{align}
\end{subequations}
We will express the nonnegativity requirements of \eqref{eq:basicmupos} and \eqref{eq:basicxipos} using the notion of the polar convex cone.
\begin{definition}[Polar convex cone]
\label{def:Polar}
For any real inner product space $\mathcal V$ of finite dimension, for any $X\subseteq \mathcal V$, the \emph{polar} convex cone\footnote{See \app~\ref{app:convex} for definitions of convex and conic sets.} $\polar{X}{\mathcal V}$ is defined as
\begin{equation}
\polar{X}{\mathcal V} := \big\{y\in\mathcal V\st \forall x\in X,\ \scal{x}{y}{\mathcal V}\geq 0\big\}.
\end{equation}
\end{definition}
We may now formulate the following theorem which links the existence of \vgrSC{a classical}\vgr{an operationally noncontextual ontological} model to the existence of specific mathematical primitives. The proof is presented in \app~\ref{app:BasicCriterion}. Such a representation is a generalization of the frame representation of quantum mechanics introduced in~\cite{Ferrie08}.
\begin{restatable}
{theorem}{ThBasicCriterion}
\label{prop:basiccriterion}
Given \resources{} that lead to the reduced space $\mainr$ (\cref{def:mainr}), there exists \vgrSC{a classical}\vgr{an operationally noncontextual ontological} model with ontic state space $\ospace$ (\cref{def:classicalmodel}) if and only if there exist mappings $F$, $\sigma$ with ranges
\begin{subequations}
\label{eq:simpleranges}
\begin{align}
F\funcrange{\ospace}{\polar{\reds}{\mainr}}, \label{eq:RangeF}\\
\sigma\funcrange{\ospace}{\polar{\rede}{\mainr}}, \label{eq:RangeSigma}
\end{align}
\end{subequations}
satisfying the normalization condition
\begin{equation}
\forall \hv\in\ospace\st 
\otrace{\sigma(\hv)}  =  1 \label{eq:normsigma}
\end{equation}
as well as the consistency requirement: for all 
$ r,s\in\mainr,$
\begin{equation}
\label{eq:consistency}
\scal{r}{s}{\mainr} = \inthv{\scal{r}{F(\hv)}{\mainr}
\scal{\sigma(\hv)}{s}{\mainr}}.
\end{equation}
\end{restatable}

This theorem will in particular prove useful to determine the \unitsep{} criterion in the next section. For completeness, as proven in \app~\ref{app:BasicCriterion}, we have the alternative expressions $\polars = \mainr\cap(\bigpolar{\sets}{\lh})$, and $\polare = \mainr\cap(\bigpolar{\sete}{\lh})$. 
Furthermore, equation \eqref{eq:normsigma} is equivalent to a trace condition since $\otrace{\sigma(\hv)} = \tr{\hil}{\sigma(\hv)}$.

We are now in a position to easily give a prescription for including or not coarse-grained effects in the set $\sete$: namely, it does not make any difference. A proof is given in \app~\ref{app:BasicCriterion}.

\begin{restatable}[Incoherent coarse-grainings]{prop}{PropCoarseGrainings}
\label{prop:coarsegrainings}
Consider any prepare-and-measure scenario \resources{}. Suppose that there exist $\{E_k\in\sete\}_{k=1}^N$ where $N \in \mathbb N \cup \{+\infty\}$ such that $\sum_{k=1}^N E_k \leq \id\hil$ and the sum $\sum_{k=1}^N E_k$ may not be in $\sete$. Then, the scenario \resources{} and the extended scenario
\begin{equation}
\Big(\sets,\sete_{\textup{ext}} := 
\textup{conv}\big(\sete\cup\big\{\textstyle\sum_{k=1}^N E_k\big\}\big)\Big)
\end{equation}
define the same reduced space $\mainr = \proj{\spane}{\spans} = \proj{\myspan{\sete_{\textup{ext}}}}{\spans}$ and same polar cone $\polare = \polar{\proj{\mainr}{\sete_\textup{ext}}}{\mainr}$.
\end{restatable}

This proves that the scenarios \resources{} and $(\sets,\sete_\textup{ext})$ are completely equivalent as inputs to \cref{prop:basiccriterion}. In particular, any \vgrSC{classical}\vgr{operationally noncontextual ontological} model for any one of these two scenarios is valid for the other too.
Furthermore, one can easily apply \cref{prop:coarsegrainings} recursively to account for the inclusion of multiple coarse-grained effects.

\section{\Unitsep{} and cardinality bounds}
\label{sec:ExistenceClassicalModel}

In this section, we will derive a more powerful criterion, referred to as \unitsep, for the existence of \vgrSC{a classical}\vgr{an operationally noncontextual ontological} model.
This criterion was inspired by the no-go theorems of~\cite{Spekkens08,Ferrie08}.

\subsection{Mathematical preliminaries}

We now introduce the mathematical tools that will be needed to state our results.
The main two notions are a notion of generalized separability
as well as a generalized \choijamiolkowsky{} isomorphism.

\subsubsection{Generalized separability}

Consider the tensor product space $\mainr\otimes\mainr$ with $\mainr$ as in \cref{def:mainr}. It is a real inner product vector space | its inner product is defined for product operators as follows: for all  $a,b,x,y\in\mainr$,
\begin{equation}
\label{eq:ScalarTensor}
\scal{a\otimes b}{x\otimes y}{\mainr\otimes\mainr}
:= \scal{a}{x}{\mainr}\scal{b}{y}{\mainr}.
\end{equation}
To obtain the complete inner product, extend this expression by linearity.
We now define the minimal tensor product of two sets as:
\begin{definition}
\label{def:minimal tensor product}
Given any two sets $X,Y\subseteq \mainr$, the minimal tensor product set $X\oprod Y\subseteq \mainr\otimes\mainr$ is defined as
\begin{equation}
X\oprod Y := \{x\otimes y\st x\in X, y\in Y\}.
\end{equation}
\end{definition}
This allows to define the two following sets which are of primordial importance:
\begin{definition}[Generalized product operators]
\label{def:prodes}
The set of generalized product operators is defined to be
\begin{equation}
\prodes := \polars \oprod \polare.
\end{equation}
\end{definition}

Recall that the convex hull $\myconv{X}$ of a set $X$ is the set of all convex combinations of finitely many elements of $X$, as defined in \app~\ref{app:convex}.

\begin{definition}[Generalized separable operators]
\label{def:sepes}
The set of generalized separable operators is defined to be
\begin{equation}
\sepes := \myconv{\prodes}.
\end{equation}
\end{definition}

Referring to the definitions introduced in \app~\ref{app:convex}, $\prodes$ is a cone and $\sepes$ is a convex cone. More details on the structure of $\prodes$ and $\sepes$ are presented in \app~\ref{app:sepes}.

\subsubsection{\choijamiolkowsky{} isomorphism}

We will make use of a simple generalization of the \choijamiolkowsky{} isomorphism~\cite{Jamiolkowsky72}.
Let $\lin\mainr$ be the space of linear maps from $\mainr$ to $\mainr$. The \choijamiolkowsky{} isomorphism maps each linear map in $\lin\mainr$ to an element of $\mainr\otimes\mainr$.

\begin{definition}
\label{def:Choi}
For any $\Phi\in\lin\mainr$, let $\choi\Phi\in\mainr\otimes\mainr$ be the \choijamiolkowsky{} operator defined uniquely by the relations
\begin{equation}
\label{eq:ChoiRelation}
\forall r,s\in\mainr\st 
\scal{r}{\Phi(s)}{\mainr} = \scal{\choi\Phi}{r\otimes s}{\mainr\otimes\mainr}.
\end{equation}
\end{definition}

The proof of uniqueness, of bijectivity and explicit coordinate solutions are derived in \app~\ref{app:choi}. The following lemma is also proven in \app~\ref{app:choi}:

\begin{lemma}
\label{prop:choiid}
For any orthonormal basis of $\mainr$ $\{R_i\in\mainr\}_{i=1}^{\dim(\mainr)}$\emph{:}
\begin{equation}
\choi{\id\mainr} = \sum_{i=1}^{\dim(\mainr)} R_i\otimes R_i.
\end{equation}
\end{lemma}

\subsection{The \unitsep{} criterion}

Starting from \cref{prop:basiccriterion}, we may now derive an alternative criterion for the existence of \vgrSC{a classical}\vgr{an operationally noncontextual ontological} model. First, we make an assumption for the types of ontic spaces that we are considering.

\begin{definition}[\vgr{Riemann integrability}]
\label{def:RiemannIntegrableModel}
\vgrSC{A classical}\vgr{An operationally noncontextual ontological} model with ontic space $\ospace$ as introduced in \cref{def:classicalmodel} is \emph{Riemann integrable} if and only if there exist
\begin{subequations}
\begin{align}
\Delta&\funcrange{\mathbb N\times\mathbb N}{\realpos}, \label{eq:RangeDelta} \\
\hvdis&\funcrange{\mathbb N\times\mathbb N}{\ospace},
\end{align}
\end{subequations}
such that
\begin{multline}
\label{eq:RiemannSum}
\forall \red\rho\in\reds,\forall \red E\in\rede\st
\inthv{\ostate{\red\rho}{\hv}\orep{\red E}{\hv}} \\
=
\lim_{N\rightarrow\infty}
\sum_{k=1}^N \Delta_{N,k}\,
\mu\big(\red\rho,\hvdis_{N,k}\big)
\xi\big(\red E,\hvdis_{N,k}\big).
\end{multline}
\end{definition}
$N$ can be thought of as being a number of subsets that form a discrete partition of $\ospace$, while $k$ is a discrete index running over all such subsets and $\hv^{\text{(dis)}}\in\ospace$ is a value in that subset of $\ospace$.
Riemann integrable \vgrSC{classical}\vgr{operationally noncontextual ontological} models include in particular:
\begin{myitem}
\item \vgrSC{classical}\vgr{operationally noncontextual ontological} models equipped with discrete, finite ontic spaces $\ospace$, which means that $\ospace$ is isomorphic to  $\{1,\dots,N\}$ for some $N\in\mathbb N$;
\item \vgrSC{classical}\vgr{operationally noncontextual ontological} models equipped with discrete, countable infinite ontic spaces $\ospace$, which means that $\ospace$ is isomorphic to $\mathbb N$;
\item \vgrSC{classical}\vgr{operationally noncontextual ontological} models equipped with a continuous ontic space $\ospace$ isomorphic to $\mathbb R^d$ for some $d\in\mathbb N$ such that for all $\red\rho\in\reds$, $\red E\in\rede$, the real function $\ostate{\red\rho}{\cdot}\orep{\red E}{\cdot}\funcrange{\mathbb R^d}{\realpos}$ is Riemann integrable. Such operationally noncontextual ontological models are reasonable physically because they may be seen as describing a system with finitely many continuous degrees of freedom such as position and momentum of finitely many particles.
\end{myitem}

\begin{restatable}[Main theorem: \unitsep]{theorem}{ThMainCriterion}
\label{th:maincriterion}
The prepare-and-measure scenario \resources{} admits a Riemann integrable \vgrSC{classical}\vgr{operationally noncontextual ontological} model (\cref{def:RiemannIntegrableModel})  if and only if:
\begin{equation}
\label{eq:ChoiInSepes}
\choi{\id\mainr} \in\sepes.
\end{equation}
\end{restatable}

This formulation is useful because it allows one to derive properties of the \vgrSC{classical}\vgr{operationally noncontextual ontological} model when it exists: the main theoretical application is described in \cref{sec:cardinality} where the cardinality of the ontic space $\card\ospace$ is shown to be constrained by the dimension of the reduced space $\mainr$.
Note that $\choi{\id\mainr}$ is easy to compute, whereas $\sepes$ is harder to characterize. Still, well-known algorithmic results from convex analysis make the \unitsep{} criterion decidable as described in \cref{sec:algo}.

\begin{proof}[Proof overview]
The complete proof is given in \app~\ref{app:criterion}. In one direction, the goal is to show that if there exists \vgrSC{a classical}\vgr{an operationally noncontextual ontological} model for \resources, then the ontic mappings $F$ and $\sigma$ from \cref{prop:basiccriterion} satisfy:
\begin{equation}
\label{eq:integralhv}
\choi{\id\mainr} = \inthv{F(\hv)\otimes\sigma(\hv)}.
\end{equation}
The assumption of Riemann integrability allows to prove that \eqref{eq:integralhv} implies $\choi{\id\mainr}\in\sepes$.

For the other direction, the idea is to show that if $\choi{\id\mainr}\in\sepes$ holds, then there exists a decomposition of the form
\begin{equation}
\choi{\id\mainr} = \sum_{i=1}^n F_i\otimes\sigma_i
\end{equation}
for $F_i\in\polar\reds\mainr$ and $\sigma_i\in\polar\rede\mainr$
which yields a valid Riemann integrable model allowing to compute quantum statistics as follows: for all $\red\rho\in\reds$, $\red E\in\rede$,
\begin{gather}
\scal{\red\rho}{\red E}{\mainr} = 
\sum_{i=1}^n \scal{\red\rho}{F_i}{\mainr}\scal{\sigma_i}{\red E}{\mainr}. \qedhere
\end{gather}
\end{proof}

\subsection{Explicit examples}
\label{subsec:example}

\subsubsection{\vgr{Operationally noncontextual} example}

It is useful to pause and consider an explicit example scenario, which is taken to be the stabilizer rebit scenario as presented in~\cite{Spekkens19}.
We will return to the general case of solving the \unitsep{} criterion in \cref{sec:algo}. 
The Hilbert space is that of a qubit, $\hil \simeq \mathbb{C}^2$, and we define
\begin{equation}
X = \big\{\zerostate,\onestate,\plusstate,\minusstate\big\},
\end{equation}
where $\ket{\pm} = (\ket{0}\pm\ket{1})/\sqrt{2}$. 
The state and effect spaces are
\begin{subequations}
\begin{align}
\sets &:= \myconv{X}, \\
\sete &:= \myconv{\{0_\hil,\id{\hil}\}\cup X},
\end{align}
\end{subequations}
Since $\spans = \spane$ in this case, we have that $\mainr = \spans$, and hence it is sufficient to find an orthonormal basis of $\spans$, which can be taken to be made out of the following three elements:
\begin{subequations}
\begin{align}
R_1 &:= \zerostate, \\
R_2 &:= \onestate, \\
R_3 &:= \frac{1}{\sqrt{2}}(2\plusstate - \zerostate - \onestate).
\end{align}
\end{subequations}
These satisfy $\scal{R_i}{R_j}{\mainr} \equiv \tr{\hil}{R_iR_j} = \delta_{ij}$ for $i,j=1,2,3$. We can now express $\reds$ and $\rede$ as convex hulls of three dimensional vectors of the form $(a,b,c)^T \equiv a R_1 + b R_2 + c R_3 \in\mainr$, through the definition of the set $\bar X$ composed of the four elements
\begin{subequations}
\begin{align}
\proj{\mainr}{\zerostate} &= (1,0,0)^T, \\ 
\proj{\mainr}{\onestate} &= (0, 1, 0)^T, \\ 
\proj{\mainr}{\plusstate} &= \frac{1}{2}(1, 1, \sqrt{2})^T, \\
\proj{\mainr}{\minusstate} &= \frac{1}{2}(1, 1, -\sqrt{2})^T.
\end{align}
\end{subequations}
We then have
\begin{subequations}
\begin{align}
\reds &:= \myconv{\bar X}, \\
\rede &:= \textup{conv}\big(\{\proj{\mainr}{0_\hil} = (0,0,0)^T, \nonumber\\
&\hspace{1.3cm}\proj{\mainr}{\id\hil}=(1,1,0)^T\}\cup \bar X\big).
\end{align}
\end{subequations}
It can be verified that the polar cones are equal to the same convex cone,\footnote{See \cref{def:ConicHull} for the definition of the conic hull, \vgr{denoted $\textup{coni}(\cdot)$}.}
\begin{equation}
\polar{\rede}{\mainr} = \polar{\reds}{\mainr} = \textup{conv}\circ\textup{coni}(\bar Y)
\end{equation}
where $\bar Y = \{y_1, y_2, y_3, y_4\}$ with
\begin{subequations}
\begin{align}
y_1 &= (\sqrt{2}, 0, 1)^T, \\
y_2 &= (\sqrt{2}, 0, -1)^T, \\
y_3 &= (0, \sqrt{2}, 1)^T, \\
y_4 &= (0, \sqrt{2},-1)^T.
\end{align}
\end{subequations}
This allows to write (see \cref{prop:ResolutionSepes} for more details)
\begin{equation}
\sepes = \textup{conv}\circ\textup{coni}\big(\{y_i\otimes y_j\}_{i,j=1}^4\big).
\end{equation}
The Choi state of the identity is
\begin{multline}
\choi{\id\mainr} = (1,0,0,\ 0,1,0,\ 0,0,1)^T \\
\equiv \textstyle\sum_{i=1}^3 R_i\otimes R_i.
\end{multline}
The unit separability criterion is then verified, i.e., $\choi{\id\mainr} \in\sepes$, since it can be verified explicitly that
\begin{equation}
\label{eq:explicitdecomprebit}
\choi{\id\mainr} = \sum_{i=1}^4 \frac{1}{4} y_i \otimes y_i \in \sepes.
\end{equation}
This proves \vgrSC{this scenario to be classical}\vgr{that this scenario admits an operationally noncontextual ontological model}, in agreement to the existing literature~\vgr{\cite{Spekkens19}}. Furthermore, from the explicit decomposition of \eqref{eq:explicitdecomprebit} over four elements of $\prodes$, one obtains the explicit operators $\{F_i = y_i/(2\sqrt{2})\}_{i=1}^4$ and $\{\sigma_i = y_i / \sqrt{2}\}_{i=1}^4$ of \cref{prop:basiccriterion} giving rise to ontic distributions over $\ospace \simeq \{1,2,3,4\}$. For instance, for the $\zerostate$ state,
\begin{multline}
\ostate{\proj{\mainr}{\zerostate}}{\lambda = 1} = \scal{F_{\lambda = 1}}{\proj{\mainr}{\zerostate}}{\mainr} \\
= \begin{pmatrix}\frac{1}{2} \\ 0 \\ \frac{1}{2\sqrt{2}} \end{pmatrix}\cdot\begin{pmatrix}
1 \\ 0 \\ 0 
\end{pmatrix} = \frac{1}{2}.
\end{multline}
Similarly, one obtains the other components so that $\mu(\proj{\mainr}{\zerostate}) \equiv (1/2,1/2,0,0)$, in agreement with~\cite{Spekkens19}.

\subsubsection{\vgr{Operationally contextual} example}
\label{subsec:non-classical example}

The simplest class of scenarios which \vgrSC{are non-classical}\vgr{do not admit an operationally noncontextual ontological model} are those for which $\sets$ and $\sete$ are all states and effects allowed within a given finite-dimensional Hilbert space $\hil \simeq \mathbb{C}^d$. It is clear that $\spans = \spane = \lh$ in this case, so that $\mainr = \lh$. The conical hulls of $\sets$ and $\sete$ are both equal to the set $\lhplus$ of all positive semi-definite matrices within $\lh$. 
The set of semi-definite matrices, as a convex cone, is ``self-dual" in the sense that its polar cone $\polar{\lhplus}{\lh}$ equals $\lhplus$. Thus, we obtain that
\begin{equation}
\polars = \polare = \lhplus,
\end{equation}
and it follows that
\begin{equation}
\sepes = \myconv{\lhplus \oprod \lhplus} = \qsep,
\end{equation}
where $\qsep$ is simply the usual set of unnormalized separable bipartite quantum states. The Choi state of the identity, given $\mainr = \lh$, can be written as follows. First, we choose an arbitrary basis of $\hil \simeq \mathbb{C}^d$ that we write as $\{\ket{j}\}_{j=1}^d$. Then, the orthonormal basis of $\mainr = \lh$ is taken to be the union of elements of the form $\{\ketbra{j}\}_{j}$, $\{(\ketbraa{j}{k}+\ketbraa{k}{j})/\sqrt{2}\}_{j\neq k}$ and $\{(i\ketbraa{j}{k} - i\ketbraa{k}{j})/\sqrt{2}\}_{j\neq k}$. One then obtains, using \cref{prop:choiid},
\begin{equation}
\choi{\id\mainr} = \sum_{ij} \left|i\right>\!\left<j\right| \otimes \left|i\right>\!\left<j\right|.
\end{equation}
Using the partial transpose criterion~\cite{peres}, one can see that $\choi{\id\mainr}$ must be entangled and not separable,\footnote{The partial transpose of $\choi{\id\mainr}$ reads in this case $\sum_{ij} \ketbraa{i}{j}\otimes\ketbraa{j}{i}$ which has a negative eigenvalue for the eigenvector $\ket{1}\otimes\ket{2} - \ket{2}\otimes\ket{1}$.} which proves that $\choi{\id\mainr}\notin \qsep = \sepes$, and hence any such scenario \vgrSC{is non-classical}\vgr{does not admit an operationally noncontextual ontological model}. This argument is mathematically the same as that of~\cite{Ferrie08}, and the same result is proven by slightly different means in~\cite{Spekkens08}.

\subsection{Ontic space cardinality}
\label{sec:cardinality}

In this section, we will show two simple bounds for the cardinality $\card\ospace$ of the ontic state space. For our purposes, it suffices to distinguish two cases: either $\card\ospace < \infty$ which means that $\ospace$ is a finite set consisting of $\card\ospace$ many elements, or $\card\ospace = \infty$ which means that $\ospace$ is countable infinite or uncountable.
Then, one can show a lower and upper bound for the size of the ontic space as in the following theorem. The proof is given in \app~\ref{app:card}.

\begin{restatable}[Ontic space cardinality bounds]{theorem}{ThCardinalityBounds}
\label{th:CardinalityBounds}
For any \resources{} that admits \vgrSC{a classical}\vgr{an operationally noncontextual ontological} model with ontic state space $\ospace$, it holds that either $\ospace$ is an infinite set, or it is discrete and respects
\begin{equation}
\dim(\mainr) \leq  \card\ospace.
\end{equation}
Furthermore, if \resources{} admits a Riemann integrable \vgrSC{classical}\vgr{operationally noncontextual ontological} model (\cref{def:RiemannIntegrableModel}), then there exists \vgrSC{a classical}\vgr{an operationally noncontextual ontological} model for \resources{} with discrete ontic space $\ospace_{\textup{min}}$ which satisfies
\begin{equation}
\dim(\mainr) \leq \card{\ospace_{\textup{min}}} \leq \dim(\mainr)^2 \leq\dim(\lh)^2.\label{eq:bounds on ontic space}
\end{equation}
\end{restatable}
Recall that $\dim(\lh) = \dim(\hil)^2$: the dimension of the quantum Hilbert space thus plays an important role in determining the maximal cardinality of the ontic space.
Furthermore, \cref{prop:coarsegrainings,prop:ancilla} together with \cref{th:CardinalityBounds} prove the bounds on the number of ontic space and the minimal ontic space cardinality to be invariant under the addition of incoherent coarse-grainings in the scenario, and also under a shift of representation of the operational (quantum) primitives.

\paragraph*{Addendum.} After the first version of this work was released, related bounds on the number of ontic states in ontological models of operational theories were produced under different assumptions. The work of~\cite{shahandeh_contextuality_2020} is also concerned with prepare-and-measure scenarios, and our \resources{} would be termed a subGPT there. A stronger upper bound was given in their theorem 3 that can be written as $\card{\ospace} = \dim(\mainr)$: the discrepancy between this bound and that of equation \eqref{eq:bounds on ontic space} comes from additional constraints related to the uniqueness of possible classical models, effectively building up a more constrained alternative to both standard Spekkens noncontextuality and the present \vgrSC{classical}\vgr{operationally noncontextual ontological} model.
In~\cite{schmid_structure_2020}, the standard notion of Spekkens noncontextuality is probed in a framework that includes preparations and measurements but also features a proper treatment of transformations. One key result of their work is that including the requirement of the noncontextual ontological representation of transformations strengthens the bound of \eqref{eq:bounds on ontic space} down to, loosely speaking,\footnote{There would be more to say about tomographic completeness assumptions: we use $\mainr$ as the representative vector space by analogy with our work, but for the proper description see~\cite{schmid_structure_2020}.}  $\card{\Lambda} = \dim(\mainr)$.

\subsection{Alternative reduced spaces}
\label{sec:altr}

We have defined the reduced space in \cref{def:mainr} as
\begin{equation}
\mainr = \proj{\myspan{\sete}}{\myspan{\sets}}.
\end{equation}
However, one could ask whether an alternative definition of the reduced space would preserve the same physical motivation for the \vgrSC{classical}\vgr{operationally noncontextual ontological} model while implying a possibly distinct notion of classicality for the prepare-and-measure scenario \resources.
Such an alternative definition could for instance be obtained from swapping the roles of $\sets$ and $\sete$ in the definition of $\mainr$, thus leading to a potential alternative reduced space $\proj{\spans}{\spane}$.

In this section, we will define and motivate a generalized class of reduced spaces from which one can construct alternative \vgrSC{classical}\vgr{operationally noncontextual ontological} models, and prove that the corresponding notions of classicality are all equivalent. To start with, consider the following class of reduced spaces:
\begin{definition}
\label{def:AltReducedSpace}
An alternative reduced space is any real, finite dimensional inner product space $\altr\mainr$ together with two linear maps $f,g\funcrange{\lh}{\altr\mainr}$ that satisfy 
\begin{subequations}
\begin{align}
\label{eq:AltScal}
\nonumber \forall \rho\in\sets,\forall E\in\sete\st & \\
\scal{\rho}{E}{\lh} &= \scal{f(\rho)}{g(E)}{\altr\mainr}\!, \\
\myspan{f(\sets)} &= \altr\mainr, \label{eq:spanfs}\\
\myspan{g(\sete)} &= \altr\mainr. \label{eq:spange}
\end{align}
\end{subequations}
\end{definition}

The fact that both maps $f,g$ have their image in the same vector space allows one to preserves the symmetry between the treatment of states and effects. The real inner product structure of any $\altr\mainr$ is a simple mathematical choice. We will return to the validity of the choice of finite dimensionality of $\altr\mainr$ later.
Equation  \eqref{eq:spange} is motivated by the fact that for any $\rho\in\sets$, the probabilities $\{\scal{\rho}{E}{\lh} : E\in\sete\}$ do not necessarily fully determine $\rho$.
On the other hand, with equation \eqref{eq:spange} and the non-degeneracy of the inner product at hand,
the probabilities $\{\scal{f(\rho)}{g(E)}{\altr\mainr}: E\in\sete\}$ completely determine $f(\rho)$. Thus, $f(\rho)$ is a good primitive to devise a noncontextual model that only resolves the degrees of freedom that are resolved by $g(E)$. This argument can be repeated swapping each $\rho,$ $\sets$ and $f$ with $E$, $\sete$ and $g$ respectively to motivate analogously equation \eqref{eq:spanfs}.
The inner product bilinearity and equations \eqref{eq:spanfs}, \eqref{eq:spange} imply that if \eqref{eq:AltScal} is to hold then $f,g$ have to be linear maps.

Without attempting to fully characterize the set of solutions to \cref{def:AltReducedSpace},
we prove that while $\mainr$ as defined in \cref{def:mainr} is indeed a valid solution to \cref{def:AltReducedSpace}, it is not the only such solution.
The proof is given in \app~\ref{app:altr}.

\begin{restatable}{prop}{PropScalarAlt}
\label{prop:scalaralt}
The choice
\begin{subequations}
\begin{align}
\altr\mainr &:= \mainr = \proj{\spane}{\spans}, \\
f(\cdot) &:=\proj{\mainr}{\cdot}, \\ g(\cdot) &:=\proj{\mainr}{\cdot}
\end{align}
\end{subequations}
yields a valid alternative reduced space in \cref{def:AltReducedSpace}; and so does the swapped version
\begin{subequations}
\begin{align}
\altr\mainr &:= \proj{\spans}{\spane} =: \mainr', \\ 
f(\cdot) &:= \proj{\mainr'}{\cdot}, \\ 
g(\cdot) &:= \proj{\mainr'}{\cdot}.
\end{align}
\end{subequations}
\end{restatable}
We required the dimension of $\altr\mainr$ to be finite: in fact, \cref{def:AltReducedSpace} allows to prove the following proposition, see \app~\ref{app:altr} for a proof.
\begin{restatable}{prop}{PropAltDimensions}
\label{prop:AltDimensions}
It holds that for any reduced space $\altr\mainr$ (\cref{def:AltReducedSpace}),
$
\dim(\altr\mainr) = \dim(\mainr)$ where $\mainr$ is defined in \cref{def:mainr}.
\end{restatable}
Recall that the vector space inclusion $\mainr\subseteq\lh$ bounds the dimension of $\mainr$ and thus also bounds that of any $\altr\mainr$: $\dim(\altr\mainr) \leq \dim(\lh) = \dim(\hil)^2$.

The \vgrSC{classical}\vgr{operationally noncontextual ontological} model is defined for a given alternative reduced space in \app~\cref{def:altclassicalmodel} by analogy to the \vgrSC{classical}\vgr{operationally noncontextual ontological} model formulated for $\mainr$ in \cref{def:classicalmodel}.
It turns out that if one were to use any alternative reduced space, one would derive equivalent results to those already obtained.
In addition, a result that holds formulated in $\mainr$ is usually equivalent to that formulated in any $\altr\mainr$.
Most importantly we have the following equivalence, proven in \app~\ref{app:equivalence}:

\begin{restatable}[Equivalence of reduced spaces]{theorem}{ThEquivalenceReducedSpaces}
\label{th:EquivalenceReducedSpaces}
Given any \resources, consider $\mainr$ and any $\altr\mainr$  constructed from \resources.
There exists \vgrSC{a classical}\vgr{an operationally noncontextual ontological} model with ontic space $\ospace$ constructed on $\mainr$ (\cref{def:classicalmodel}) if and only if
there exists \vgrSC{a classical}\vgr{an operationally noncontextual ontological} model constructed on $\altr\mainr$  (\app~\cref{def:altclassicalmodel}) with the same ontic space $\ospace$.
\end{restatable}

Note that the ontic primitives of the models in $\mainr$ and a given $\altr\mainr$ may be different, in particular they may belong to distinct vector spaces; but the ontic space that underlies the \vgrSC{classical}\vgr{operationally noncontextual ontological} model is the same in either case. The implications of \cref{th:EquivalenceReducedSpaces} are the following:
\begin{myitem}
\item saying that the scenario \resources{} admits \vgrSC{a classical}\vgr{an operationally noncontextual ontological} model is a statement which can be made regardless of which reduced space one chooses to use;
\item in the case of Riemann integrable \vgrSC{classical}\vgr{operationally noncontextual ontological} models (\cref{def:RiemannIntegrableModel}), the generic case is that the ontic space is discrete as stated in \cref{th:CardinalityBounds}. Then, according to \cref{th:EquivalenceReducedSpaces}, any choice of alternative reduced space $\altr\mainr$ will yield ontic spaces of the same cardinality as those of $\mainr$.
\end{myitem}

Our choice to use $\mainr$ rather than another alternative reduced space $\altr\mainr$ is without significance.
Some additional equivalences between the alternative reduced spaces that are relevant for the algorithmic evaluation of the \unitsep{} criterion will be provided in \cref{sec:Computational equivalence of reduced spaces}.

\section{Algorithmic formulation, witnesses and certifiers}

\label{sec:algo}

The content of this section is organized as follows. First, we describe general results from convex analysis and introduce the vertex enumeration problem in \cref{sec:venum}.
Then, we describe general theoretical results that hold for an arbitrary scenario \resources{} in \cref{sec:GeneralAspects}: these results help characterize the set of separable operators $\sepes$ appearing in the \unitsep{} criterion of \cref{th:maincriterion}.
In \cref{sec:PolyhedralCase}, we specialize to the to-be-defined polyhedral scenarios for which the \unitsep{} criterion can be verified exactly.
In \cref{sec:PolyhedralApprox}, we show how to certify the \vgrSC{classicality or witness the non-classicality of}\vgr{existence or non-existence of an operationally noncontextual operational model for} an arbitrary scenario \resources{} using the results of the polyhedral case, and discuss the convergence of the resulting hierarchies of algorithmic tests.
In \cref{sec:Computational equivalence of reduced spaces}, we show the equivalence between the computational complexities of the algorithm formulated in $\mainr$ and that formulated in an alternative reduced space.

\subsection{Vertex enumeration}
\label{sec:venum}

Let us introduce some notation. A review of the main convex analysis definitions is  presented in \app~\ref{app:convex}, and the proofs of the propositions of this section are presented in \app~\ref{app:ConvexConeResolution}.
 
For any finite dimensional real inner product space $\mathcal V$, let $X\subseteq \mathcal V$ be an arbitrary set.
The conic hull $\mycone{X}$ is the set of elements of the form $\lambda x \in\mathcal V$ for any $\lambda\in\realpos$ and $x\in X$. 
A convex cone $\mathcal C\subseteq\mathcal V$ is one that equals its convex hull and also its conic hull: $\mathcal C = \myconv{\mathcal C} = \mycone{\mathcal C}$.
A half-line is the conic hull of a single element of the vector space. An extremal half-line of $\mathcal C$ is a half-line whose elements cannot be expressed as the average of linearly independent elements of $\mathcal C$. 
The set of extremal half-lines of a convex cone $\mathcal C$ is denoted $\coneextr{\mathcal C}$.

\begin{restatable}[Pointed cone]{definition}{DefPointedCone}
\label{def:PointedCone}
Let $\mathcal C\subseteq\mathcal V$ be a convex cone. $\mathcal C$ is said to be a pointed cone if
\begin{myitem}
\item $\mathcal C$ is closed; \label{item:ClosedPointedCone}
\item $\mathcal C \neq \emptyset$ and $\mathcal C\neq \{0\}$; \label{item:NonTrivalPointedCone}
\item there exists a linear function $L\funcrange{\mathcal V}{\mathbb R}$ such that for all $c\in\mathcal C\setminus\{0\}$, $L(c) > 0$. \label{item:LinPointedCone}
\end{myitem}
\end{restatable}

The following proposition guarantees the representation of pointed cones as the convex hull of their extremal half-lines.

\begin{restatable}{prop}{PropResolutionPointedCone}
\label{prop:ResolutionPointedCone}
If $\mathcal C\subseteq \mathcal V$ is a pointed cone, then it holds that
\begin{equation}
\textstyle \mathcal C = \myconv{\bigcup_{\halfl\in\coneextr{\mathcal C}} \halfl}.
\end{equation}
\end{restatable}

We will also need the representation of the polar cone (\cref{def:Polar}) as the convex hull of its extremal half-lines: the following definition and proposition will be useful for that purpose.

\begin{restatable}[Spanning cone]{definition}{DefSpanningCone}
\label{def:SpanningCone}
A convex cone $\mathcal C\subseteq\mathcal V$ is a spanning cone in $\mathcal V$ if
\begin{myitem}
\item $\mathcal C$ is closed; \label{item:SpanningCone(i)}
\item $\mathcal C \neq \mathcal V$;
\label{item:ClosedSpanningCone}
\item $\myspan{\mathcal C} = \mathcal V$.
\label{item:SpanningSpanningCone}
\end{myitem}
\end{restatable}
Notice that the spanning cone property depends on the vector space $\mathcal V$ in which one embeds $\mathcal C$.

\begin{restatable}{prop}{PropResolutionPolarCone}
\label{prop:ResolutionPolarCone}
If $\mathcal C\subseteq\mathcal V$ is a spanning cone, then the polar cone $\polar{\mathcal C}{\mathcal V}\subseteq\mathcal V$ (\cref{def:Polar}) is a pointed cone, which implies by \cref{prop:ResolutionPointedCone} that
\begin{equation}
\textstyle \polar{\mathcal C}{\mathcal V} = \myconv{\bigcup_{\halfl\in\coneextr{\polar{\mathcal C}{\mathcal V}}} \halfl}.
\end{equation}
\end{restatable}

We see that if $\mathcal C\subseteq\mathcal V$ is a spanning pointed cone in $\mathcal V$, both $\mathcal C$ and the polar $\polar{\mathcal C}{\mathcal V}$ may be represented as the convex hull of their respective extremal half-lines. This defines the so-called vertex enumeration problem\footnote{We follow the denomination given in the literature, e.g., in~\cite{Avis2000}, but we define the vertex enumeration problem even if the cone has infinitely many extremal half-lines.}:

\begin{definition}[Vertex enumeration problem]
\label{def:venum}
For $\mathcal C\subseteq \mathcal V$ a spanning pointed cone, the vertex enumeration problem consists in obtaining the extremal half-lines of $\polar{\mathcal C}{\mathcal V}$ from the extremal half-lines of $\mathcal C$. We denote the vertex enumeration map $\venum{\mathcal V}{\cdot}$\textup{:}
\begin{equation}
\venum{\mathcal V}{\coneextr{\mathcal C}} := \coneextr{\polar{\mathcal C}{\mathcal V}}.
\end{equation}
\end{definition}

The last proposition that will prove useful is the following half-space representation of a convex cone, starting from its extremal half-lines\footnote{A half-space $H\subseteq\mathcal V$ is the geometric interpretation of a homogeneous linear inequality solution set $H := \{v\in\mathcal V\st L_H(v)\geq0\}$ where $L_H$ is a linear functional. By Riesz' representation \cref{th:Riesz},
there exists a unique $v_H\in\mathcal V$ such that $L_H(\cdot) = \scal{v_H}{\cdot}{\mathcal V}$. It is then clear that $H = \polar{\{v_H\}}{\mathcal V}=\polar{[\mycone{v_H}]}{\mathcal V}$.
Thus, the polar cone of a half-line is a half-space and conversely.}.

\begin{restatable}{prop}{PropHyperspaceRepresentation}
\label{prop:HyperspaceRepresentation}
A solution to the vertex enumeration problem allows one to represent a spanning pointed cone $\mathcal C\subseteq\mathcal V$ as the intersection of half-spaces:
\begin{equation}
\textstyle \mathcal C = \bigcap_{\halfl\in\venum{\mathcal V}{\coneextr{\mathcal C}}} \polar{ \halfl}{\mathcal V}.\label{eq:vertex intersection hyper}
\end{equation}
\end{restatable}

\subsection{General aspects of the algorithm}
\label{sec:GeneralAspects}

The proofs of the propositions of this section are presented in \app~\ref{app:algo}.

For the purpose of determining the structure of $\sepes$, it turns out that rather than considering the convex sets $\rede$ and $\reds$, the main objects of interest are the convex cones $\mycone{\proj{\mainr}{\closure\sets}}$ and $\mycone{\proj{\mainr}{\closure\sete}}$, where $\closure X$ denotes the closure of $X$. Indeed:

\begin{restatable}{prop}{PropPointedConeExpression}
\label{prop:PointedConeExpression}
\begin{subequations}
\begin{align}
\polar{\reds}{\mainr} &= \polar{[\mycone{\proj{\mainr}{\closure\sets}}]}{\mainr}, \label{eq:PolarPRS} \\
\polar{\rede}{\mainr} &= \polar{[\mycone{\proj{\mainr}{\closure\sete}}]}{\mainr}. \label{eq:PolarPRE}
\end{align}
\end{subequations}
\end{restatable}
These expressions are useful due to the fact that the vertex enumeration problem is well-defined for the sets $\mycone{\proj{\mainr}{\closure\sets}}$ and $\mycone{\proj{\mainr}{\closure\sete}}$:
\begin{restatable}{prop}{PropPointedSpanningPRSE}
\label{prop:PointedSpanningPRSE}
$\mycone{\proj{\mainr}{\closure\sets}}$ and $\mycone{\proj{\mainr}{\closure\sete}}$ are spanning pointed cones in $\mainr$.
\end{restatable}
Together with \cref{prop:PointedConeExpression}, this shows that applying the vertex enumeration map to $\mycone{\proj{\mainr}{\closure\sets}}$ and $\mycone{\proj{\mainr}{\closure\sete}}$ will yield the extremal half-lines of $\polar{\reds}{\mainr}$ and $\polar{\rede}{\mainr}$:
\begin{subequations}
\label{eq:VertexEnumSE}
\begin{align}
\coneextr{\polar\reds\mainr} &= \venum{\mainr}{\coneextr{\mycone{\proj{\mainr}{\closure\sets}}}},  \\
\coneextr{\polar\rede\mainr} &= \venum{\mainr}{\coneextr{\mycone{\proj{\mainr}{\closure\sete}}}}. 
\end{align}
\end{subequations}
Knowing the extremal half-lines of $\polar\reds\mainr$ and $\polar\rede\mainr$, the characterization of $\sepes$ as the convex hull of its extremal half-lines is readily obtained. First consider the following proposition:
\begin{restatable}{prop}{PropSepesSpanningPointed}
\label{prop:SepesSpanningPointed}
$\sepes$ is a spanning pointed cone in $\mainr\otimes\mainr$.
\end{restatable}
This proposition together with \cref{prop:ResolutionPointedCone} guarantees that we may represent $\sepes$ as the convex hull of its extremal half-lines.
%
%
If $\halfl_1$ and $\halfl_2$ are half-lines in $\mainr$, then $\halfl_1\oprod\halfl_2$ is a half-line in $\mainr\otimes\mainr$. The following proposition makes explicit the extremal half-lines of $\sepes$ (recall \cref{def:minimal tensor product}):
\begin{restatable}{prop}{PropResolutionSepes}
\label{prop:ResolutionSepes}
It holds that
\begin{multline}
\coneextr{\sepes} = \left\{
\halfl_1 \oprod \halfl_2\st 
\halfl_1 \in \coneextr{\polar\reds\mainr},\right. \\
\left. \halfl_2 \in \coneextr{\polar\rede\mainr}
\right\}.
\label{eq:sepesextremallines}
\end{multline}
\end{restatable}
Thus, knowing the extremal half-lines of $\polars$ and $\polare$ is equivalent to knowing the extremal half-lines of $\sepes$.

Once the extremal half-lines of $\sepes$ are known, there are two options to determine whether the Choi state $\choi{\id\mainr}$ belongs to $\sepes$. The first, which is rather computationally intensive, is to run the vertex enumeration procedure on $\sepes$ to go from its extremal half-line representation to its supporting half-space representation. It then suffices to verify whether $\choi{\id\mainr}$ belongs to all the corresponding half-spaces. Alternatively, for a better computational efficiency, one may formulate the test of whether $\choi{\id\mainr} \in \sepes$ as a linear program, once the extremal half-lines of $\sepes$ are known. An explicit such linear program computes a value $t^*$ as follows:
\begin{subequations}
\begin{align}
t^* := \max_{t,\{\lambda_i \in \mathbb{R}\}_i} &t \\
\textup{s.t. }
\forall i:\ \lambda_i \geq&\ t, \\
\choi{\id\mainr} =& \sum_i \lambda_i l_i,
\end{align}
\end{subequations}
where $i$ is an index for the extremal half-lines $\{\mathfrak{l}_i\}_i$ of $\sepes$ as in \cref{prop:ResolutionSepes}, and for each $i$ we picked an arbitrary $0 \neq l_i \in \mathfrak{l}_i$.
This linear program is always feasible, since for $t = -\infty$, it amounts to finding a real (not necessarily nonnegative) linear combination of the extremal half-lines of $\sepes$ that equals $\choi{\id\mainr}$, which is always possible since $\sepes$ spans $\mainr\otimes\mainr$ to which $\choi{\id\mainr}$ belongs.
If $t^* \geq 0$, then the solution coefficients $\{\lambda_i\}_i$ of the linear program are nonnegative and they yield a valid conic combination the extremal half-lines of $\sepes$ that equals $\choi{\id\mainr}$. Overall, we that the linear program always terminates with a value $t^*$, and if $t^* < 0$, this proves that the scenario \vgrSC{is non-classical}\vgr{does not admit an operationally noncontextual ontological model}, while $t^* \geq 0$ proves the scenario to \vgrSC{be classical}\vgr{admit an operationally noncontextual ontological model}.\footnote{When $t^* < 0$, the absolute value of $t^*$ could be used as a heuristic, scenario-dependent ``measure" of non-classicality.}

The authors were made aware of the fact that a vertex enumeration-based routine was already proposed in~\cite{elie_algo_2017}, 
allowing to test algorithmically the existence of a different but related classical model.
One important technical difference is that the authors of~\cite{elie_algo_2017} were looking at a reduced set of inequalities, i.e., in the light of the current results, they were looking at a lower dimensional projection of the convex cone $\sepes$. For this reason, \cref{prop:ResolutionSepes} did not apply to their setup, which is why in this case it was crucial to remove the redundant elements out of the set on the right-hand side of equation \eqref{eq:sepesextremallines} to optimize the runtime of the algorithm, as was emphasized in their work.

\subsection{Solvable cases: polyhedral scenarios}
\label{sec:PolyhedralCase}

In this section, we describe the case of polyhedral scenarios, for which the \unitsep{} criterion may be evaluated in finite time.

\begin{definition}[Polyhedral scenarios]
\label{def:PolyhedralScenario}
The prepare-and-measure scenario \resources{} is said to be a \emph{polyhedral scenario} if the convex cones $\mycone{\proj{\mainr}{\closure\sets}}$ and $\mycone{\proj{\mainr}{\closure\sete}}$ have finitely many extremal half-lines:
\begin{subequations}
\label{eq:lptest}
\begin{align}
N_\sets := \card{\coneextr{\mycone{\proj{\mainr}{\closure\sets}}}} < \infty, \\
N_\sete := \card{\coneextr{\mycone{\proj{\mainr}{\closure\sete}}}} < \infty.
\end{align}
\end{subequations}
\end{definition}

A sufficient condition for \resources{} to form a polyhedral scenario is that $\sets$ is the convex hull of finitely many quantum states, and $\sete$ is the convex hull of finitely many quantum effects.
The motivation for the name is that convex cones that are generated by finitely many extremal half-lines are special cases of the well-known polyhedral convex cones~\cite{ConvexAnalysis}.

In the vertex enumeration problem, if $\mathcal C\subset\mathcal V$ is a spanning pointed cone that has finitely many extremal half-lines, i.e., if $\mathcal C$ is polyhedral, then $\polar{\mathcal C}{\mathcal V}$ will have finitely many extremal half-lines, as described in, e.g., section 4.6 of~\cite{ConvexAnalysis}.
Efficient algorithms to solve the vertex enumeration in this case exist in the literature such as the reverse search approach of~\cite{Avis2000}.

Thus, for a polyhedral scenario \resources, the first vertex enumeration problems, i.e., those of equations \eqref{eq:VertexEnumSE}, will each produce a finite number of extremal half-lines. Let there be $M_\sets\in\mathbb N$ extremal half-lines of $\polar\reds\mainr$, and $M_\sete\in\mathbb N$ for $\polar\rede\mainr$. These will form, via \cref{prop:ResolutionSepes}, the $M_\sets\cdot M_\sete$ extremal half-lines of $\sepes$.
The linear program of \eqref{eq:lptest} has $1 + M_\sets\cdot M_\sete$ variables and $\dim(\mainr)^2 + M_\sets\cdot M_\sete$ constraints, which are both finite only for such polyhedral scenarios, and thus, for these polyhedral scenarios the linear program \eqref{eq:lptest} can be implemented.

\subsection{Polyhedral approximations}
\label{sec:PolyhedralApprox}

In the general case where \resources{} is not a polyhedral scenario (\cref{def:PolyhedralScenario}), or where \resources{} is a polyhedral scenario but the runtime of the previous algorithm is prohibitively long due to, e.g., a large number of extremal half-lines, one may still choose any polyhedral inner or outer approximation of the relevant cones, yielding \vgrSC{either classicality certifiers or non-classicality witnesses}\vgr{certificates for either the existence or non-existence of an operationally noncontextual ontological model} as described in the following sections.

\subsubsection{\vgr{Certifying the existence of an operationally noncontextual ontological model}}

First, consider an outer approximation of the input cones: let $\couts,\coute\subseteq\mainr$ be spanning pointed cones (\cref{def:PointedCone,def:SpanningCone}) in $\mainr$ such that
\begin{subequations}
\label{eq:OuterApprox}
\begin{align}
\mycone{\proj{\mainr}{\closure\sets}} &\subseteq \couts, \\
\mycone{\proj{\mainr}{\closure\sete}} &\subseteq \coute,
\end{align}
\end{subequations}
and such that $\card{\coneextr\couts},\card{\coneextr\coute} < \infty$. Such cones always exist: let us give a constructive example. Consider the hyperspace description of $\mycone{\proj{\mainr}{\closure\sets}}$, $\mycone{\proj{\mainr}{\closure\sets}}$ as in \cref{prop:HyperspaceRepresentation}.
If one keeps a finite set of at least $\dim(\mainr)$ hyperspaces, the resulting cones will be spanning pointed
cones outer approximations of $\mycone{\proj{\mainr}{\closure\sets}}$, $\mycone{\proj{\mainr}{\closure\sets}}$ with finitely many extremal half-lines. 
The algorithm described in the previous section may be run in exactly the same way as in the polyhedral case, with $\couts$ and $\coute$ replacing $\mycone{\proj{\mainr}{\closure\sets}}$ and $\mycone{\proj{\mainr}{\closure\sete}}$ as inputs to the algorithm.
Let $\sepin$ be the cone that the algorithm will characterize:
\begin{equation}
\label{eq:InnerSepes}
\sepin := \myconv{\polar{[\couts]}{\mainr}
\oprod
\polar{[\coute]}{\mainr}
}.
\end{equation}
Using \cref{lem:SubPolar} and equations \eqref{eq:OuterApprox}, it can be shown that
\begin{equation}
\label{eq:SepoutSubsetSepes}
\sepin \subseteq \sepes,
\end{equation}
which justifies the reversed superscript of $\sepin$: outer conic approximations in \eqref{eq:OuterApprox} yield an inner approximation of $\sepes$ in \eqref{eq:SepoutSubsetSepes}.
Hence, if the algorithm obtains that $\choi{\id\mainr} \in \sepin$, it also holds that $\choi{\id\mainr}\in\sepes$, and hence this proves the scenario we started with to \vgrSC{be classical}\vgr{admit an operationally noncontextual ontological model}. On the other hand, if $\choi{\id\mainr}\notin \sepin$, the outer approximation is inconclusive.
One may then, for example, use refined polyhedral outer approximations ${\couts}{'}\subset \couts$, $\coute{'}\subset\coute$ that are subsets of the previous ones but still satisfy \eqref{eq:OuterApprox} to obtain a finer inner approximation of $\sepes$, and repeat the procedure. The convergence of this hierarchy of finer approximations will be discussed shortly but let us first describe the outer approximations of $\sepes$.

\subsubsection{\vgr{Certifying the non-existence of an operationally noncontextual ontological model}}
\label{subsec:non-classicality witnesses}

In parallel to attempting to certify \vgrSC{the classicality of}\vgr{the existence of an operationally noncontextual ontological model for} \resources{} by using outer approximations, one may also consider inner approximations to the input cones: choose spanning pointed cones $\cins\!,\, \cine\subseteq\mainr$ such that
\begin{subequations}
\label{eq:InnerApproxCones}
\begin{align}
\cins&\subseteq \mycone{\proj{\mainr}{\closure\sets}}, \\
\cine&\subseteq \mycone{\proj{\mainr}{\closure\sete}},
\end{align}
\end{subequations}
and such that $\card{\coneextr\cins},\card{\coneextr\cine} < \infty$. Such cones always exist. For example, consider the extremal half-line description of $\mycone{\proj{\mainr}{\closure\sets}}$, $\mycone{\proj{\mainr}{\closure\sets}}$ as in \cref{prop:ResolutionPointedCone}.
By keeping a finite set of at least $\dim(\mainr)$ extremal half-lines, the resulting cones will be spanning pointed cones inner approximations of $\mycone{\proj{\mainr}{\closure\sets}}$, $\mycone{\proj{\mainr}{\closure\sets}}$ with finitely many extremal half-lines.
The algorithm may be run in that case as well, with $\cins$, $\cine$ as inputs rather than $\mycone{\proj{\mainr}{\closure\sets}}$, $\mycone{\proj{\mainr}{\closure\sets}}$. Let $\sepout$ be the cone that the algorithm will characterize:
\begin{equation}
\sepout := \myconv{\polar{[\cins]}{\mainr}
\oprod
\polar{[\cine]}{\mainr}}.
\end{equation}
Using \cref{lem:SubPolar} and equations \eqref{eq:InnerApproxCones}, it can be shown that
\begin{equation}
\label{eq:OuterSepes}
\sepes\subseteq \sepout,
\end{equation}
which again justifies the reversed superscripts. This time, if the algorithm concludes that $\choi{\id\mainr} \notin \sepout$, this implies $\choi{\id\mainr} \notin \sepes$ and hence the scenario \vgrSC{is non-classical}\vgr{does not admit an operationally noncontextual ontological model}. If the algorithm concludes that $\choi{\id\mainr} \in\sepout$, the inner approximation is inconclusive and one should use a refined inner approximation in \eqref{eq:InnerApproxCones}.

\subsubsection{Comments on convergence}

It is important to realize that finer and finer approximations will have more and more extremal half-lines, and will yield computationally harder instances of vertex enumeration and linear programming.
The procedure of repeatedly refining the inner or outer approximations will in principle converge to a definite answer provided that $\choi{\id\mainr}$ is in an interior or exterior point of the closed  (\cref{prop:sepclosed}) convex cone $\sepes$.
An alternative approach to refining the polyhedral approximations would be to change the inner or outer approximations randomly while keeping the number of extremal half-lines fixed. This procedure would have the merit of probing more of the structure of \resources{} while keeping the computational complexity fixed, but there is no guarantee for the convergence of this approach.
If $\choi{\id\mainr}$ lies on the boundary of the closed (\cref{prop:sepclosed}) cone $\sepes$, then most inner approximations in the non-polyhedral case will not allow to prove the scenario to \vgrSC{be classical}\vgr{admit an operationally noncontextual ontological model}. However, such cases can reasonably be considered as edge cases since there will exists small deformations of the inputs \resources{} which will make $\choi{\id\mainr}$ an interior or exterior point. Hence, if one is interested in families of prepare-and-measure scenarios, it is most likely that such cases will not occur for most scenarios.

\subsubsection{Connections with quantum entanglement}

The present algorithm may be recast as a basic algorithm to treat the usual problem of verifying the entanglement of a given bipartite state. Let us give the key ideas to relate the two procedures. Let the convex cone of positive semi-definite matrices be $\lhplus\subset \lh$.
The convex cone of unnormalized separable quantum states is $\qsep := \myconv{\lhplus\oprod\lhplus}$.
If a state $\Omega\in\linherm{\hil\otimes\hil}$ belongs to $\qsep$, it is said to be separable, else it is said to be entangled.

To recast the problem of determining whether $\Omega$ is entangled or not to an application of the algorithm described in the previous sections, consider the following main identifications. First, in the previous algorithm, replace $\mainr$ with $\lh$. Then, the input cones $\mycone{\proj{\mainr}{\closure\sets}}$ and
$\mycone{\proj{\mainr}{\closure\sete}}$ are both replaced with $\lhplus$. The tensor product cone $\qsep$ is related to $(\lhplus,\lhplus)$ in the same way that $\sepes$ is related to $(\mycone{\proj{\mainr}{\closure\sets}},
\mycone{\proj{\mainr}{\closure\sete}})$ (see \cref{def:sepes}).
For this identification to work, one needs to recall the basic result that $\bigpolar{\lhplus}{\lh} = \lhplus$. 
Then, the state $\Omega\in\linherm{\hil\otimes\hil}$ replaces $\choi{\id\mainr}$. 
Characterizing whether $\Omega\in\qsep$ can thus be reduced to a non-polyhedral instance of the previous algorithm, due to the infinite number of extremal half-lines of $\lhplus$: the set of extremal half-lines of $\lhplus$ is equal to the set of all half-lines $\mycone{\ketbra\psi}$ with $\ket\psi\in\hil$.
Thus, it makes sense to use inner and outer approximations 
as described in \cref{sec:PolyhedralApprox} to decide the separability of the state.
Due to the complexity of vertex enumeration in the general case, there exist more efficient algorithms in the literature to produce entanglement witnesses such as the SDP hierarchy of~\cite{ImprovedSDPHierarchy}.

\subsection{Computational equivalence: changing reduced spaces and quantum primitives}
\label{sec:Computational equivalence of reduced spaces}

\subsubsection{Changing reduced spaces}

In \cref{sec:altr}, it was shown that the \vgrSC{classicality of}\vgr{existence of an operationally noncontextual ontological model for} \resources{} is a concept that is independent of whether one chooses to work with the initial reduced space $\mainr$ (\cref{def:mainr}) or with any alternative reduced space $\altr\mainr$ (\cref{def:AltReducedSpace}).
The previous sections suggested an algorithmic procedure to verify the \vgrSC{classicality of}\vgr{existence of an operationally noncontextual ontological model for} \resources{} through an evaluation of the unit separability criterion, \cref{th:maincriterion}.
One may ask whether it is simpler to execute this algorithmic procedure when working in $\mainr$ or any other $\altr\mainr$. 
As stated in \cref{sec:PolyhedralCase}, the computational complexity of the algorithm depends on 1) the dimension of the ambient vector space, but by \cref{prop:AltDimensions} these are the same in $\mainr$ and any $\altr\mainr$, and 2) the number of extremal half-lines of the relevant cones and their dual cones. The following propositions will prove the equivalence of number of extremal half-lines of these cones built in $\mainr$ or any other $\altr\mainr$.

\begin{definition}
\label{def:IsoCone}
Given any two finite dimensional real inner product spaces $\mathcal U$, $\mathcal V$ such that $\dim(\mathcal U) = \dim(\mathcal V)$, two convex cones $\mathcal C\subseteq\mathcal U$ and $\mathcal D \subseteq \mathcal V$ are said to be isomorphic, denoted $\mathcal C \sim \mathcal D$, if and only if there exists an invertible linear map $\Phi\funcrange{\mathcal U}{\mathcal V}$ such that
\begin{equation}
\Phi(\mathcal C) = \mathcal D.
\end{equation}
\end{definition}

Applying this definition to the relevant cones in our setup, we obtain:

\begin{restatable}{prop}{PropAltrAlgoEquiv}
\label{prop:AltrAlgoEquiv}
Choosing any alternative reduced space $\altr\mainr$ with associated mappings $f,g$ (\cref{def:AltReducedSpace}), it holds that:
\begin{subequations}
\begin{align}
\mycone{\proj{\mainr}{\closure\sets}}) &\sim \mycone{f(\closure\sets)}, 
\label{eq:IsoPrs}\\
\mycone{\proj{\mainr}{\closure\sete}}) &\sim \mycone{g(\closure\sete)},\label{eq:IsoPre} \\
\sepes &\sim \altr\sepes,  \label{eq:IsoSepes}
\end{align}
\end{subequations}
where
\begin{equation}
\altr\sepes := \myconv{
\bigpolar{f(\sets)}{\altr\mainr}
\oprod
\bigpolar{g(\sete)}{\altr\mainr}
}. 
\end{equation}
\end{restatable}

The following proposition will allow one to assert the computational equivalence of $\mainr$ and $\altr\mainr$:

\begin{restatable}{prop}{PropIsoCones}
\label{prop:IsoCones}
Given any two finite dimensional real inner product spaces $\mathcal U$, $\mathcal V$ such that $\dim(\mathcal U) = \dim(\mathcal V)$, any two convex cones $\mathcal C\subseteq\mathcal U$ and $\mathcal D \subseteq \mathcal V$ such that $\mathcal C\sim\mathcal D$ have the following properties:
\begin{myitem}
\item there is a one-to-one correspondence between the extremal half-lines of $\mathcal C$ and those of $\mathcal D$;
\item the same holds for the extremal half-lines of the polar cones due to $\polar{\mathcal C}{\mathcal U} \sim \polar{\mathcal D}{\mathcal V}$.
\end{myitem}
\end{restatable}

\Cref{prop:AltrAlgoEquiv,prop:IsoCones}, proven in \app~\ref{app:AlgorithmicEquivalence}, prove that all the cones involved in the algorithm that verifies the \unitsep{} criterion will yield vertex enumeration problems of the same complexity because this complexity depends primarily on the number of extremal half-lines as described in \cref{sec:PolyhedralCase}.

\subsubsection{Changing quantum primitives}

We now expand on \cref{prop:ancilla} to show that not only are statistically equivalent scenarios equivalent for the existence of \vgrSC{a classical}\vgr{an operationally noncontextual ontological} model of a certain type,
but also that the computational complexity of verifying the unit separability criterion is the same for either one of the two descriptions. The proof is given in \app~\ref{app:AlgorithmicEquivalence}.

\begin{restatable}{prop}{PropAncillaComp}
\label{prop:ancillacomp}
For the scenarios \resources{} and $(\tilde \sets,\tilde\sete)$ and the corresponding reduced spaces $\mainr$ and $\tilde\mainr$ defined in \cref{prop:ancilla}, it holds that
\begin{align}
\mycone{\proj{\mainr}{\sets}} &\sim 
\mycone{P_{\tilde\mainr}(\tilde\sets)}, \label{eq:ancillacones}\\
\mycone{\proj{\mainr}{\sete}} &\sim 
\mycone{P_{\tilde\mainr}(\tilde\sete)}, \label{eq:ancillaconee}\\
\sepes &\sim 
\textup{Sep}(\tilde\sets,\tilde\sete), \label{eq:ancillaconesepes}
\end{align}
which proves by \cref{prop:IsoCones} the computational equivalence of starting with either \resources{} or $(\tilde \sets,\tilde\sete)$.
\end{restatable}

\section{Connections with generalized probabilistic theories}
\label{sec:Connection}

\subsection{Generalized probabilistic reformulation}
\label{sec:GPT}

Although we formulated the \vgrSC{classical}\vgr{operationally noncontextual ontological} model of \cref{def:classicalmodel} for quantum primitives, the fact that the sets $\sets$, $\sete$ originate from the Hilbert space of the quantum system is not crucial for the \vgrSC{present classical}\vgr{operationally noncontextual ontological} model construction. 

Instead, rather than considering the vector space $\lh$ equipped with the Hilbert-Schmidt inner product, consider any real inner product space $\mathcal V$ of finite dimension:
\begin{subequations}%
\label{eq:GPTsubst}
\vspace{-0.4cm}
\begin{equation}
\lh \rightarrow \mathcal V.
\end{equation}
Then, replace $\sets$ by $\Omega\subseteq\mathcal V$ and $\sete$ by $\mathcal E\subseteq \mathcal V$, following standard notation~\cite{Spekkens19,Janotta_2014}:
\begin{align}
\sets &\rightarrow \Omega, \\
\sete &\rightarrow \mathcal E.
\end{align}
The probability that an effect $E\in\mathcal E$ occurs upon measuring a state $\rho\in\Omega$ is given by the inner product $\scal{\rho}{E}{\mathcal V}$ by analogy with the usual Hilbert-Schmidt inner product probability rule of quantum mechanics.
The properties required for $\Omega$ and $\mathcal E$ that are necessary for the results of this \doc{} are the following. $\Omega$ and $\mathcal E$ must be nonempty, bounded convex sets such that for all $s\in\Omega$ and $e\in\mathcal E$: $\scal{s}{e}{\mathcal V} \geq 0$. It must be that $0\in\mathcal V$ belongs to $\mathcal E$. There must exist $u\in\mathcal E$ such that for all $s\in\Omega$: $\scal{s}{u}{\mathcal V} = 1$ --- this $u$ replaces $\id\hil$:
\begin{equation}
\id\hil \rightarrow u.
\end{equation}
We also require that for all $e\in\mathcal E$, there exists a completion $\{e_k\in\mathcal E\}_k$ such that $e + \sum_k e_k = u$.
\end{subequations}

All the results of this \doc{} can then easily be rederived in this generalized setting: once the prepare-and-measure scenario is defined, the derivations only rely on the axiomatic properties of the state and effect sets together with the basic real, finite-dimensional inner product space structure which are assumed both in the quantum setting and this generalized setting.
As an illustration, the reduced space of \cref{def:mainr} obtained under the substitution \eqref{eq:GPTsubst} is
\begin{equation}
\mainr_{G} := \proj{\myspan{\mathcal E}}{\myspan{\Omega}} \subseteq \mathcal V.
\end{equation}

\subsection{Connections with simplex-embeddability}
\label{sec:SimplexEmbeddability}

Recently, a similar approach to the contextuality of arbitrary tomographically complete (i.e., $\myspan{\Omega} = \myspan{\mathcal E} = \mathcal V$)
prepare-and-measure scenarios was presented in~\cite{Spekkens19}. In this section, we will relate the present results to their work.
For tomographically complete generalized probabilistic theories, the reduced space under the substitution \eqref{eq:GPTsubst} is simply the vector space $\mathcal V$, since
\begin{equation}
\label{eq:TomographicallyCompleteReducedSpace}
\mathcal R_G = \proj{\myspan{\mathcal E}}{\myspan{\Omega}} = \proj{\mathcal V}{\mathcal V} = \mathcal V.
\end{equation}

The definition 1 in~\cite{Spekkens19} (reproduced in \app~\cref{def:Simplex}) of simplex-embeddability, which is formulated only for tomographically complete prepare-and-measure scenarios, turns out to yield a mathematically equivalent criterion to \unitsep{} when the latter is restricted to such tomographically complete prepare-and-measure scenarios
(see \app~\ref{app:connection} for the proof):

\begin{restatable}{prop}{PropSimplexImpliesClassical}
\label{prop:SimplexImpliesClassical}
Any tomographically complete generalized probabilistic theory $(\mathcal V,\Omega,\mathcal E)$ is simplex-embeddable in $d$ dimensions in the sense of definition 1 of~\cite{Spekkens19}, if and only if the tomographically complete prepare-and-measure scenario $(\Omega,\mathcal E)$ admits \vgrSC{a classical}\vgr{an operationally noncontextual ontological} model in the sense of \cref{def:classicalmodel} (under the substitution \eqref{eq:GPTsubst}) with a discrete ontic space of finite cardinality $d$.
\end{restatable}

Now consider an arbitrary tomographically complete generalized probabilistic theory denoted $G:=(\mathcal V,\Omega,\mathcal E)$.
Then, let $b(G)\in\mathbb N$ be such that if $G$ is simplex-embeddable, then it is also simplex-embeddable in at most $b(G)$ dimensions. It was asked in~\cite{Spekkens19} whether there existed such a bound.
\Cref{prop:SimplexImpliesClassical} proves as a corollary the existence of this bound:

\begin{restatable}{corollary}{CorollaryDimSimplex}
\label{corollary:DimSimplex}
For any tomographically complete generalized probabilistic theory $G= (\mathcal V, \Omega, \mathcal E)$ for which there exists $d\in\mathbb N$
such that $G$ is simplex-embeddable in $d$ dimensions, it holds that $G$ is also simplex-embeddable in $d_{\textup{min}}\in\mathbb N$ dimensions with
\begin{equation}
\dim(\mathcal V) \leq d_\textup{min} \leq \dim(\mathcal V)^2,
\end{equation}
i.e., $b(G) = b(\mathcal V,\Omega,\mathcal E) = \dim(\mathcal V)^2$.
\end{restatable}
\begin{proof}
If $G=(\mathcal V,\Omega,\mathcal E)$ is simplex-embeddable, then by \cref{prop:SimplexImpliesClassical}, the prepare-and-measure scenario $(\Omega,\mathcal E)$ admits a finite, discrete \vgrSC{classical}\vgr{operationally noncontextual ontological} model which is a special case of Riemann integrable \vgrSC{classical}\vgr{operationally noncontextual ontological} models (\cref{def:RiemannIntegrableModel} under the substitution \eqref{eq:GPTsubst}).
By \cref{th:CardinalityBounds} under the substitution \eqref{eq:GPTsubst}, there also exists \vgrSC{a classical}\vgr{an operationally noncontextual ontological} model with a minimal ontic space cardinality $\card\ospace =: d_{\text{min}}$ such that $\dim(\mathcal V)\leq d_{\text{min}} \leq \dim(\mathcal V)^2$ where we used \eqref{eq:TomographicallyCompleteReducedSpace} to substitute $\mainr$ in \cref{th:CardinalityBounds} with $\mathcal V$.
Again by \cref{prop:SimplexImpliesClassical}, this means that the generalized probabilistic theory $G$ is simplex-embeddable in $d_{\text{min}}$ dimensions.
\end{proof}

In~\cite{Spekkens19}, by leveraging arguments of~\cite{Spekkens18}, it was shown that if a tomographically complete generalized probabilistic theory is such that $\mathcal E$  admits finitely many extremal points, then there exists such a bound $b(G)$,
and the analysis of~\cite{Spekkens18} also suggests that a similar bound holds if the set of states $\Omega$ has finitely many extremal points.
However, this bound which is the number of extremal points of the polytope defined in the ``Characterization P1 of the noncontextual
measurement-assignment polytope" of~\cite{Spekkens18}, depends on the set $\mathcal E$ and does not have a clear behavior as the number of extremal points of $\mathcal E$ grows --- it could in principle diverge.
For fixed $\mathcal V$, however, the upper bound $\dim(\mathcal V)^2$ of \cref{corollary:DimSimplex} remains constant for arbitrary choice of $(\Omega,\mathcal E)$, even with infinitely many extremal points.

We now turn to applying the results of~\cite{Spekkens19} to the original framework of this \doc{}.
In~\cite{Spekkens19}, an argument is given about the need for so-called  ``dimension mismatches''.
This useful argument can be rephrased in our setup as a proof that the lower bound $\dim(\mainr)$ in \cref{th:CardinalityBounds} is not always tight, i.e., there exist  \resources{} that admits a Riemann integrable \vgrSC{classical}\vgr{operationally noncontextual ontological} model with minimal ontic state space cardinality
\begin{equation}
d_{\min} = \dim(\mainr)+1.
\end{equation}
While not always tight, it is easy to see from the simplex-embeddability criterion of~\cite{Spekkens19} that there exist \resources{} such that the lower bound in \cref{th:CardinalityBounds} is saturated.
These considerations raise the open question of whether the upper bound $\dim(\mainr)^2$ in \cref{th:CardinalityBounds} is tight, i.e., whether there exist \resources{} such that the minimal ontic space has cardinality $\dim(\mainr)^2$.

\section{Conclusion}

After introducing the prepare-and-measure scenario \resources{} and the reduced space $\mainr$, \vgrSC{a noncontextual ontological model was formulated as in}\vgr{the operationally noncontextual ontological model was defined in} \cref{prop:basiccriterion}. \vgrSC{A new classicality criterion, \unitsep}\vgr{The \unitsep{} criterion for the existence of an operationally noncontextual ontological model} was extracted in \cref{th:maincriterion}. This criterion allowed to extract properties for the size of the ontic space $\ospace$, with most importantly the new bound $\card{\ospace}\leq\dim(\mainr)^2$ in \cref{th:CardinalityBounds} on page~\pageref{th:CardinalityBounds}.
The algorithmic formulation of the criterion was discussed in \cref{sec:algo}, allowing one to \vgrSC{evaluate numerically the (non-)classicality of}\vgr{numerically certify the existence or non-existence of an operationally noncontextual ontological model for} a given scenario \resources. Connections with generalized probabilistic theories were given in \cref{sec:Connection}, with most importantly the ontic space cardinality bounds translating as dimension bounds for  simplex-embeddability as in \cref{corollary:DimSimplex}.
Future directions of research include most importantly \vgrSC{the application of the classicality criterion to modern protocols in quantum information theory}\vgr{investigating interesting quantum protocols through the prism of whether or not they admit an operationally noncontextual ontological model}.
\vgrSC{Such applications will hopefully uncover links between this notion of non-classicality and the efficiency of quantum protocols.}\vgr{In particular, linking the non-existence of an operationally noncontextual ontological model to potential quantum advantages would be most interesting.}

\tocless\acknowledgments

We would like to thank Roberto Baldijão, Nuriya Nurgalieva, Martin Pl\'avala and Elie Wolfe for
fruitful discussions. M.W.\ acknowledges support from the Swiss National Science Foundation (SNSF) via an AMBIZIONE Fellowship (PZ00P2\_179914).
This work was supported as a part of NCCR QSIT, a National Centre of Competence (or Excellence) in Research, funded by the Swiss National Science Foundation (grant number 51NF40-185902).

\bibliographystyle{quantum}
\bibliography{gshv_bibliography}




\newpage
\setcounter{tocdepth}{2}
\tableofcontents

\twocolumngrid

\appendix



\section{Review of convex analysis}

\label{app:convex}

Throughout this section, we assume that $\mathcal V$ is a finite dimensional real inner product space. For a more detailed review, see~\cite{ConvexAnalysis}.

\begin{definition}[Convex set]
A set $X\subseteq \mathcal V$ is convex if and only if for all $0\leq\lambda\leq 1$, for all $x_1,x_2\in X$,
\begin{equation}
(\lambda x_1 + (1-\lambda) x_2) \in X.
\end{equation}
\end{definition}

\begin{definition}[Convex hull]
For any set $X\subseteq \mathcal V$, the convex hull $\myconv{X}$ is defined as
\begin{multline}
\myconv{X} :=
\textstyle \big\{\sum_{i=1}^n \lambda_i x_i\st 
n\in\mathbb N, \\
\textstyle \lambda_i\in\realpos,\sum_{i=1}^n \lambda_i = 1, x_i\in X\big\}.
\end{multline}
\end{definition}

It holds that  $\myconv X$ is the smallest convex set that contains $X$.

\begin{definition}[Extreme points]\label{def:extreme points}
Let $X\subseteq \mathcal V$ be a convex set. $x\in X$ is an extreme point of $X$ if and only if, given $\lambda\in\, ]0,1[$ and $x_1,x_2\in X$ such that
\begin{equation}
x = \lambda x_1 + (1-\lambda)x_2,
\end{equation}
one necessarily has $x_1 = x_2 = x$.
The set of extremal points of $X$ is denoted $\convextr{X}$.
\end{definition}

\begin{definition}[Conic set]
A set $X\subseteq \mathcal V$ is a cone, or a conic set, if and only if for all $\lambda \geq 0$, for all $x\in X$,
\begin{equation}
\lambda x \in X.
\end{equation}
\end{definition}

\begin{definition}[Conic hull]
\label{def:ConicHull}
For any set $X\subseteq \mathcal V$, the conic hull $\mycone{X}$ is defined as
\begin{equation}
\mycone X := \big\{ \lambda x\st \lambda\in\realpos, x\in X\big\}.
\end{equation}
\end{definition}

It holds that $\mycone X$ is the smallest conic set that contains $X$.

\begin{definition}[Extremal half-lines, see section~4.4 in~\cite{ConvexAnalysis}]
\label{def:ExtremalHalfLines}
Let $\mathcal C\subseteq\mathcal V$ be a convex cone. A vector $c_0\in\mathcal C$, $c_0\neq 0$ is an extremal direction of $\mathcal C$ if and only if, for all $d_1,d_2\in\mathcal C$ that satisfy
\begin{equation}
c_0 = d_1 + d_2,
\end{equation}
$d_1$ and $d_2$ are linearly dependent. It is then easy to show that for any $\lambda\in\mathbb R$, $\lambda > 0$, $\lambda c_0$ is also an extremal direction of $\mathcal C$. The half-line $\mycone{c_0}$ is said to be an extremal half-line of $\mathcal C$, i.e., all nonzero elements of the extremal half-line are extremal directions.
The set of all extremal half-lines of $\mathcal C$ is denoted $\coneextr{\mathcal C}$. 
\end{definition}

Note that $\coneextr{\mathcal C}$ is a set of set of points of $\mathcal C$.

\begin{lemma}
\label{lem:SubPolar}
Consider two sets $X\subseteq Y\subseteq \mathcal V$. It holds that
\begin{equation}
\polar{Y}{\mathcal V} \subseteq \polar{X}{\mathcal V}.
\end{equation}
\end{lemma}
\begin{myproof}
Let $v\in\polar{Y}{\mathcal V}$. Then, consider any $x\in X$, and we will show that $\scal{v}{x}{\mathcal V} \geq 0$. But since $X\subseteq Y$, it holds that $x\in Y$. Thus by \cref{def:Polar} of the polar cone and since $v\in\polar{Y}{\mathcal V}$, it holds that $\scal{x}{v}{\mathcal V} \geq0$. Thus, $v\in\polar{X}{\mathcal V}$.
\end{myproof}

\section{Presentation of the \vgr{operationally noncontextual ontological} model}

\subsection{The prepare-and-measure scenario}
\label{app:qexp}

Let us give an example of how convex mixtures of effects may be obtained from probabilistic mixtures.
Suppose that $E_1$ and $E_2$ belong to $\sete$, the set of quantum effects that are accessible in the lab, and also that $\id\hil - E_1$ and $\id\hil - E_2$ belong to $\sete$.

Suppose that one associates to $E_i$ the outcome $+1$ and to $\id\hil-E_i$ the outcome $-1$. Then, if one measures $\{E_1,\id\hil-E_1\}$ with probability $\lambda\in[0,1]$ and $\{E_2,\id\hil-E_2\}$ with probability $1-\lambda$, then effectively the effect associated with obtaining outcome $+1$ is described by $E(\lambda) := \lambda E_1 + (1-\lambda) E_2$, while the outcome $-1$ has associated quantum effect $\id\hil - E(\lambda)$. This shows that if one allows for such probabilistic mixtures, effectively the set of effects becomes convex.

\subsection{The reduced space}
\label{app:mainr}

In this section, we present general results about the reduced space that are used in the main text as well as in the following appendices.

\begin{definition}
\label{def:Proj}
Let $\mathcal V$ be a real inner product space of finite dimension, and let $\mathcal X\subseteq \mathcal V$ be a vector subspace. Let $\{x_i\in\mathcal X\}_{i=1}^{\dim(\mathcal X)}$ be an orthonormal basis with respect to the inner product of $\mathcal V$. Then, we define the projection over $\mathcal X$ as:
\begin{equation}
\forall v\in\mathcal V\st
\proj{\mathcal X}{v} := \sum_{i=1}^{\dim(\mathcal X)} \scal{x_i}{v}{\mathcal V} x_i.
\end{equation}
The projection of a set $S\subseteq \mathcal V$ is defined as the set of projected elements of $S$, i.e.,
\begin{equation}
\proj{\mathcal X}{S} := \{\proj{\mathcal X}{s}\st s\in S\}.
\end{equation}
\end{definition}

\begin{lemma}
\label{prop:basicproj}
Let $\mathcal V$ be a real inner product space of finite dimension, and let $\mathcal X\subseteq \mathcal V$ be a vector subspace equipped with the inner product inherited from $\mathcal V$. Then, for all $v\in\mathcal V$, for all $x\in\mathcal X$:
\begin{equation}
\scal{v}{x}{\mathcal V} = \scal{\proj{\mathcal X}{v}}{x}{\mathcal X}.
\end{equation}
\end{lemma}
\begin{myproof}
Let $\{X_i\in\mathcal X\}_i$ be an orthonormal basis of $\mathcal X$.
Extend this basis to an orthonormal basis of $\mathcal V$ of the form $\{X_i\}_i\cup\{V_j\}_j$.
Due to the orthogonality relations, we have $\scal{X_i}{V_j}{\mathcal V} = 0$ for all $i,j$, which also implies
\begin{equation}
\forall x\in\mathcal X,\forall j\st\scal{x}{V_j}{\mathcal V} = 0.
\end{equation}
Thus, using the completeness relation $v= \sum_i \scal{X_i}{v}{\mathcal V}X_i + \sum_j \scal{V_j}{v}{\mathcal V}V_j$, we have
\begin{align}
\scal{v}{x}{\mathcal V} &= \sum_i \scal{X_i}{v}{\mathcal V}\scal{X_i}{x}{\mathcal V}  + \sum_j \scal{V_j}{v}{\mathcal V}\scal{V_j}{x}{\mathcal V} \nonumber\\
&= \scal{\textstyle \sum_i \scal{X_i}{v}{\mathcal V}X_i }{x}{\mathcal V} \nonumber\\
&= \scal{\proj{\mathcal X}{v}}{x}{\mathcal V} \nonumber\\
&= \scal{\proj{\mathcal X}{v}}{x}{\mathcal X}.
\end{align}
\end{myproof}

\begin{lemma}
\label{prop:projsub}
For any real inner product space $\mathcal V$ of finite dimension, and for any vector subspaces $\mathcal X_1\subseteq \mathcal X_2\subseteq \mathcal V$,
\begin{equation}
\forall v\in\mathcal V\st \proj{\mathcal X_1}{v} = \proj{\mathcal X_1}{\proj{\mathcal X_2}{v}}.
\end{equation}
\end{lemma}
\begin{myproof}
Choose an orthonormal basis $\{X_i\}_i$ of $\mathcal X_1$ and extend it to an orthonormal basis $\{X_i\}_i\cup\{S_j\}_j$ of $\mathcal X_2$. Then,
\begin{multline}
\proj{\mathcal X_1}{\proj{\mathcal X_2}{v}} \\
= \sum_i \scal{X_i}{v}{\mathcal V}\proj{\mathcal X_1}{X_i} + \sum_j \scal{S_j}{v}{\mathcal V}\proj{\mathcal X_1}{S_j}.
\end{multline}
But thanks to $\proj{\mathcal X_1}{X_i} = X_i$ for all $i$ and $\proj{\mathcal X_1}{S_j} = 0$ for all $j$, the claim follows.
\end{myproof}

\Cref{prop:reducedscalar} is now proven as a special case of the following proposition.

\begin{prop}
\label{prop:reducedscalargen}
Let $\mathcal V$ be any real inner product space of finite dimension, and let $\mathcal X\subseteq\mathcal V$ and $\mathcal Y\subseteq\mathcal V$ be vector subspaces thereof equipped with the inner product inherited from $\mathcal V$. Let
\begin{equation}
\mathcal Z := \proj{\mathcal Y}{\mathcal X}.
\end{equation}
$\mathcal Z$ is a vector subspace which we equip with the inner product inherited from $\mathcal V$.
Then, for all $x\in\mathcal X$, for all $y\in\mathcal Y$,
\begin{equation}
\scal{x}{y}{\mathcal V} = \scal{\proj{\mathcal Z}{x}}{\proj{\mathcal Z}{y}}{\mathcal Z}.
\end{equation}
\end{prop}
\begin{myproof}
By \cref{prop:basicproj}, and due to $y\in\mathcal Y$,
\begin{equation}
\scal{x}{y}{\mathcal V} = \scal{\proj{\mathcal Y}{x}}{y}{\mathcal Y}
\end{equation}
But now $\proj{\mathcal Y}{x} \in \mathcal Z$, so that
\begin{equation}
\label{eq:tempproofappmainr}
\scal{x}{y}{\mathcal V} = \scal{\proj{\mathcal Y}{x}}{\proj{\mathcal Z}{y}}{\mathcal Z}.
\end{equation}
Then, thanks to \cref{prop:projsub}, and using
$
\mathcal Z =\proj{\mathcal Y}{\mathcal X} \subseteq \mathcal Y,
$
for any $x\in\mathcal X$ it holds that
\begin{equation}
\proj{\mathcal Z}{x} = \proj{\mathcal Z}{\proj{\mathcal Y}{x}}.
\end{equation}
But $\proj{\mathcal Y}{x}\in\mathcal Z$, so that it actually holds that
\begin{equation}
\proj{\mathcal Z}{x} = \proj{\mathcal Y}{x},
\end{equation}
and then equation \eqref{eq:tempproofappmainr} becomes
\begin{gather}
\scal{x}{y}{\mathcal V} = \scal{\proj{\mathcal Z}{x}}{\proj{\mathcal Z}{y}}{\mathcal Z}.\qedhere
\end{gather}
\end{myproof}

\begin{lemma}
\label{prop:spansgen}
Let $\mathcal V$ be any real inner product space of finite dimension. Let $\mathcal X\subseteq V$ and $\mathcal Y\subseteq V$ be vector subspaces. Let $\mathtt x\subseteq\mathcal X$ be any spanning set of $\mathcal X$, and let $\mathtt y\subseteq \mathcal Y$ be any spanning set of $\mathcal Y$. Let
\begin{equation}
\mathcal Z := \proj{\mathcal Y}{\mathcal X}.
\end{equation}
Then,
\begin{subequations}
\begin{align}
\myspan{\proj{\mathcal Z}{\mathtt x}} &= \mathcal Z,\\
\myspan{\proj{\mathcal Z}{\mathtt y}} &= \mathcal Z.
\end{align}
\end{subequations}
\end{lemma}
\begin{myproof}
First, consider:
\begin{multline}
\myspan{\proj{\mathcal Z}{\mathtt x}} = \proj{\mathcal Z}{\myspan{\mathtt x}} = \proj{\mathcal Z}{\mathcal X} \\
= \proj{\mathcal Z}{\proj{\mathcal Y}{\mathcal X}} = \proj{\mathcal Y}{\mathcal X} = \mathcal Z,
\end{multline}
where we used $\mathcal Z\subseteq \mathcal Y$ and \cref{prop:projsub}, to conclude $ \proj{\mathcal Z}{\mathcal X} = \proj{\mathcal Z}{\proj{\mathcal Y}{\mathcal X}}$.
Now, consider
\begin{equation}
\myspan{\proj{\mathcal Z}{\mathtt y}} = \proj{\mathcal  Z}{\mathcal Y}.
\end{equation}
Let $\{Z_i\}_i$ be an orthonormal basis of $\mathcal Z$, and extend it to an orthonormal basis $\{Z_i\}_i\cup\{Y_j\}_j$ of $\mathcal Y$ (indeed, $\mathcal Z$ is a subset of $\mathcal Y$). Then,
\begin{multline}
\textstyle \proj{\mathcal Z}{\mathcal Y} = \proj{\mathcal Z}{\left\{\sum_i \alpha_i Z_i + \sum_j \beta_j Y_j\st\alpha_i,\beta_j\in\mathbb R\right\}} \\
\textstyle = \big\{\sum_i \alpha_i Z_i\st \alpha_i\in\mathbb R\big\} = \mathcal Z. \qedhere
\end{multline}
\end{myproof}

\Cref{prop:spansgen} can be specialized as follows.

\begin{corollary} 
\label{prop:spans}
The projected states and effects span the whole reduced space:
\begin{subequations}
\begin{align}
\myspan{\reds} = \mainr, \\
\myspan{\rede} = \mainr.
\end{align}
\end{subequations}
\end{corollary}

\begin{lemma}
\label{lem:traceone}
For all $\red\rho\in\reds$,
\begin{equation}
\otrace{\red\rho} = 1.
\end{equation}
\end{lemma}
\begin{myproof}
For all $\red\rho\in\reds$, there exists $\rho\in\sets$ such that $\red\rho= \proj{\mainr}{\rho}.$ Then,
\begin{equation}
\otrace{\red\rho} = \scal{\proj{\mainr}{\rho}}{\proj{\mainr}{\id\hil}}{\mainr}.
\end{equation}
Using \cref{prop:reducedscalar},
\begin{equation}
\otrace{\red\rho} = \scal{\rho}{\id\hil}{\lh}
= \tr{\hil}{\rho}.
\end{equation}
The claim then follows from the fact that for all $\rho\in\sets\subseteq\states$, $\tr{\hil}{\rho} = 1$.
\end{myproof}

We now prove the following lemma which will be used to derive the \vgrSC{main classicality}\vgr{\unitsep{}} criterion \vgr{for the existence of an operationally noncontextual ontological model}, \cref{th:maincriterion}.

\begin{lemma}
\label{lem:nulltrace}
For all $\sigma\in\polar\rede\mainr$, the trace of $\sigma$ satisfies
\begin{equation}
\otrace{\sigma}\geq 0
\end{equation}
with equality if and only if $\sigma = 0$.
\end{lemma}
\begin{myproof}
Due to  $\proj{\mainr}{\id\hil}\in\proj{\mainr}{\sete}$, for all $\sigma\in\polar\rede\mainr$ it holds that
\begin{equation}
\otrace{\sigma} \geq 0.
\end{equation}
Now consider the case when $\sigma\in\polar\rede\mainr$ and $\otrace{\sigma} = 0$. For all $E\in\sete$, there exists $\{E_k\in\sete\}_k$ such that $\id\hil = E + \sum_k E_k$, so that
\begin{align}
0 &= \scal{\sigma}{\proj{\mainr}{\id\hil}}{\mainr} \nonumber\\
&= \scal{\sigma}{\proj{\mainr}{E}}{\mainr} + \sum_k \scal{\sigma}{\proj{\mainr}{E_k}}{\mainr}.
\end{align}
However, all the terms $\scal{\sigma}{\proj{\mainr}{E}}{\mainr}$ and $\scal{\sigma}{\proj{\mainr}{E_k}}{\mainr}$ are nonnegative and sum to zero, which implies that they are all zero. This shows that for all $E\in\sete$,
\begin{equation}
\scal{\sigma}{\proj{\mainr}{E}}{\mainr} = 0.
\end{equation}
But thanks to \cref{prop:spans} and the non-degeneracy of the inner product, this means that $\sigma = 0$. The other direction is trivial.
\end{myproof}

\begin{lemma}
\label{lem:emptysets}
It is impossible with the assumptions of the main text that $\proj{\mainr}{\sets} = \{0\}$ or that $\proj{\mainr}{\sete} = \{0\}$. This implies that it is impossible that $\polar\reds\mainr= \{0\}$ or that $\polar\rede\mainr = \{0\}$. As a corollary, the cone $\prodes$ is never the trivial cone $\{0\in\mainr\otimes\mainr\}$, nor is the convex cone $\sepes$ the trivial convex set $\{0\in\mainr\otimes\mainr\}$.
\end{lemma}
\begin{myproof}
Suppose that $\proj{\mainr}{\sets} = \{0\}$. 
This implies, for any $\red \rho\in\proj{\mainr}{\sets}$, that $\otrace{\red \rho} = 0$, which is a contradiction to \cref{lem:traceone}. 

Suppose now $\proj{\mainr}{\sete} =\{0\}$. Then, since $\sets\neq\emptyset$ according to \cref{def:available states}, choose any $\red\rho\in\proj{\mainr}{\sets}$. Then, the fact that $\proj{\mainr}{\id\hil}\in\rede$ leads to $\otrace{\red\rho} = \scal{\red\rho}{0}{\mainr} = 0$ which again violates \cref{lem:traceone}.

Then, note that $\proj{\mainr}{\sete} \subseteq \polar\reds\mainr$. Thus, if $\polar\reds\mainr = \{0\}$, also $\proj{\mainr}{\sete} = \{0\}$, since $\proj{\mainr}{\sete}\neq \emptyset$ according to \cref{def:available effects}. But $\rede = \{0\}$ has been shown to be impossible. 

For the other case, note that $\reds\subseteq \polar\rede\mainr$, so that if $\polar\rede\mainr = \{0\}$, then also $\proj{\mainr}{\sets} = \{0\}$ since $\proj{\mainr}{\sets}\neq \emptyset$ according to \cref{def:available states}. This has been shown to be impossible.
\end{myproof}

\PropBadClassicalModel* 

\begin{myproof}
The item (i) is trivial: any \vgrSC{classical}\vgr{operationally noncontextual ontological} model
is also a valid minimal Spekkens noncontextual ontological model where the ontic primitives happen not to depend on the part $\rho - \proj{\mainr}{\rho}$ of the states $\rho\in\sets$ and the part $E - \proj{\mainr}{E}$ of the effects $E\in\sete$. 

For the item (ii), we first choose any prepare-and-measure scenario $(\tilde\sets,\tilde\sete)$ on some finite dimensional Hilbert space $\tilde\hil$ and where $\tilde\sets$ and $\tilde\sete$ take the form
\begin{subequations}
\begin{align}
\tilde\sets &= \textup{conv}\big(\big\{\tilde\rho_i \in \linherm{\tilde\hil} \big\}_{i=1}^N\big), \\
\tilde\sete &= \textup{conv}\big(\big\{\tilde E_j \in \linherm{\tilde\hil}\big\}_{j\in J}\big)
\end{align}
\end{subequations}
for some finite $N\in \mathbb N$ and some discrete or continuous range $J$; this $(\tilde\sets,\tilde\sete)$ is assumed to satisfy \cref{def:available states,def:available effects}. We assume that this prepare-and-measure scenario does not admit \vgrSC{a classical}\vgr{an operationally noncontextual ontological} model --- one can pick such a scenario from, e.g., the example of \cref{subsec:non-classical example}, with a sufficiently fine-grained polyhedral inner approximation (in the sense of \cref{subsec:non-classicality witnesses}) of the state space $\sets$.
Let us introduce the Hilbert space
\begin{equation}
\hil := \tilde\hil\otimes \mathbb C^N.
\end{equation}
Despite the tensor product structure, we think of this Hilbert space as describing one single system whose Hilbert space happens to be isomorphic to such a tensor product structure.
Let $\{\ket{\phi_i}\}_{i=1}^N$ be an orthonormal basis of $\mathbb C^N$.
We define, for all $i=1,\dots,N$ and $j\in J$ the states and effects
\begin{subequations}
\begin{align}
\rho_i &:= \tilde\rho_i \otimes \ketbra{\phi_i}, \\
E_j &:= \tilde E_j \otimes \id{N}.
\end{align}
\end{subequations}
The scenario \resources{} where
\begin{subequations}
\begin{align}
\sets &:= \textup{conv}\big(\big\{\rho_i \in \linherm{\hil} \big\}_{i=1}^N\big), \\
\sete &:= \textup{conv}\big(\big\{ E_j\in\linherm{\hil}\big\}_{j\in J}\big)
\end{align}
\end{subequations}
satisfies \cref{def:available states,def:available effects} and we claim that it admits a minimal Spekkens noncontextual ontological model but no \vgrSC{classical}\vgr{operationally noncontextual ontological} model.
To see that it does not admit any \vgrSC{classical}\vgr{operationally noncontextual ontological} model,
we first note
that for all $i=1,\dots,N$, for all $j\in J$, 
\begin{equation}
\tr{\hil}{\rho_i E_j} = \textup{Tr}_{\tilde\hil}[\tilde \rho_i \tilde E_j].
\end{equation}
Thus, \cref{prop:ancilla}
applies here and
implies that since $(\tilde\sets,\tilde\sete)$ does not admit \vgrSC{a classical}\vgr{an operationally noncontextual ontological} model,
also $(\sets,\sete)$ does not admit such a model.

However, we can construct a minimal Spekkens noncontextual ontological model for $(\sets,\sete)$. Define $\Lambda := \{1,\dots,N\}$. For all $\lambda\in\Lambda$, $\rho \in \sets$ and all $E \in \sete$, 
\begin{subequations}
\begin{align}
\ostate{\rho}{\lambda} &:= \frac{\tr{\hil}{\rho \rho_\lambda}}{\textup{Tr}_{\tilde\hil}[\tilde\rho_\lambda ^2]}, \\
\orep{E}{\lambda} &:= \tr{\hil}{E \rho_\lambda}.
\end{align}
\end{subequations}
Since all the matrices involved are nonnegative, the nonnegativity requirements \eqref{eq:basicmupos} and \eqref{eq:basicxipos} (under the replacement $P_\mainr \rightarrow \id\lh$ of \eqref{eq:standard NCHV}) are verified. 
By linearity of the trace, it is also clear that the adapted versions of \eqref{eq:convexlinearityrho} and \eqref{eq:convexlinearityxi} are also verified.
To see that the normalization of $\mu$ holds, note that for all $\rho\in\sets$ there exist $\{p_i\in\realpos\}_{i=1}^N$ with $\sum_{i=1}^N p_i = 1$ such that $\rho = \sum_{i=1}^N p_i \rho_i$. Then, one can compute
\begin{align}
\sum_{\lambda\in\Lambda} \ostate{\rho}{\lambda} &= \sum_{\lambda,i = 1}^N p_i
\frac{\tr{\hil}{\rho_i \rho_\lambda}}{\textup{Tr}_{\tilde\hil}[\tilde\rho_\lambda^2]} \nonumber\\
&= \sum_{\lambda,i = 1}^N p_i \delta_{i\lambda} = 1.
\end{align}
so that \eqref{eq:NormMu} is verified, and so is clearly the normalization requirement \eqref{eq:explicitnormalizationxi} (again under the same replacement) given the linearity of $\xi$. It remains to compute, for any $\rho = \sum_{i=1}^N p_i \rho_i$ as previously defined and $E\in\sete$,
\begin{align}
\sum_{\lambda=1}^N \ostate{\rho}{\lambda} \orep{E}{\lambda} &=
\sum_{i,\lambda=1}^N p_i \delta_{i\lambda} \tr{\hil}{E\rho_\lambda} \nonumber\\
&= \tr{\hil}{E \rho},
\end{align}
thus reproducing the adapted version of \eqref{eq:consistmuxi}. This concludes the proof. 
\end{myproof}

\subsection{An example correspondence between reduced space projection and partial trace}
\label{subsec:example partial trace}

In \cref{subsection:discussion of gshv}, we mentioned how taking a reduced space projection and a partial trace can be thought of as similar way of restricting the focus to a system.
Consider a system $A$, with Hilbert space $\hil_A$, for which we would like to build a noncontextual ontological model, and a system $B$, with Hilbert space $\hil_B$ which is an auxiliary system that can be thought of as being ``close by" with respect to system $A$.
Suppose that an agent prepares states $\{\rho^{(A)}_i \in \linherm{\hil_A}\}_i$ on system $A$, but each time they do so, there is a certain influence on the auxiliary system $B$ so that the agent effectively prepares the joint states $\{\rho^{(A)}_i \otimes \sigma^{(B)}_i \in \linherm{\hil_A \otimes \hil_B}\}_i$.
The measurements of the agent are however strictly restricted to system $A$, and hence the effects are of the form $\{E^{(A)}_j\otimes \id{\hil_B}\}_j$.

We offer two ways of formulating a noncontextual ontological model for system $A$. The first is to use \cref{def:classicalmodel}, with the reduced space prescription implying that the ontic state distributions and ontic response functions only depend on $\{\rho_i^{(A)}\}_i$ and $\{E_j^{(A)}\}_j$ (this will be formalized in \cref{prop:partial trace}).
Alternatively, one may start from the basics of Spekkens noncontextual ontological models. A potential ontological model for the joint system of $A$ and $B$ could in principle be allowed to depend on both the quantum primitives of $A$ and those of $B$.
However, a Spekkens noncontextual ontological model for system $A$ should only depend on the reduced quantum state of $A$. Taking the partial trace on the system $B$ implements this reduction, and we see that the reduced space projection and the partial trace implement the same kind (depending on whether the prepare-and-measure scenario reduced to system $A$ is tomographically complete) of restriction for the dependency of the ontic primitives in this case.
Formally, one may for instance write down the following proposition.

\begin{prop}
\label{prop:partial trace}
Let $I,J \subseteq \mathbb{N}$. We consider a bipartite Hilbert space $\hil^{(AB)} = \hil^{(A)} \otimes \hil^{(B)}$, a set of density matrix $\{\rho_i^{(AB)} \in \linherm{\hil^{(AB)}}\}_{i\in I}$ and of effects $\{E_j^{(A)}\in\linherm{\hil^{(A)}}\}_{j\in J} \supseteq \{0^{(A)},\id{}^{(A)}\}$. We define
\begin{subequations}
\label{eq:app scenario ab}
\begin{align}
\sets^{(AB)} &= \myconv{\big\{\rho^{(AB)}_i\big\}_{i\in I}}, \\
\sete^{(AB)} &= \myconv{\{E^{(A)}_j \otimes \id{}^{(B)}\}_{j\in J}},
\end{align}
\end{subequations}
as well as
\begin{subequations}
\label{eq:app scenario a}
\begin{align}
\sets^{(A)} &= \myconv{\big\{\tr{B}{\rho^{(AB)}_{i}}\big\}_{i \in I}}, \\
\sete^{(A)} &= \myconv{\big\{E^{(A)}_j\big\}_{j\in J}}.
\end{align}
\end{subequations}
$\sete^{(AB)}$ and $\sete^{(A)}$ are assumed to satisfy \cref{def:available effects}.
We further assume that 
\begin{equation}
\label{eq:app scenario assumption}
\myspan{\sets^{(A)}} = \myspan{\sete^{(A)}} = \lh.
\end{equation}
Then the scenario $(\sets^{(AB)},\sete^{(AB)})$ admits \vgrSC{a classical}\vgr{an operationally noncontextual ontological} model (\cref{def:classicalmodel}) if and only if the scenario $(\sets^{(A)},\sete^{(A)})$ admits a minimal Spekkens noncontextual ontological model (as defined in \cref{prop:badclassicalmodel}).
\end{prop}
\begin{proof}
The scenarios $(\sets^{(AB)},\sete^{(AB)})$ and $(\sets^{(A)},\sete^{(A)})$ satisfy the assumptions of \cref{prop:ancilla}, and hence, there exists \vgrSC{a classical}\vgr{an operationally noncontextual ontological} model for one if and only if there exists \vgrSC{a classical}\vgr{an operationally noncontextual ontological} model with identical ontic space $\ospace$ for the other. Furthermore, when considering $(\sets^{(A)},\sete^{(B)})$, and thanks to the assumption of equation \eqref{eq:app scenario assumption}, the projection over the reduced space $\proj{\mainr}{\cdot}$ equals the identity map $\id{\lh}(\cdot)$. Thus, a minimal Spekkens noncontextual ontological model (defined in \cref{prop:badclassicalmodel}) is actually equivalent to \vgrSC{a classical}\vgr{an operationally noncontextual ontological} for the scenario $(\sets^{(A)},\sete^{(A)})$.
\end{proof}

\subsection{Linear extensions to the ontic mappings}
\label{app:LinExt}

We now prove \cref{prop:extensions} as follows: \cref{prop:extensionmu} proves how the extension is built for $\mu$, while \cref{prop:extensionxi} considers the extension for $\xi$.
Both \cref{prop:extensionmu,prop:extensionxi} will make use of the following lemma. Note that convex and conic sets are defined in \app~\ref{app:convex}.

\begin{lemma}
\label{lem:extcone}
Let $\mathcal V$ be a real inner product space of finite dimension, and let $\mathcal C\subseteq \mathcal V$ be a convex cone such that
\begin{equation}
\label{eq:spancv}
\myspan{\mathcal C} = \mathcal V.
\end{equation}
Let $f\funcrange{\mathcal C}{\mathbb R}$ be any function that satisfies the following two properties\footnote{Observe that $\forall c_1,c_2\in\mathcal C$, $\alpha\in\realpos$, $\alpha c_1\in\mathcal C,$ $c_1+c_2\in\mathcal C$ since $\mathcal C$ is convex and conic.}:
\begin{subequations}
\begin{align}
\forall \alpha\in\realpos,\forall c\in\mathcal C\st& f(\alpha c) = \alpha f(c), \label{eq:fposscal}\\
\forall c_1,c_2\in\mathcal C\st& f(c_1+c_2) = f(c_1) + f(c_2). \label{eq:fposlin}
\end{align}
\end{subequations}
Then, there exists a unique function
\begin{equation}
g\funcrange{\mathcal V}{\mathbb R}
\end{equation}
that is linear and that satisfies
\begin{equation}
\restrict{g}{\mathcal C} = f, \label{eq:g on f 1}
\end{equation}
which means that
\begin{equation}
\forall c\in\mathcal C\st 
g(c) = f(c).\label{eq:g on f 2}
\end{equation}
\end{lemma}
\begin{myproof}

First, we will show that $f$ satisfies the following: for all $\alpha_i\in\mathbb R$, for all $c_i\in\mathcal C$ such that $\sum_i \alpha_i c_i \in\mathcal C$,
\begin{equation}
\label{eq:wantedposlin}
\textstyle f\big(\sum_i \alpha_i c_i\big) = \sum_i \alpha_i f(c_i).
\end{equation}

To start with, let $c_1,c_2\in\mathcal C$ be such that $c_1-c_2\in\mathcal C$. Then, using equation \eqref{eq:fposlin},
\begin{equation}
f\left((c_1-c_2) + c_2\right) = f(c_1-c_2)  + f(c_2),
\end{equation}
which is equivalent to
\begin{equation}
\label{eq:templinsub}
f(c_1-c_2) = f(c_1) - f(c_2).
\end{equation}

Then, let $I=\{1,\dots,n\}$ for some $n\in\mathbb N$, let $\{\alpha_i\in\mathbb R\}_{i\in I}$ and $\{c_i\in\mathcal C\}_{i\in I}$ be such that $\sum_{i\in I} \alpha_i c_i \in\mathcal C$. Then, let\footnote{The sign function $\sgn{x}$ is $-1$ if $x < 0$, $0$ if $x=0$ and $1$ if $x> 0$. The absolute value is $|x| = \sgn x x.$}
\begin{subequations}
\begin{align}
I_+ &:= \{i\in I\st \sgn{\alpha_i} = +1\}, \\
I_0 &:= \{i\in I\st \alpha_i = 0\}, \\
I_- &:= \{i\in I\st \sgn{\alpha_i} = -1\}.
\end{align}
\end{subequations}
Using these,
\begin{multline}
\textstyle f\left(\sum_{i\in I}\alpha_i c_i\right) \\
= \textstyle f\left(\sum_{i\in I_+}|\alpha_i| c_i -\sum_{j\in I_-} |\alpha_j|c_j\right).
\end{multline}
Clearly, both sums $\sum_{i\in I_\pm} |\alpha_i| c_i$ belong to $\mathcal C$. Using equation \eqref{eq:templinsub},
\begin{multline}
\textstyle f\left(\sum_{i\in I}\alpha_i c_i\right)  \\
= \textstyle f\left(\sum_{i\in I_+}|\alpha_i| c_i \right) - f\left(\sum_{j\in I_-} |\alpha_j|c_j\right).
\end{multline}
Then, using repeatedly equation \eqref{eq:fposlin} to expand the sums, as well as equation \eqref{eq:fposscal} to extract the positive factors, and bringing back the trivial summands $\{i\in I_0\}$, we obtain equation \eqref{eq:wantedposlin}.

We can now easily extend $f$ to a  linear map $g$ whose domain is $\mathcal V$.
To do so, choose a basis $\{S_i\in\mathcal C\}_{i=1}^{\dim(\mathcal V)}$ of $\mathcal V$.
This is always possible thanks to the assumption \eqref{eq:spancv}.
For any $v\in\mathcal V$ and $i\in I$, let $s_i(v)\in\mathbb R$ be the coordinate of $v$ in the basis $\{S_i\}_i$.
Of course, $s_i\funcrange{\mathcal V}{\mathbb R}$ is linear for each $i$. Then, define for all $v\in\mathcal V$:
\begin{equation}
g(v) := \sum_i s_i(v) f(S_i).
\end{equation}
This choice for $g$ is unique: indeed, if $g|_{\mathcal C} = f$ is to hold, then in particular $g$ has to agree with $f$ on the basis elements $\{S_i\}_i$, but the action of a linear map on a basis completely determines its action on the whole space.
Thanks to equation \eqref{eq:wantedposlin}, it is then easy to see that indeed
\begin{equation}
\restrict{g}{\mathcal C} = f.
\end{equation}
Explicitly, for all $c\in\mathcal C$,
\begin{multline}
\textstyle g(c) = g\left(\sum_i s_i(c) S_i\right) = \sum_i s_i(c) f(S_i) \\
\textstyle = f\left(\sum_i s_i(c) S_i\right) = f(c). \qedhere
\end{multline}
\end{myproof}

\begin{prop}
\label{prop:extensionmu}
Let $\hv\in\ospace$ be arbitrary. Starting from the convex-linear mapping
\begin{equation}
\mu(\cdot,\hv)\funcrange{\reds}{\realpos},
\end{equation}
there exists a unique linear extension
\begin{equation}
\mu_{\textup{ext}}(\cdot,\hv)\funcrange{\mainr}{\mathbb R}.\label{eq:m comment}
\end{equation}
\end{prop}
\begin{myproof}
Throughout this proof, we omit the fixed argument $\hv$. Recalling $\reds\subseteq \mycone{\reds}\subseteq \mainr$, let us first look for the intermediate function
\begin{equation}
\mu_{\text{cone}}\funcrange{\mycone{\reds}}{\mathbb R},
\end{equation}
that satisfies, for all $\alpha\in\realpos$,  for all $\red\rho\in\reds$,
\begin{equation}
\label{eq:requirementmualpha}
\mu_{\text{cone}}(\alpha \red\rho) = \alpha\mu(\red\rho).
\end{equation}

Equation \eqref{eq:requirementmualpha} is a necessary condition for linearity which can be formulated given the restriction of the domain to a conic set.
Recall \cref{lem:traceone}:
\begin{equation}
\forall \red\rho \in\reds\st \tr{\hil}{\red\rho} = \otrace{\red\rho} = 1.
\end{equation}
Thus, the right-hand side of equation \eqref{eq:requirementmualpha} may be rewritten as follows when $\alpha \neq 0$:
\begin{equation}
\mu_{\text{cone}}(\alpha\red\rho) = \tr{\hil}{\alpha\red\rho}\mu\left(\frac{\alpha\red\rho}{\tr{\hil}{\alpha\red\rho}}\right).
\end{equation}
This shows that the unique choice for $\mu_{\text{cone}}$ is the following: for all $r\in\mycone{\reds}$,
\begin{equation}
 \mu_{\text{cone}}(r) := \left\{
\begin{aligned}
&0 &&\text{if } r= 0, \\
&\tr{\hil}{r}\mu\left(\frac{r}{\tr{\hil}{r}}\right) &&\text{else}.
\end{aligned}
\right.
\end{equation}
One important property that $\mu_{\text{cone}}$ satisfies is that it agrees with $\mu$ when the argument is in $\reds$.

Let us now show that $\mu_{\text{cone}}$ satisfies the assumptions of \cref{lem:extcone}. Clearly, using \cref{prop:spans},
\begin{equation}
\myspan{\mycone{\proj{\mainr}{\sets}}} = \mainr.\label{eq:span text prop D 2}
\end{equation}
Furthermore, since $\sets$ is convex and $\proj{\mainr}{\cdot}$ is linear, $\proj{\mainr}{\sets}$ is convex. Therefore $\mycone{\proj{\mainr}{\sets}}$ is a convex cone. This property together with equation \eqref{eq:span text prop D 2} allows to verify equation \eqref{eq:spancv}.
Let us now prove that $\mu_{\text{cone}}$ satisfies \eqref{eq:fposscal}: let $\alpha \in\realpos$, and $r\in\mycone{\reds}$. If either $\alpha = 0$ or $r=0$, then clearly $\mu_{\text{cone}}(\alpha r) = \alpha\mu_{\text{cone}}(r)$. If both $\alpha \neq 0$ and $r\neq0$, then
\begin{multline}	
\mu_{\text{cone}}(\alpha r) = \tr{\hil}{\alpha r}\mu\left(\frac{\alpha r}{\tr{\hil}{\alpha r}}\right) \\
= \alpha \tr{\hil}{r}\mu\left(\frac{r}{\tr{\hil}{r}}\right) = \alpha\mu_{\text{cone}}(r).
\end{multline}
Thus $\mu_{\text{cone}}$ satisfies \eqref{eq:fposscal}. 
Now let $r,s\in\mycone{\reds}$, and we will verify \eqref{eq:fposlin}. If $r=0$ or $s=0$, or both, then trivially $\mu_{\text{cone}}(r+s) = \mu_{\text{cone}}(r) + \mu_{\text{cone}}(s)$. Otherwise if $r,s\neq0$,
\begin{multline}
\mu_{\text{cone}}(r+s) = \tr{\hil}{r+s}\mu\left(\frac{r+s}{\tr{\hil}{r+s}}\right) \\
= 
\tr{\hil}{r+s}\mu\left(\frac{r}{\tr{\hil}{r+s}}
+\frac{s}{\tr{\hil}{r+s}} \right) \\
=  \tr{\hil}{r+s}\mu\left(p\frac{r}{\tr{\hil}{r}}
+(1-p) \frac{s}{\tr{\hil}{s}}\right) 
\end{multline}
where we defined for brevity $p=\tr{\hil}{r}/\tr{\hil}{r+s}$. By the convex-linearity of $\mu$ as in equation \eqref{eq:convexlinearityrho}, however, this becomes
\begin{multline}
\mu_{\text{cone}}(r+s)
=\\
\tr{\hil}{r+s}\left( p\mu\left(\frac{r}{\tr{\hil}{r}}\right) + (1-p)\mu\left(\frac{s}{\tr{\hil}{s}}\right)\right) \\
= \tr{\hil}{r}\mu\left(\frac{r}{\tr{\hil}{r}}\right) + \tr{\hil}{s}\mu\left(\frac{s}{\tr{\hil}{s}}\right) \\
= \mu_{\text{cone}}(r) + \mu_{\text{cone}}(s).
\end{multline}

Thus, $\mu_{\text{cone}}$ fully satisfies the assumptions of \cref{lem:extcone}. This shows that there exists a unique linear map
\begin{equation}
\mu_{\text{ext}}\funcrange{\mainr}{\mathbb R}
\end{equation}
such that
\begin{equation}
\restrict{\mu_{\text{ext}}}{\mycone{\reds}} = \mu_{\text{cone}}.
\end{equation}
But then, by subset inclusion, and because $\mu_{\text{cone}}$ extends $\mu$,
\begin{multline}
\restrict{\mu_{\text{ext}}}{\reds} = \restrict{\left(\restrict{\mu_{\text{ext}}}{\mycone{\reds}}\right)}{\reds} \\
= \restrict{\mu_{\text{cone}}}{\reds} 
 = \mu. \qedhere
\end{multline}
\end{myproof}

\begin{prop}
\label{prop:extensionxi}
Let $\hv\in\ospace$ be arbitrary. Starting from the convex-linear mapping
\begin{equation}
\xi(\cdot,\hv)\funcrange{\rede}{\realpos},
\end{equation}
there exists a unique linear extension
\begin{equation}
\xi_{\text{ext}}(\cdot,\hv)\funcrange{\mainr}{\mathbb R}.
\end{equation}
\end{prop}
\begin{myproof}
Again, we omit the fixed argument $\hv\in\ospace$ throughout the proof.
First, note that $\xi(0)=0$.\footnote{$0\in\mainr$ is automatically in the domain of $\xi$ since $0\in\lh$ is always an allowed effect within $\sete$ as required in \cref{def:available effects}.} This follows easily from \eqref{eq:explicitnormalizationxi}.

Then, we show the following property of $\xi$. 
If there exists $\red E\in\rede$ and $\alpha\in\realpos$ such that also $\alpha\red E\in\rede$, then,
\begin{equation}
\label{eq:tempscalxi}
\xi(\alpha \red E) = \alpha \xi(\red E).
\end{equation}
Without loss of generality assume that $\alpha \leq 1$ (indeed, if $\alpha > 1$, one may simply interchange the role of $\alpha \red E$ and $\red E$). Then,
\begin{multline}
\xi(\alpha \red E )
= \xi(\alpha \red E + (1-\alpha)\cdot 0)
= \alpha \xi(\red E) + (1-\alpha) \xi(0) \\
= \alpha \xi(\red E),
\end{multline}
where we used the convex linearity of $\xi$ as in equation \eqref{eq:convexlinearityxi}.

Let us now look for the intermediate extension
\begin{equation}
\xi_{\text{cone}}\funcrange{\mycone{\rede}}{\mathbb R}
\end{equation}
that satisfies, for all $\alpha\in\realpos$, for all $\red E\in\rede$,
\begin{equation}
\label{eq:wantedxipos}
\xi_{\text{cone}}(\alpha\red E) = \alpha \xi(\red E).
\end{equation}
Equation \eqref{eq:wantedxipos} is a necessary condition for linearity which can be formulated on a conic domain.
The only way to define this extension is clearly the following: for all $r\in\mycone{\rede}$, by \cref{def:ConicHull} of the conic hull there always exist $\alpha_r\in\realpos$ and $\red E_r\in\rede$ such that $r=\alpha_r\red E_r$, and then define
\begin{equation}
\label{eq:defxicone}
\xi_{\text{cone}}(r) := \alpha_r \xi(\red E_r).
\end{equation}
This definition is meaningful because it does not depend on the $\alpha_r,\red E_r$ that one chooses. Indeed, suppose that instead of decomposing $r=\alpha_r \red E_r$, one chooses instead $r=\beta_r\red F_r$ where $\beta_r\in\realpos$, $\red F_r\in\rede$. If $r=0$ the present discussion is irrelevant, so that we may assume that $\alpha_r,\beta_r > 0$. Then, the value one obtains with the alternative decomposition $r= \beta_r\red F_r$ is
\begin{equation}
\label{eq:TempBetaF}
\beta_r \xi(\red F_r) = \beta_r \xi\left(\frac{1}{\beta_r}r\right) = \beta_r \xi\left(\frac{\alpha_r}{\beta_r}\red E_r\right).
\end{equation}
Using equation \eqref{eq:tempscalxi} applied to $\red E_r$ and $\red F_r = (\alpha_r/\beta_r)\red E_r$, which both belong to $\rede$, \eqref{eq:TempBetaF} becomes
\begin{equation}
\beta_r \xi(\red F_r) = \alpha_r \xi(\red E_r).
\end{equation}
This proves that $\xi_{\text{cone}}$ as in \eqref{eq:defxicone} is well-defined. Also, it is clear that equation \eqref{eq:wantedxipos} is verified. It is then easy to see that $\xi_{\text{cone}}$ satisfies the first assumption \eqref{eq:fposscal} of  \cref{lem:extcone}:
\begin{multline}
\label{eq:XiConeScalPos}
\forall \alpha\in\realpos,\forall r\in\mycone{\rede}\st \\
\xi_{\text{cone}}(\alpha r) 
 = \alpha \xi_{\text{cone}}(r).
\end{multline}
Also, thanks to \cref{prop:spans}, the span assumption \eqref{eq:spancv} of  \cref{lem:extcone} is verified in this case: indeed,
\begin{equation}
\myspan{\mycone{\rede}} = \mainr.
\end{equation}
Let us verify the last assumption \eqref{eq:fposlin} of  \cref{lem:extcone}: let $r,s\in\mycone{\rede}$, and we will show that
\begin{equation}
\xi_{\text{cone}}(r+s) = \xi_{\text{cone}}(r)+\xi_{\text{cone}}(s).
\end{equation}
Let $\alpha_r,\alpha_s\in\realpos$, $\red E_r,\red E_s\in\rede$ be such that
\begin{subequations}
\begin{align}
r&= \alpha_r\red E_r,\\
s &= \alpha_s \red E_s.
\end{align}
\end{subequations}
Then, using equation \eqref{eq:XiConeScalPos},
\begin{multline}
\xi_{\text{cone}}(r+s)
= \xi_{\text{cone}}(\alpha_r \red E_r + \alpha_s\red E_s) \\
\textstyle = (\alpha_r + \alpha_s)\xi_{\text{cone}}\left(\frac{\alpha_r}{\alpha_r+\alpha_s}\red E_r + \frac{\alpha_s}{\alpha_r + \alpha_s}\red E_s\right).
\end{multline}
Note that by the convexity of $\sete$ and of $\rede$, $\left(\frac{\alpha_r}{\alpha_r+\alpha_s}\red E_r + \frac{\alpha_s}{\alpha_r + \alpha_s}\red E_s\right)\in\rede$, so that in fact
\begin{multline}
\xi_{\text{cone}}(r+s) \\
\textstyle = (\alpha_r + \alpha_s)\xi\left(\frac{\alpha_r}{\alpha_r+\alpha_s}\red E_r + \frac{\alpha_s}{\alpha_r + \alpha_s}\red E_s\right).
\end{multline}
Using the convex-linearity property \eqref{eq:convexlinearityxi} of $\xi$,
\begin{multline}
\xi_{\text{cone}}(r+s) \\
\textstyle = \alpha_r \xi(\red E_r) + \alpha_s \xi(\red E_s) = \xi_{\text{cone}}(r) + \xi_{\text{cone}}(s).
\end{multline}
Thus also the assumption \eqref{eq:fposlin} is verified and we may apply \cref{lem:extcone} to conclude that there exists a unique linear map
\begin{equation}
\xi_{\text{ext}}\funcrange{\mainr}{\mathbb R}
\end{equation}
that agrees with $\xi_{\text{cone}}$ on $\mycone{\rede}$, and thus also that agrees with $\xi$ of $\rede$.
\end{myproof}

\begin{theorem}[Riesz' representation theorem, see theorem~4.47 in~\cite{LinearAlgebra}]
\label{th:Riesz}
Let $\mathcal V$ be an arbitrary real inner product space of finite dimension. For any linear map $f\funcrange{\mathcal V}{\mathbb R}$, there exists a unique $F\in\mathcal V$ such that
\begin{equation}
\forall v\in\mathcal V\st
f(v) = \scal{F}{v}{\mathcal V}.
\end{equation}
\end{theorem}

\subsection{Basic criterion for the existence of \vgr{an operationally noncontextual ontological} model}
\label{app:BasicCriterion}

We now restate and prove the basic criterion for the existence of \vgrSC{a classical}\vgr{an operationally noncontextual ontological} model.

\ThBasicCriterion*

\begin{myproof}

We start from \cref{def:classicalmodel} of the \vgrSC{classical}\vgr{operationally noncontextual ontological} model, bearing in mind the extended ontic state mapping and extended ontic response function mapping introduced in \cref{prop:extensions}, as well as their representation as scalar products in equation \eqref{eq:OnticScalarReprs}. By construction, the desired convex-linearity requirements in equations \eqref{eq:convexlinearityrho} and \eqref{eq:convexlinearityxi} are automatically verified as a special case of the linearity of the scalar products in \eqref{eq:OnticScalarReprs}. Let us constrain the mappings $F\funcrange{\ospace}{\mainr}$ and $\sigma\funcrange{\ospace}{\mainr}$ by imposing the relevant nonnegativity constraints \eqref{eq:basicmupos} and \eqref{eq:basicxipos}:
\begin{subequations}
\begin{align}
\forall\hv\in\ospace,\forall \red\rho\in\reds\st&&
\scal{\red\rho}{F(\hv)}{\mainr} &\geq 0, \\
\forall \hv\in\ospace,\forall \red E\in\rede\st&&
\scal{\sigma(\hv)}{\red E}{\mainr} &\geq 0.
\end{align}
\end{subequations}
Using the \cref{def:Polar} of the polar cone, this is equivalent to
\begin{subequations}
\begin{align}
\forall\hv\in\ospace\st
F(\hv)\in\polar\reds\mainr, \\
\forall \hv\in\ospace\st
\sigma(\hv)\in\polar\rede\mainr.
\end{align}
\end{subequations}
This proves that the nonnegativity of the ontic primitives \eqref{eq:basicmupos} and \eqref{eq:basicxipos} is equivalent to the ranges of $F$ and $\sigma$ as in equations \eqref{eq:simpleranges}.

The consistency requirement \eqref{eq:consistmuxi} in the \cref{def:classicalmodel} of the \vgrSC{classical}\vgr{operationally noncontextual ontological} model reads: 
\begin{multline}
\label{eq:tempconsist}
\forall \red\rho\in\reds,\forall\red E\in\rede\st \\
\scal{\red\rho}{\red E}{\mainr} = \inthv{\scal{\red\rho}{F(\hv)}{\mainr}
\scal{\sigma(\hv)}{\red E}{\mainr}}.
\end{multline}
Due to $\myspan{\reds} = \myspan{\rede} = \mainr$ (proven in \cref{prop:spans}), it is clear that \eqref{eq:tempconsist} implies, and is implied by, the consistency requirement \eqref{eq:consistency} of \cref{prop:basiccriterion}.

Let us now show that the normalization of $\sigma$ as in equation \eqref{eq:normsigma} is implied by the \cref{def:classicalmodel} of the \vgrSC{classical}\vgr{operationally noncontextual ontological} model. This is easy to see: starting from the normalization \eqref{eq:explicitnormalizationxi}, we have in particular that $\xi(\proj{\mainr}{\id\hil}{,\hv}) = 1$ for all $\hv\in\ospace$. This translates as equation \eqref{eq:normsigma}.

Let us now prove that the normalization of $\sigma$ as in equation \eqref{eq:normsigma}
implies the full normalization of the ontic response function \eqref{eq:explicitnormalizationxi}:
\begin{multline}
\forall \hv\in\ospace,\forall K\in\mathbb N\cup\{+\infty\}, \\
\forall \left\{E_k\in\sete\st 
\textstyle \sum_{k=1}^K E_k = \id\hil\right\}\st \\
\sum_{k=1}^K \xi(\proj{\mainr}{E_k},\hv)
= \sum_{k=1}^K \scal{\sigma(\hv)}{\proj{\mainr}{E_k}}\mainr \\
= \scal{\sigma(\hv)}{\proj{\mainr}{\id\hil}}{\mainr}
= 1.
\end{multline}
The normalization of the ontic states as in equation \eqref{eq:NormMu} reads: for any $\red\rho\in\reds$,
\begin{multline}
\inthv{\mu(\red\rho,\hv)} 
= \inthv{\scal{\red\rho}{F(\hv)}{\mainr}} \\
= \inthv{\scal{\red\rho}{F(\hv)}{\mainr}\otrace{\sigma(\hv)}} \\
= \scal{\red\rho}{\proj{\mainr}{\id\hil}}{\mainr} = 1.
\end{multline}
We used first the normalization \eqref{eq:normsigma} of $\sigma$, then the consistency requirement \eqref{eq:consistency} and finally \cref{lem:traceone} to conclude.

Overall, we have shown that \cref{def:classicalmodel} implies the structure of \cref{prop:basiccriterion}, and that the latter suffices to recover a valid \vgrSC{classical}\vgr{operationally noncontextual ontological} model as in \cref{def:classicalmodel}.
\end{myproof}

The following general lemma proves the alternative expressions  $\polars = \mainr\cap\bigpolar{\sets}{\lh}$ and $\polare = \mainr\cap\bigpolar{\sete}{\lh}$.

\begin{lemma}
Let $\mathcal V$ be a finite dimensional real inner product space. Let $X\subseteq\mathcal V$ be any set. Let $\mathcal U\subseteq \mathcal V$ be a vector subspace of $\mathcal V$ equipped with the inner product inherited from $\mathcal V$. It holds that:
\begin{equation}
\label{eq:LemmaPuX}
\polar{\proj{\mathcal U}{X}}{\mathcal U}
=
\mathcal U \cap \polar{X}{\mathcal V}.
\end{equation}
\end{lemma}
\begin{myproof}
Let us prove that $\polar{\proj{\mathcal U}{X}}{\mathcal U}
\subseteq
\mathcal U \cap \polar{X}{\mathcal V}$. Let $u\in\polar{\proj{\mathcal U}{X}}{\mathcal U}$. Then, $u\in\mathcal U$ so it suffices to verify $u\in\polar{X}{\mathcal V}$. For all $x\in X$, using \cref{prop:basicproj},
\begin{equation}
\scal{u}{x}{\mathcal V} = \scal{u}{\proj{\mathcal U}{x}}{\mathcal U} \geq 0,
\end{equation}
where we used $u\in\polar{\proj{\mathcal U}{X}}{\mathcal U}$ to conclude. Thus, it holds that $u\in\mathcal U\cap\polar{X}{\mathcal V}$.

Let us now prove that $\mathcal U\cap\polar{X}{\mathcal V}\subseteq\polar{\proj{\mathcal U}{X}}{\mathcal U}$. Let $u'\in\mathcal U\cap\polar{X}{\mathcal V}$. For all $\red x\in\proj{\mathcal U}{X}$, choose $x\in X$ such that $\red x = \proj{\mathcal U}{x}$. Then, using \cref{prop:basicproj},
\begin{equation}
\scal{u'}{\red x}{\mathcal U} = \scal{u'}{x}{\mathcal V} \geq 0,
\end{equation}
where we used $u'\in\polar{X}{\mathcal V}$ and $x\in X$ to conclude.
\end{myproof}

\PropAncilla*

\begin{myproof}
Define
$$
\mainr = \proj{\spane}{\spans} \textup{ and } \tilde\mainr = \proj{\myspan{\tilde\sete}}{\myspan{\tilde\sets}}.
$$
Equation \eqref{eq:ancillastats} together with \cref{prop:reducedscalar} proves that for all $k,l\in I$,
\begin{equation}
\label{eq:reducedancillastats}
\scal{\proj{\mainr}{\rho_k}}{\proj{\mainr}{E_l}}{\mainr}
=
\big<\proj{\tilde\mainr}{\tilde\rho_k},P_{\tilde\mainr}(\tilde E_l)\big>_{\tilde\mainr}.
\end{equation}
To prove that $\mainr$ and $\tilde\mainr$ have the same dimension, we use that $\{P_\mainr(\rho_k)\}_{k\in I}$ spans $\mainr$ and $\{P_{\tilde\mainr}(\tilde\rho_k)\}_{k\in I}$ spans $\tilde\mainr$ (\cref{prop:spans}) and we prove that, for any $J\subseteq I$ (it suffices to take $J$ a discrete, finite set), $\{P_\mainr(\rho_k)\}_{k\in J}$ is linearly dependent in $\mainr$ if and only if $\{P_{\tilde\mainr}(\tilde\rho_k)\}_{k\in J}$ is linearly dependent in $\tilde\mainr$: if there exist $\{\alpha_k\in\mathbb R\}_{k\in J}$ not all zero such that
\begin{equation}
\sum_{k\in J} \alpha_k \proj{\mainr}{\rho_k} = 0,
\end{equation}
then for all $l\in I$,
\begin{align}
0 &= \sum_{k\in J} \alpha_k \scal{\proj{\mainr}{\rho_k}}{\proj{\mainr}{E_l}}{\mainr} \nonumber \\
&\hspace{-0.33cm}\overset{\eqref{eq:reducedancillastats}}{=} 
\Big<
\sum_{k\in J} \alpha_k P_{\tilde\mainr}(\tilde\rho_k),P_{\tilde\mainr}(\tilde E_l)
\Big>_{\tilde\mainr}.
\end{align}
By \cref{prop:spans}, this implies that also $\sum_{k\in J} \alpha_k P_{\tilde\mainr}(\tilde \rho_k) = 0$ which proves the ``only if" direction;
the symmetry in the definition of the two scenarios implies the converse direction.
This in turn proves that bases of $\mainr$ and $\tilde\mainr$ have the same cardinality: if there were a basis of $\mainr$ of the form $\{\proj{\mainr}{\rho_k}\}_{k\in J_0}$ with $J_0$ a discrete set consisting of $d\in\mathbb N$ elements and a basis of $\tilde\mainr$ of the form $\{P_{\tilde\mainr}(\tilde \rho_k)\}_{k\in \tilde J_0}$ with $\tilde J_0$ a discrete set of $\tilde d\in\mathbb N$ elements with $\tilde d > d$, then we would obtain a contradiction because these $\tilde d$ elements would have to be linearly dependent.

Now suppose that the scenario \resources{} admits \vgrSC{a classical}\vgr{an operationally noncontextual ontological} model with ontic space $\ospace$. This is equivalent to the requirement that there exist $\{F(\lambda) \in \polars\}_{\hv\in\ospace}$ and $\{ \sigma_\hv\in\polare\}_{\hv\in\ospace}$ as in \cref{prop:basiccriterion}. 
Using \cref{prop:spans}, it holds that there exist two sets of indices $J$ and $K$, each with $\dim(\mainr)$ many indices, and real numbers $\{f_j(\hv) \in \mathbb R\}_{j\in J,\hv\in\ospace}$, $\{s_j(\hv)\in\mathbb R\}_{j\in K,\hv\in\ospace}$ such that
\begin{subequations}
\begin{align}
F(\hv) &= \sum_{j\in J} f_j(\hv) \proj{\mainr}{E_j}, \\
\sigma(\hv) &= \sum_{j\in K} s_j(\hv) \proj{\mainr}{\rho_j}.
\end{align}
\end{subequations}
To define the \vgrSC{classical}\vgr{operationally noncontextual ontological} model for the scenario $(\tilde\sets,\tilde\sete)$, it suffices to define
\begin{subequations}
\begin{align}
\tilde F(\hv) &:= \sum_{j\in J} f_j(\hv) P_{\tilde\mainr}(\tilde E_j) \in \tilde\mainr, \\
\tilde \sigma(\hv) &:= \sum_{j\in K} s_j(\hv) P_{\tilde\mainr}(\tilde\rho_j) \in \tilde\mainr.
\end{align}
\end{subequations}
Equation \eqref{eq:reducedancillastats} implies by linearity that for all $k\in I$ and for all $\hv\in\ospace$,
\begin{subequations}
\begin{align}
\big<P_{\tilde\mainr}(\tilde\rho_k),\tilde F(\hv)\big>_{\tilde\mainr}
=
\scal{\proj{\mainr}{\rho_k}}{F(\hv)}{\mainr}, \\
\big<\tilde\sigma(\hv),P_{\tilde\mainr}(\tilde E_k)\big>_{\tilde\mainr}
=
\scal{\sigma(\hv)}{\proj{\mainr}{E_k}}{\mainr}.
\end{align}
\end{subequations}
From there on, it is trivial to see that the primitives $\{\tilde F(\hv)\}_{\hv\in\ospace}$ and $\{\tilde \sigma(\hv)\}_{\hv\in\ospace}$ satisfy the constraints of \cref{prop:basiccriterion}, and the symmetry in the definition of \resources{} versus $(\tilde\sets,\tilde\sete)$ allows one to conclude that the converse direction also holds.
\end{myproof}

\PropCoarseGrainings*

\begin{myproof}
First, it is clear that $\spane = \myspan{\sete_{\textup{ext}}}$ so that the equality of the reduced spaces $\mainr = \proj{\spane}{\spans}$ is clear. We now have to prove that $\polare = \polar{\proj{\mainr}{\sete_\textup{ext}}}{\mainr}$. Since $\sete \subseteq \sete_\textup{ext}$, by \cref{lem:SubPolar}, it holds that $\polar{\proj{\mainr}{\sete_\textup{ext}}}{\mainr} \subseteq \polare$. It remains to prove the reverse inclusion, which simply follows from the fact that if $\sigma\in\polare$, then also $\scal{\sigma}{\proj{\mainr}{\sum_{k=1}^N E_k}}{\mainr} \geq 0$ by linearity.
\end{myproof}

\section{\Unitsep{} and cardinality bounds}

\subsection{Generalized separability}
\label{app:sepes}

\subsubsection{Review of elementary analysis}

Let us first state some elementary results about convergence, sequences and closed sets. A more complete description can be found in~\cite{FunctionalAnalysis} for example.
Let $\mathcal V$ be a finite dimensional real inner product space. $\mathcal V$ is a complete normed space equipped with the norm induced by the inner product:
\begin{equation}
\forall v\in\mathcal V\st
\norm{v}{\mathcal V} := \sqrt{\scal{v}{v}{\mathcal V}}.
\end{equation}
A sequence $(v_k\in\mathcal V)_{k\in\mathbb N}$ is convergent if and only if there exists $v^*\in\mathcal V$ such that
\begin{equation}
\lim_{k\rightarrow\infty}v_k = v^*,
\end{equation}
which is a short hand notation to state that
\begin{equation}
\lim_{k\rightarrow\infty} \norm{v_k - v^*}{\mathcal V} = 0.
\end{equation}

Note that as a special case of the definition of a continuous function~\cite{FunctionalAnalysis}, any function $f\funcrange{\mathcal V}{\mathbb R}$ that is continuous has the property that for any convergent sequence $(v_k\in\mathcal V)_{k\in\mathbb N}$, it holds that
\begin{equation}
\textstyle \lim_{k\rightarrow\infty} f(v_k) = f(\lim_{k\rightarrow\infty} v_k).
\end{equation}

We state without proof the following lemmas. Their proofs are either simple exercises or stated explicitly in~\cite{FunctionalAnalysis}.

\begin{lemma}
\label{lem:NormScalContinuous}
The norm $\norm{\cdot}{\mathcal V}\funcrange{\mathcal V}{\mathbb R}$ is continuous, and for every fixed $v_0\in\mathcal V$, the scalar products $\scal{\cdot}{v_0}{\mathcal V}\funcrange{\mathcal V}{\mathbb R}$ and $\scal{v_0}{\cdot}{\mathcal V}\funcrange{\mathcal V}{\mathbb R}$ are also continuous.
\end{lemma}

\begin{lemma}
\label{lem:ClosedEquiv}
Any subset $X\subseteq \mathcal V$ is closed if and only if, for any sequence $(x_k\in X)_{k\in\mathbb N}$ that converges to $x^*\in\mathcal V$, the limit $x^*$ belongs to $X$. 
\end{lemma}

\begin{lemma}
\label{lem:ConvergentImpliesBounded}
Any convergent sequence $(v_k\in\mathcal V)_{k\in\mathbb N}$ is also a bounded sequence. This means that there exists a finite constant $C\in\mathbb R$ such that
\begin{equation}
\forall k\in\mathbb N\st 
\norm{v_k}{\mathcal V} \leq C.
\end{equation}
\end{lemma}

\begin{lemma}
\label{lem:SubsquenceLimit}
If a sequence $(v_k\in\mathcal V)_{k\in\mathbb N}$ converges, than for any subsequence defined by the strictly increasing set of indices $\{k_l\}_{l\in\mathbb N}\subseteq \mathbb N$, it holds that
\begin{equation}
\lim_{l\rightarrow\infty}v_{k_l}= \lim_{k\rightarrow\infty} v_k.
\end{equation}
\end{lemma}

\begin{lemma}
\label{lem:SumOfSequences}
Let $N\in\mathbb N$. For each $n=1,\dots, N$, let $(v_k^{(n)}\in\mathcal V)_{k\in\mathbb N}$ be a sequence that converges to $V^{(n)}\in\mathcal V$. Then, it holds that
\begin{equation}
\lim_{k\rightarrow\infty} \sum_{n=1}^N v_k^{(n)} = \sum_{n=1}^N V^{(n)}.
\end{equation}
\end{lemma}

\begin{lemma}
\label{lem:ProductOfSequences}
Let $N\in\mathbb N$. For each $n=1,\dots, N$, let $\mathcal V^{(n)}$ be an arbitrary real inner product space of finite dimension.
Let $(v_k^{(n)}\in\mathcal V^{(n)})_{k\in\mathbb N}$ be a real sequence that converges to $V^{(n)}\in\mathcal V^{(n)}$. Then, it holds that the limit of the tensor product equals the tensor product of the limits:
\begin{equation}
\lim_{k\rightarrow\infty} \bigotimes_{n=1}^N v_k^{(n)} = \bigotimes_{n=1}^N V^{(n)}.
\end{equation}
\end{lemma}
\begin{myproof}[Proof overview]
The first thing to show is that the limit of the product of two convergent sequences in $\mathbb R$ is equal to the product of the limits of the sequences. Then, generalize to any number of real sequences by recursion. Finally, expend the tensor products in any basis of the underlying vector spaces and apply the result derived for the real sequence case.
\end{myproof}

\begin{theorem}[Bolzano-Weierstrass theorem]
\label{th:bolzano}
Any bounded sequence $(v_k\in \mathcal V)_{k\in\mathbb N}$, where $\mathcal V$ is any finite-dimensional real inner product space, admits a convergent subsequence $(v_{k_l})_{l\in\mathbb N}$. Specifically, if the sequence $(v_k)_{k\in\mathbb N}$ satisfies, for some constant $C\in\mathbb R$ independent of $k$,
\begin{equation}
\forall k\in\mathbb N\st \norm{v_k}{\mathcal V} \leq C,
\end{equation}
then there exists a strictly increasing subset of indices, denoted $\{k_l\}_{l\in\mathbb N}\subseteq \mathbb N$, and there exists $v^*\in\mathcal V$ such that the subsequence $(v_{k_l})_{l\in\mathbb N}$ converges to $v^*$:
\begin{equation}
\lim_{l\rightarrow\infty} v_{k_l} = v^*.
\end{equation}
\end{theorem}
\begin{myproof}
We specialized the more general theorem~6.21 in~\cite{FunctionalAnalysis} according to the needs of the present matter.
\end{myproof}

\subsubsection{Generalized product operators}

Referring to the \cref{def:prodes} of the generalized product state set $\prodes$, let us first verify the following lemma.

\begin{lemma}
\label{prop:cclosedset}
$\prodes$ is a closed set.
\end{lemma}
\begin{proof}
Consider any sequence $(d_k\in\prodes)_{k\in\mathbb N}$. By definition of $\prodes$, there exist sequences
\begin{subequations}
\begin{align}
\left(a_k\in\polar\reds\mainr\right)_{k\in\mathbb N}, \\
\left(b_k\in\polar\rede\mainr\right)_{k\in\mathbb N}, 
\end{align}
\end{subequations}
such that $d_k = a_k\otimes b_k$ for all $k\in\mathbb N$.
Then, for all $k\in\mathbb N$, there always exist $m_k\in\mathcal R$ and $n_k\in\mathcal R$ such that
\begin{subequations}
\label{eq:tempnormmn}
\begin{align}
\hspace{1cm} a_k &= \norm{a_k}{\mainr}m_k, &\norm{m_k}{\mainr}&= 1, \label{eq:tempnormm}\\
b_k &= \norm{b_k}{\mainr}n_k, 
&\norm{n_k}{\mainr}&=1.\label{eq:tempnormn}
\end{align}
\end{subequations}
Now suppose that this sequence $(d_k)_{k\in\mathbb N}$ is convergent and converges to $d^*\in\mathcal R\otimes\mathcal R$. We want to show that $d^*\in\prodes$. We know
\begin{equation}
d^* = \lim_{k\rightarrow\infty} d_k = \lim_{k\rightarrow\infty} (\norm{a_k}{\mainr}\cdot\norm{b_k}{\mainr})(m_k\otimes n_k).
\end{equation}
The norm $\norm{\cdot}{\mainr\otimes\mainr}$ being continuous according to \cref{lem:NormScalContinuous}, it holds that
\begin{align}
\norm{d^*}{\mainr\otimes\mainr} &= \lim_{k\rightarrow\infty}\norm{d_k}{\mainr\otimes\mainr} \nonumber\\
&= \lim_{k\rightarrow\infty}(\norm{a_k}{\mainr}\cdot\norm{b_k}{\mainr}),
\end{align}
where we used $\norm{m_k\otimes n_k}{\mainr\otimes\mainr} = \norm{m_k}{\mainr}\cdot\norm{n_k}{\mainr} = 1$ according to \eqref{eq:tempnormmn}. This shows that the real sequence $(\norm{a_k}{\mainr}\cdot\norm{b_k}{\mainr}\in\mathbb R)_{k\in\mathbb N}$ converges to $\norm{d^*}{\mainr\otimes\mainr}$. Next, consider the sequence $(m_k\in\mainr)_{k\in\mathbb N}$: it is bounded in norm thanks to its normalization \eqref{eq:tempnormm}. By the Bolzano-Weierstrass \cref{th:bolzano}, we may extract a convergent subsequence with indices $\{j_l\in\mathbb N\}_{l\in\mathbb N}\subseteq \mathbb N$. The corresponding limit is denoted $m^*\in\mainr$, that is,
\begin{equation}
\lim_{l\rightarrow\infty} m_{j_l} = m^*.
\end{equation}
The sequence $(n_{j_l}\in\mainr)_{l\in\mathbb N}$ is also bounded from \eqref{eq:tempnormn} so by the Bolzano-Weierstrass \cref{th:bolzano} we can further extract a convergent subsequence with indices
\begin{equation}
\{k_l\in\mathbb N\}_{l\in\mathbb N} \subseteq \{j_l\}_{l\in\mathbb N} \subseteq \mathbb N,
\end{equation}
and we denote the corresponding limit $n^*\in\mainr$. With this further refinement of indices, both subsequences $(m_{k_l})_{l\in\mathbb N}$ (using \cref{lem:SubsquenceLimit}) and $(n_{k_l})_{l\in\mathbb N}$ converge in $\mainr$, i.e.,
\begin{subequations}
\begin{align}
\lim_{l\rightarrow\infty} m_{k_l} &= m^*,\\ \lim_{l\rightarrow\infty} n_{k_l} &= n^*.
\end{align}
\end{subequations}
Using \cref{lem:ProductOfSequences} applied to $\mathbb R\otimes \mainr\otimes\mainr$ to commute the limit and the product, and using \cref{lem:SubsquenceLimit} for the subsequence $(d_{k_l})_{l\in\mathbb N}$, we obtain:
\begin{multline}
d^* = \lim_{l\rightarrow\infty} d_{k_l} \\
= \left(\lim_{l\rightarrow \infty} \norm{a_{k_l}}{\mainr}\cdot\norm{b_{k_l}}{\mainr}\right)
\left(\lim_{l\rightarrow\infty}m_{k_l}\right)\otimes \left(\lim_{l\rightarrow\infty}n_{k_l}\right) \\
= \norm{d^*}{\mainr\otimes\mainr}(m^*\otimes n^*).
\end{multline}
To show $d^*\in\prodes$, it only remains to show that
\begin{subequations}
\begin{align}
m^* \in\polar\reds\mainr, \\
n^* \in\polar\rede\mainr.
\end{align}
\end{subequations}
First, note that for all $l\in\mathbb N$, due to $a_{k_l}\in\polar\reds\mainr$, it holds that for all $\red \rho\in\reds$:
\begin{equation}
\label{eq:temppolarm}
 0 \le \scal{a_{k_l}}{\red\rho}{\mainr} = \norm{a_{k_l}}{\mainr}\scal{m_{k_l}}{\red\rho}{\mainr}.
\end{equation}
For the indices $\{l\in\mathbb N\st \norm{a_{k_l}}{\mainr} > 0\}$, equation \eqref{eq:temppolarm} implies that $m_{k_l}\in\polar\reds\mainr$.
For the remaining indices $\{l\in\mathbb N\st \norm{a_{k_l}}{\mainr} = 0\}$, we can make an arbitrary choice in \eqref{eq:tempnormm} when we write $0 = a_{k_l} = 0\cdot m_{k_l}$: choose any normalized $m_{k_l}\in\polar\reds\mainr$ for these indices. This is always possible thanks to \cref{lem:emptysets}. This shows that for all $l\in\mathbb N$, $m_{k_l}\in\polar\reds\mainr$. Using \cref{lem:NormScalContinuous}, we may commute the scalar product with the limit to obtain, for all $\red\rho\in\reds$,
\begin{equation}
\scal{m^*}{\red\rho}{\mainr}
= \lim_{l\rightarrow\infty}\scal{m_{k_l}}{\red\rho}{\mainr} \geq 0.
\end{equation}
To conclude, we used that $\realpos$ is a closed interval of $\mathbb R$. This proves that $m^*\in\polar\reds\mainr$.
By an entirely analogous reasoning we obtain that $n^* \in\polar\rede\mainr$. This proves that for any converging sequence $(d_k\in\prodes)_{k\in\mathbb N}$, we have
\begin{gather}
\lim_{k\rightarrow\infty}d_k \in\prodes,
\end{gather}
which according to \cref{lem:ClosedEquiv} proves that $\prodes$ is closed.
\end{proof}

\subsubsection{Generalized separable operators}

Let us first prove the following proposition.

\begin{prop}[Specialized Carathéodory's theorem for convex cones]
\label{prop:finitesum}
For all $\Omega\in\sepes$, there exist $n\in\{1,\dots,\dim(\mainr)^2\}$ and families
\begin{subequations}
\begin{align}
\big\{F_i\in\polar\reds\mainr\big\}_{i=1}^{n}, \\
\big\{\sigma_i\in\polar\rede\mainr\big\}_{i=1}^n,
\end{align}
\end{subequations}
which satisfy
\begin{equation}
\Omega=\sum_{i=1}^n F_i\otimes\sigma_i.
\end{equation}
\end{prop}

\begin{myproof}
This proposition is the content of Carathéodory's theorem for convex cones as presented in theorem~4.3.2 in~\cite{ConvexAnalysis}. For completeness, we present a proof with the notation adapted to the context of this \doc.

Suppose that there exists $\Omega \in\sepes$ for which the shortest convex decomposition over $\prodes$ is of length $n \geq \dim(\mainr)^2+1$:
\begin{equation}
\label{eq:tempdecompomega}
\Omega= \sum_{i=1}^n F_i \otimes \sigma_i.
\end{equation}
Because the space $\mainr\otimes\mainr$ is of dimension $\dim(\mainr)^2$, any family of $n\geq \dim(\mainr)^2+1$ elements of $\mainr\otimes\mainr$ has to be linearly dependent: this is the case of the set $\{F_i\otimes\sigma_i\in\mainr\otimes\mainr\}_{i=1}^{n}$. This implies that there exist scalars $\{\alpha_i\in\mathbb R\}_{i=1}^n$ not all zero such that
\begin{equation}\label{eq:alphai0}
\sum_{i=1}^n \alpha_i(F_i\otimes \sigma_i) = 0.
\end{equation}
Suppose that for all $i=1,\dots,n$: $\alpha_i \leq0$. Because not all $\alpha_i$ are zero, there must exist $i$ such that $\alpha_i < 0$. In that case, replace all $\alpha_i$ by their opposite $-\alpha_i$ so that there now exists $i$ such that $\alpha_i > 0$.

Thus, without loss of generality, there must exists $i$ such that $\alpha_i > 0$.
We now can assert that $\max_j\alpha_j > 0$. Now, consider the following alternative decomposition of $\Omega$ where we subtracted a multiple of 0 in the form of \eqref{eq:alphai0} from the initial decomposition \eqref{eq:tempdecompomega}:
\begin{multline}
\Omega = \sum_{i=1}^n (F_i \otimes \sigma_i) - \frac{1}{\max_j\alpha_j}\sum_{i=1}^n \alpha_i (F_i\otimes \sigma_i) \\
= \sum_{i=1}^n \left(1-\frac{\alpha_i}{\max_j\alpha_j}\right) (F_i\otimes \sigma_i).
\end{multline}
Define
\begin{equation}
\theta_i := 1-\frac{\alpha_i}{\max_j \alpha_j}.
\end{equation}
For all $i=1,\dots,n$ we have
$
\theta_i \geq 0.
$
Now clearly, for $j_0$ such that $\max_j \alpha_j = \alpha_{j_0}$, we have that $\theta_{j_0} = 0$ which means we can rewrite $\Omega$ as a shorter positive linear combination of elements of $\prodes$:
\begin{equation}
\sum_{i\in\{1,\dots,n\}\setminus \{j_0\}} \theta_i(F_i\otimes \sigma_i).
\end{equation}
This yields the contradiction, and we conclude that any element of $\sepes$ can be written as a convex combination of at most $\dim(\mathcal R)^2$ elements of $\prodes$.
\end{myproof}

\begin{prop}
\label{prop:sepclosed}
$\sepes$ is a closed convex cone.
\end{prop}

Note that this is not entirely obvious. $\sepes$ is the convex hull of a closed \emph{unbounded} set, namely, $\prodes$. In general, the convex hull of a closed unbounded set may not be closed.

\begin{myproof}
Consider any converging sequence $(\Omega_k\in\sepes)_{k\in\mathbb N}$ with limit $\Omega^*\in\mainr\otimes\mainr$.
Let $I=\{1,\dots,\dim(\mathcal R)^2\}$.
Note that by \cref{prop:finitesum}, for all $k\in\mathbb N$, there exists a decomposition of $\Omega_k$ as $\dim(\mathcal R)^2$ elements of $\prodes$\footnote{One may have to pad shorter decompositions with $0\in\prodes$.} which we write as $\{d^{(k)}_i\in\prodes\}_{i\in I}$:
\begin{equation}
\Omega_k = \sum_{i\in I} d^{(k)}_i.
\end{equation}
Let us prove that for all $i\in I$, the sequence  $(d_i^{(k)})_{k\in\mathbb N}$ is bounded.
We have to show that there exists a finite upper bound $\lambda_i\in\mathbb R$ independent of $k$ such that
\begin{equation}
\norm{d_i^{(k)}}{\mainr\otimes\mainr} \leq \lambda_i.
\end{equation}

Let $\{R_m\in\mainr\otimes\mainr\}_{m\in I}$ be an orthonormal basis of $\mainr\otimes\mainr$.
Thanks to \cref{prop:spans}, we can also pick $\dim(\mainr)^2$ elements of the form
\begin{equation}
\left\{\red \rho_p\otimes\red E_p\st \red\rho_p \in\reds,\red E_p\in\rede\right\}_{p\in I}
\end{equation}
to obtain a basis of $\mainr\otimes\mainr$, although this basis will in general not be an orthonormal one.
The two bases $\{R_m\}_{m\in I}$ and $\{\red \rho_p\otimes\red E_p\}_{p\in I}$ are related by an invertible change of basis: there exists a $\dim(\mainr)^2\times \dim(\mainr)^2$ real, invertible matrix $Q$ with components $\{Q_{mp}\in\mathbb R\}_{m,p\in I}$ such that:

\begin{subequations}
\begin{align}
\forall m\in I\st 
R_m = \sum_{p\in I} Q_{mp} (\red \rho_p\otimes\red E_p), \\
\forall p\in I\st
\red \rho_p\otimes\red E_p = \sum_{m\in I} Q^{-1}_{pm} R_m.
\end{align}
\end{subequations}
Expanding the norm in the orthonormal basis $\{R_m\}_{m\in I}$, it holds that
\begin{align}
\nonumber &\norm{d_i^{(k)}}{\mainr\otimes\mainr}^2 \\
\nonumber &=  \scal{d_i^{(k)}}{
\sum_{m\in I}\scal{R_m} {d_i^{(k)}}{\mainr\otimes\mainr} R_m
}{\mainr\otimes\mainr} \\
\nonumber &= \sum_{m\in I} \scal{d_i^{(k)}}{R_m}{\mainr\otimes\mainr}^2 \\
&= \sum_{m\in I}\left(\sum_{p\in I}Q_{mp} \scal{d_i^{(k)}}{\red \rho_p\otimes\red E_p}{\mainr\otimes\mainr}\right)^2.
\end{align}
Then, using the triangle inequality for the absolute value:
\begin{equation}
\label{eq:TempUpperBoundD}
\norm{d_i^{(k)}}{\mainr\otimes\mainr}^2 
\leq 
\sum_{m\in I}\left(\sum_{p\in I}|Q_{mp}| \scal{d_i^{(k)}}{\red \rho_p\otimes\red E_p}{\mainr\otimes\mainr}\right)^2,
\end{equation}
where we used that due to $d_i^{(k)}\in\prodes$,
\begin{equation}
\label{eq:TempPosD}
\forall i,p\in I, \forall k\in\mathbb N\st
\scal{d_i^{(k)}}{\red\rho_p\otimes\red E_p}{\mainr\otimes\mainr} \geq 0,
\end{equation}
which allowed us to remove the absolute value off of these scalar products in \eqref{eq:TempUpperBoundD}. Then, let
\begin{equation}
\chi_1 := \sum_{m\in I}\left(\max_{p\in I}|Q_{mp}|\right)^2 \in\realpos.
\end{equation}
The upper bound \eqref{eq:TempUpperBoundD} becomes
\begin{equation}
\label{eq:TempBound2}
\norm{d_i^{(k)}}{\mainr\otimes\mainr}^2
\leq \chi_1 \left(\sum_{p\in I}\scal{d_i^{(k)}}{\red \rho_p\otimes\red E_p}{\mainr\otimes\mainr}\right)^2.
\end{equation}
Due to \eqref{eq:TempPosD}, $\forall i,p\in I, \forall k\in\mathbb N\st$
\begin{equation}
\hspace{-0.3cm}\scal{d_i^{(k)}}{\red\rho_p\otimes\red E_p}{\mainr\otimes\mainr} \leq \sum_{j\in I} \scal{d_j^{(k)}}{\red\rho_p\otimes\red E_p}{\mainr\otimes\mainr}.
\end{equation}
This allows us to upper-bound equation \eqref{eq:TempBound2} as
\begin{align}
\label{eq:TempBound3}
\nonumber \norm{d_i^{(k)}}{\mainr\otimes\mainr}^2
&\leq \chi_1 \left(\sum_{j,p\in I}\scal{d_j^{(k)}}{\red \rho_p\otimes\red E_p}{\mainr\otimes\mainr}\right)^2 \\
\nonumber &=\chi_1\left(\sum_{p\in I}\scal{\Omega_k}{\red \rho_p\otimes\red E_p}{\mainr\otimes\mainr}\right)^2 \\
\nonumber &= \chi_1 \left(\sum_{p,m\in I} Q^{-1}_{pm} \scal{\Omega_k}{R_m}{\mainr\otimes\mainr}\right)^2 \\
&\leq \chi_1 \left(\sum_{p,m\in I} |Q^{-1}_{pm}| \left|\scal{\Omega_k}{R_m}{\mainr\otimes\mainr}\right|\right)^2.
\end{align}
Let
\begin{equation}
\chi_2 := \left(\max_{m\in I} \sum_{p\in I} |Q^{-1}_{pm}| \right)^2 \in\realpos.
\end{equation}
Then, the bound becomes
\begin{align}
\label{eq:TempBound4}
\norm{d_i^{(k)}}{\mainr\otimes\mainr}^2
\leq \chi_1\chi_2 \left(\sum_{m\in I} \left|\scal{\Omega_k}{R_m}{\mainr\otimes\mainr}\right|\right)^2.
\end{align}
Note that for all $m\in I$:
\begin{equation}
\scal{\Omega_k}{R_m}{\mainr\otimes\mainr}^2 \leq \sum_{n\in I} \scal{\Omega_k}{R_n}{\mainr\otimes\mainr}^2  = \norm{\Omega_k}{\mainr\otimes\mainr}^2.
\end{equation}
The bound \eqref{eq:TempBound4} becomes
\begin{align}
\nonumber\norm{d_i^{(k)}}{\mainr\otimes\mainr}^2 &\leq \chi_1\chi_2\left(\sum_{m\in I} \norm{\Omega_k}{\mainr\otimes\mainr}\right)^2 \\
\nonumber&= \chi_1\chi_2 \card{I}^2 \norm{\Omega_k}{\mainr\otimes\mainr}^2 \\
&= \chi_1\chi_2\dim(\mainr)^4\norm{\Omega_k}{\mainr\otimes\mainr}^2.
\end{align}
The sequence $(\Omega_k)_{k\in\mathbb N}$ converges, so by \cref{lem:ConvergentImpliesBounded}, it is a bounded sequence: there exists $C\in\realpos$ such that for all $k\in\mathbb N$, $\norm{\Omega_k}{\mainr\otimes\mainr} \leq C$. We have shown that for all $i\in I$, for all $k\in\mathbb N$,
\begin{equation}
\label{eq:FinalBound}
\norm{d_i^{(k)}}{\mainr\otimes\mainr}^2 \leq \chi_1\chi_2\dim(\mainr)^4 C^2.
\end{equation}

We may now apply the Bolzano-Weierstrass \cref{th:bolzano} to the bounded sequence $(d_1^{(k)})_{k\in\mathbb N}$ to extract a first set of strictly increasing indices $\{a_l\}_{l\in\mathbb N}\subseteq \mathbb N$ such that the induced subsequence of $(d_1^{(a_l)})_{l\in\mathbb N}$ converges. Then, consider the subsequence $(d_2^{(a_l)})_{l\in\mathbb N}$.  Using \eqref{eq:FinalBound}, it is bounded as well, so that there exists a subset of strictly increasing indices 
\begin{equation}
\{b_l\}_{l\in\mathbb N} \subseteq \{a_l\}_{l\in\mathbb N} \subseteq \mathbb N
\end{equation}
such that the subsequence $(d_2^{(b_l)})_{l\in\mathbb N}$ converges. By \cref{lem:SubsquenceLimit}, the subsequence $(d_1^{(b_l)})_{l\in\mathbb N}$ converges to the same limit as $(d_1^{(a_l)})_{l\in\mathbb N}$. Repeat this procedure to obtain a new set of strictly increasing indices
\begin{equation}
\{c_l\}_{l\in\mathbb N} \subseteq \{b_l\}_{l\in\mathbb N} \subseteq \{a_l\}_{l\in\mathbb N} \subseteq \mathbb N
\end{equation}
so that the sequences $(d_1^{(c_l)})_{l\in\mathbb N}$, $(d_2^{(c_l)})_{l\in\mathbb N}$ and $(d_3^{(c_l)})_{l\in\mathbb N}$ converge, etc., and after $\dim(\mainr)^2$ steps, the process stops. We denote the final set of strictly increasing indices $\{k_l\}_{l\in\mathbb N}$, and we denote the limits as
\begin{equation}
\label{eq:limdi}
\forall i\in I\st d^*_i := \lim_{l\rightarrow\infty} d_i^{(k_l)} \in\prodes,
\end{equation}
where we used \cref{prop:cclosedset} to conclude that the limits lie in $\prodes$.
Note that the freedom in choosing the convergent subsequences from the bounded sequences is irrelevant: in any case, using \cref{lem:SubsquenceLimit}, the induced subsequence $(\Omega_{k_l})_{l\in\mathbb N}$ converges to $\Omega^*.$
Then, using \cref{lem:SumOfSequences} to commute the sum and the limit, 
\begin{multline}
\Omega^* = \lim_{k\rightarrow\infty} \Omega_k = \lim_{l\rightarrow\infty} \Omega_{k_l} = \lim_{l\rightarrow\infty}\sum_{i\in I} d_i^{(k_l)} \\ = \sum_{i\in I} d_i^*.
\end{multline}
Thanks to equation \eqref{eq:limdi}, this proves $\Omega^*\in\sepes$, and because the sequence $(\Omega_k)_{k\in\mathbb N}$ was arbitrary in $\sepes$, this proves that $\sepes$ is a closed set. 
\end{myproof}

\subsection{\choijamiolkowsky{} isomorphism}
\label{app:choi}

We now prove the consistency of the \cref{def:Choi} of the \choijamiolkowsky{} isomorphism. We restrict to the study of linear maps from $\mainr$ to $\mainr$, i.e., maps in $\lin\mainr$, but these results hold equally well should one replace $\mainr$ with any real inner product space of finite dimension.
\begin{lemma}
\label{lem:ChoiUnique}
For any $\Phi\in\lin\mainr$, if the \choijamiolkowsky{} operator $\choi\Phi$ exists, then it is unique.
\end{lemma}
\begin{myproof}
Suppose that there exist two operators $\choi\Phi,\tilde{\mathbb J}(\Phi)\in\mainr\otimes\mainr$ which satisfy the requirement of \cref{def:Choi}. We will show that $\choi\Phi = \tilde{\mathbb J}(\Phi)$. Indeed, by equation \eqref{eq:ChoiRelation}, for any $r,s\in\mainr$,
\begin{multline}
\scal{\choi\Phi - \tilde{\mathbb J}(\Phi)}{r\otimes s}{\mainr\otimes\mainr} = \scal{r}{\Phi(s)}{\mainr} - \scal{r}{\Phi(s)}{\mainr} \\
= 0.
\end{multline}
This being valid for any $r,s\in\mainr\otimes\mainr$, by the non-degeneracy of the scalar product we obtain
\begin{equation}
\choi\Phi = \tilde{\mathbb J}(\Phi).
\end{equation}
This proves that for any $\Phi\in\lin\mainr$, the \choijamiolkowsky{} operator $\choi\Phi$ is unique.
\end{myproof}

\begin{lemma}
\label{lem:ChoiConsistent}
The \cref{def:Choi} is consistent in that for all $\Phi\in\lin\mainr$, $\choi\Phi$ exists and is unique. Given an orthonormal basis $\{R_i\in\mainr\}_{i=1}^{\dim(\mainr)}$ of $\mainr$, it is given by
\begin{equation}
\label{eq:ChoiCoord}
\choi\Phi = \sum_{i=1}^{\dim(\mainr)} \Phi(R_i)\otimes R_i.
\end{equation}
\end{lemma}
\begin{myproof}
The existence may be proven as follows. One can always expand $\choi\Phi\in\mainr\otimes\mainr$ in the basis $\{R_i\otimes R_j\}_{i,j=1}^{\dim(\mainr)}$ of $\mainr\otimes\mainr$, with coefficients $\{j(\Phi)_{i,j}\in\mathbb R\}_{i,j=1}^{\dim(\mainr)}$:
\begin{subequations}
\begin{align}
\choi\Phi &= \sum_{i,j} j(\Phi)_{i,j} R_i\otimes R_j, \label{eq:TempChoiCoord}\\
j(\Phi)_{i,j} &= \scal{\choi\Phi}{R_i\otimes R_j}{\mainr\otimes\mainr}.
\end{align}
\end{subequations}
Then, using equation \eqref{eq:ChoiRelation}, for all $i,j=1,\dots,\dim(\mainr)$,
\begin{equation}
j(\Phi)_{i,j} = \scal{R_i}{\Phi(R_j)}{\mainr}.
\end{equation}
Inserting this result into \eqref{eq:TempChoiCoord},
\begin{equation}
\label{eq:TempChoiCoord2}
\choi\Phi = \sum_j\left(\sum_i \scal{R_i}{\Phi(R_j)}{\mainr} R_i\right)\otimes R_j.
\end{equation}
One recognizes the completeness relation for the basis $\{R_i\}_i$:
\begin{equation}
\label{eq:MainRCompleteness}
\textstyle \forall r\in\mainr\st
r = \sum_{i=1}^{\dim(\mainr)} \scal{R_i}{r}{\mainr}R_i.
\end{equation}
Inserting this result in equation \eqref{eq:TempChoiCoord2}, equation \eqref{eq:ChoiCoord} is readily obtained. This proves the existence of $\choi\mainr$ for any $\Phi\in\lin\mainr$, and the uniqueness follows from \cref{lem:ChoiUnique}.
\end{myproof}

\Cref{lem:ChoiConsistent} proves the expression for $\choi{\id\mainr}$ in \cref{prop:choiid} as a corollary.

\begin{lemma}
\label{lem:ChoiIsomorphism}
The \choijamiolkowsky{} mapping in \cref{def:Choi} is indeed an isomorphism. The inverse mapping, for any $\Omega\in\mainr\otimes\mainr$, is denoted $\invchoi{\Omega}\in\lin\mainr$ and is defined by the relations:
\begin{equation}
\label{eq:InvChoiRelation}
\forall r,s\in\mainr\st
\scal{r}{\invchoi{\Omega}(s)}{\mainr}
= \scal{\Omega}{r\otimes s}{\mainr\otimes\mainr}.
\end{equation}
The injectivity of $\choi\cdot$ is particularly interesting: for any two functions $\Phi_1,\Phi_2\in\lin\mainr$,
\begin{equation}
\choi{\Phi_1} = \choi{\Phi_2} \implies \Phi_1 = \Phi_2.
\end{equation}
\end{lemma}
\begin{myproof}
It suffices to prove that
\begin{subequations}
\begin{align}
\forall \Phi\in\lin\mainr\st
\invchoi{\choi\Phi} = \Phi, \\
\forall \Omega\in\mainr\otimes\mainr\st
\choi{\invchoi{\Omega}} = \Omega,
\end{align}
\end{subequations}
which follows easily from the relations \eqref{eq:ChoiRelation} and \eqref{eq:InvChoiRelation}.
\end{myproof}

The following lemma will prove useful in \app~\ref{app:criterion}.

\begin{lemma}
\label{lem:rankphi}
Let $\Phi\in\lin\mainr$. Suppose that there exists $n_\Phi$, and $a_i,b_i\in\mainr$ for $i=1,\dots,n_\Phi$ such that
\begin{equation}
\label{eq:rankphi}
\choi{\Phi} = \sum_{i=1}^{n_\Phi} a_i\otimes b_i.
\end{equation}
Then, the dimension $\rank\Phi$ of the image vector space of $\Phi$ satisfies
\begin{equation}
\rank{\Phi} \leq n_\Phi.
\end{equation}
\end{lemma}

\begin{myproof}
Let $\{R_i\in\mainr\}_{i=1}^{\dim(\mainr)}$ be an orthonormal basis of $\mainr$. Then, the linear map $\Phi\in\lin\mainr$ may be represented as a real matrix in this basis. We will be using the completeness relation of $\mainr$ in the form \eqref{eq:MainRCompleteness}:

\begin{align}
\label{eq:expansionphi}
\nonumber \Phi(r)
&= \sum_{k=1}^{\dim(\mainr)} \scal{R_k}{\Phi(r)}{\mainr} R_k \\
\nonumber &= \sum_{k=1}^{\dim(\mainr)} \scal{\choi\Phi}{R_k\otimes r}{\mainr\otimes\mainr} R_k \\
\nonumber &= \sum_{k=1}^{\dim(\mainr)}\sum_{i=1}^{n_\Phi} \scal{a_i}{R_k}\mainr\scal{b_i}{r}\mainr R_k \\
&= \sum_{i=1}^{n_\Phi} \scal{b_i}{r}\mainr a_i.
\end{align}
Clearly, this shows that the image vector subspace $\im\Phi := \{\Phi(r)\st r\in\mainr\}\subseteq\mainr$ satisfies
\begin{equation}
\im\Phi \subseteq \myspan{\{a_i\}_{i=1}^{n_\Phi}},
\end{equation}
which implies that the dimensions respect
\begin{multline}
\rank\Phi := \dim(\im\Phi) \\ \leq \dim\big(\myspan{\{a_i\}_{i=1}^{n_\Phi}}\big) \leq n_\Phi. \qedhere
\end{multline}
\end{myproof}

\subsection{The \unitsep{} criterion}
\label{app:criterion}

\ThMainCriterion*

\begin{myproof}

Suppose that there exists a Riemann integrable \vgrSC{classical}\vgr{operationally noncontextual ontological} model for \resources, so that equations \eqref{eq:simpleranges}, \eqref{eq:normsigma} and \eqref{eq:consistency} of \vgrSC{the basic classicality criterion}\cref{prop:basiccriterion} are verified. It is easy to show that the \cref{def:RiemannIntegrableModel} of a Riemann integrable \vgrSC{classical}\vgr{operationally noncontextual ontological} model implies that the right-hand side of the consistency requirement \eqref{eq:consistency} can be rewritten as:
\begin{multline}
\inthv{\scal{r}{F(\hv)}{\mainr}\scal{\sigma(\hv)}{s}{\mainr}} \\
= \lim_{N\rightarrow\infty}\sum_{k=1}^N \Delta_{N,k} \scal{r}{F\big(\hvdis_{N,k}\big)}\mainr\scal{\sigma\big(\hvdis_{N,k}\big)}{s}{\mainr}.
\end{multline}
Rewriting the right-hand side in tensor product form using the scalar product property \eqref{eq:ScalarTensor}:
\begin{multline}
\inthv{\scal{r}{F(\hv)}{\mainr}\scal{\sigma(\hv)}{s}{\mainr}} =\\
\lim_{N\rightarrow\infty} \scal{\sum_{k=1}^N \Delta_{N,k} F\big(\hvdis_{N,k}\big)\otimes\sigma\big(\hvdis_{N,k}\big)}{r\otimes s}{\mainr\otimes\mainr}.
\end{multline}
Define for all $N\in\mathbb N$:
\begin{equation}
\label{eq:DefAn}
A_N := \sum_{k=1}^N \Delta_{N,k} F\big(\hvdis_{N,k}\big)\otimes\sigma\big(\hvdis_{N,k}\big).
\end{equation}
Then, the left-hand side of equation \eqref{eq:consistency} can be rewritten using the identity map and the defining property of the \choijamiolkowsky{} isomorphism as in \eqref{eq:ChoiRelation}:
\begin{equation}
\scal{r}{s}{\mainr} = \scal{r}{\id{\mainr}(s)}{\mainr} = \scal{\choi{\id\mainr}}{r\otimes s}{\mainr\otimes\mainr}.
\end{equation}
Thus, equation \eqref{eq:consistency} is equivalent to: $\forall r,s\in\mainr,$
\begin{equation}
\label{eq:tempchoiderivation}
\scal{\choi{\id\mainr}}{r\otimes s}{\mainr\otimes\mainr} = \lim_{N\rightarrow\infty} \scal{A_N}{r\otimes s}{\mainr\otimes\mainr}.
\end{equation}
Equation \eqref{eq:tempchoiderivation} shows in particular that all the components of $A_N$ converge to the components of $\choi{\id\mainr}$, which proves the convergence of the sequence $(A_N)_{N\in\mathbb N}$:
\begin{equation}
\label{eq:ChoiLimAn}
\lim_{N\rightarrow\infty} A_N = \choi{\id\mainr}.
\end{equation}
By the definition of $A_N$ in \eqref{eq:DefAn}, the nonnegativity of $\Delta_{N,k}$ as in \eqref{eq:RangeDelta}, the domains of $F$, $\sigma$ as in \eqref{eq:simpleranges} and the \cref{def:sepes} of $\sepes$:
\begin{equation}
\forall N\in\mathbb N\st
A_N\in\sepes.
\end{equation}
\Cref{prop:sepclosed} proved that $\sepes$ is a closed set within $\mainr\otimes\mainr$, so it holds that
\begin{equation}
\lim_{N\rightarrow\infty} A_N \in\sepes.
\end{equation}
and thus also, by \eqref{eq:ChoiLimAn}, that
\begin{equation}
\choi{\id\mainr} \in\sepes.
\end{equation}
We have proven that if the scenario \resources{} admits a Riemann integrable \vgrSC{classical}\vgr{operationally noncontextual ontological} model of the form of  \cref{def:classicalmodel,def:RiemannIntegrableModel}, then $\choi{\id\mainr}\in\sepes$.

Let us consider the other direction: suppose that $\choi{\id\mainr}\in\sepes$. By \cref{prop:finitesum}, there exist $n\leq\dim(\mainr)^2$ and
\begin{subequations}
\label{eq:tildefsigma}
\begin{align}
\big\{\tilde F_i\in\polar\reds\mainr\big\}_{i=1}^n, \\
\big\{\tilde\sigma_i\in\polar\rede\mainr\big\}_{i=1}^n,
\end{align}
\end{subequations}
such that
\begin{equation}
\label{eq:decompchoi}
\choi{\id\mainr} = \sum_{i=1}^n \tilde F_i\otimes\tilde\sigma_i.
\end{equation}
If we assume that any zero element in the decomposition \eqref{eq:decompchoi} has been removed, then for all $i=1,\dots,n$ we may assume $\tilde\sigma_i\neq0$ which also implies according to \cref{lem:nulltrace} that
\begin{equation}
\label{eq:postraces} \forall i=1,\dots,n\st\otrace{\tilde\sigma_i} > 0.
\end{equation}
Let, for all $i=1,\dots,n:$
\begin{subequations}
\label{eq:correctprimitives}
\begin{align}
 F_i &:= \otrace{\tilde\sigma_i}\tilde F_i, \label{eq:corretprimitivesf}\\
 \sigma_i &:= \left(\otrace{\tilde\sigma_i}\right)^{-1}\tilde\sigma_i. \label{eq:correctprimitivessigma}
\end{align}
\end{subequations}
We now show that the $F_i$'s and $\sigma_i$'s of equations \eqref{eq:correctprimitives} constitute a valid \vgrSC{classical}\vgr{operationally noncontextual ontological} model as framed in \cref{prop:basiccriterion}.
First, the nonnegativity requirements of equation \eqref{eq:simpleranges} are verified thanks to equations \eqref{eq:tildefsigma}, \eqref{eq:postraces} and \eqref{eq:correctprimitives}:
\begin{subequations}
\begin{align}
F_i&\in\polar\reds\mainr, \\
\sigma_i&\in\polar\rede\mainr.
\end{align}
\end{subequations}
The normalization in \eqref{eq:normsigma} is verified:
\begin{equation}
\forall i=1,\dots,n\st \otrace{\sigma_i} = 1
\end{equation}
as can be seen from equation \eqref{eq:correctprimitivessigma}.
Finally, the reproduction of quantum statistics in equation \eqref{eq:consistency} is still verified: indeed, from equations \eqref{eq:decompchoi} and \eqref{eq:correctprimitives},
\begin{equation}
\choi{\id\mainr} = \sum_{i=1}^n F_i\otimes\sigma_i,
\end{equation}
which in turns implies $\forall r,s\in\mainr:$
\begin{equation}
\label{eq:TempConsistencyMainCrit}
\scal{r}{s}{\mainr} = \sum_{i=1}^n \scal{r}{ F_i}{\mainr}\scal{\sigma_i}{s}{\mainr}.
\end{equation}
It is easy to see that such a model is Riemann integrable in the sense of \cref{def:RiemannIntegrableModel}.

We have thus shown that $\choi{\id\mainr}\in\sepes$ if and only if the scenario \resources{} admits a Riemann integrable \vgrSC{classical}\vgr{operationally noncontextual ontological} model.
\end{myproof}

\subsection{Ontic space cardinality}
\label{app:card}

We now prove \cref{th:CardinalityBounds}.

\ThCardinalityBounds*

\begin{myproof}
Suppose that \resources{} admits \vgrSC{a classical}\vgr{an operationally noncontextual ontological} model with discrete ontic space $\ospace = \{1,\dots,\card\ospace\}$. Note that such a model is automatically Riemann integrable according to \cref{def:RiemannIntegrableModel}: the primitives $\Delta$ and $\hvdis$ take the form
\begin{subequations}
\begin{align}
\Delta_{N,k} &= \left\{
\begin{aligned}
&1\text{ if } k\leq\card\ospace, \\
&0\text{ else,}
\end{aligned}\right. \\
\hvdis_{N,k} &= \min\big(k,\card{\ospace}\big).
\end{align}
\end{subequations}
Building upon the proof of the \unitsep{} criterion, \cref{th:maincriterion}, we see that equations \eqref{eq:DefAn} and \eqref{eq:ChoiLimAn} taken together in the case when $\ospace$ is discrete imply
\begin{equation}
\label{eq:fakedecomp}
\choi{\id\mainr} = \sum_{i=1}^{\scard\ospace} F_i\otimes \sigma_i.
\end{equation}
Equation \eqref{eq:fakedecomp} together with 
\cref{lem:rankphi} imply
\begin{equation}
\card\ospace \geq \rank{\id\mainr} = \dim(\mainr).
\end{equation}
This proves the first part of \cref{th:CardinalityBounds}.

We now prove the second part.
If there exists a Riemann integrable \vgrSC{classical}\vgr{operationally noncontextual ontological} model for \resources, then $\choi{\id\mainr} \in\sepes$ by the \unitsep{} criterion \cref{th:maincriterion}. By \cref{prop:finitesum}, there exists a decomposition of $\choi{\id\mainr}$ over $n$ elements of $\prodes$ where $n\leq\dim(\mainr)^2$. 
Assume that $n$ is minimal, i.e., that this decomposition of $\choi{\id\mainr}$ over elements of $\prodes$ is the shortest one.
In particular this implies that there are no zero elements in the decomposition so that we are in the case considered in equation \eqref{eq:correctprimitives} in the proof of \cref{th:maincriterion}. This decomposition allows one to construct a valid \vgrSC{classical}\vgr{operationally noncontextual ontological} model of cardinality $\card\ospace = n$ (where $n\leq\dim(\mainr)^2$) as demonstrated in, e.g., equation~\eqref{eq:TempConsistencyMainCrit}. This cardinality is minimal: if there was an ontic space of cardinality $n' < n$, then equation \eqref{eq:fakedecomp} would yield a decomposition of $\choi{\id\mainr}$ over $n'<n$ elements of $\prodes$ whereas $n$ was assumed minimal. The already proven first part of \cref{th:CardinalityBounds} also proves that $n \geq \dim(\mainr)$. 
Finally, the fact that $\dim(\mainr)^2\leq \dim(\lh)^2$ follows easily from the fact that $\mainr\subseteq\lh$.
\end{myproof}

\section{Alternative reduced spaces}

\subsection{Construction of the alternative \vgr{operationally noncontextual ontological} model}
\label{app:altr}

Let us restate and prove \cref{prop:scalaralt}:

\PropScalarAlt*

\begin{myproof}
First off, the choice $\mainr=\proj{\spane}{\spans}$ together with $f,g=P_{\mainr}$ satisfies \eqref{eq:AltScal} by virtue of \cref{prop:reducedscalar}, and satisfies equations \eqref{eq:spanfs}, \eqref{eq:spange} by virtue of \cref{prop:spans}. Thus, this choice fits in as a special case of \cref{def:AltReducedSpace}.

Now consider the choice $\mainr'=\proj{\spans}{\spane}$ together with $f,g = P_{\mainr'}$. It satisfies \eqref{eq:AltScal} as a corollary of \cref{prop:reducedscalargen}, and equations \eqref{eq:spanfs} and \eqref{eq:spange} are verified as an application of \cref{prop:spansgen}.
\end{myproof}

We will make use of the following lemma.

\begin{lemma}
\label{corollary:Reflection}
Consider any choice of $\altr\mainr$, $f$ and $g$ as in \cref{def:AltReducedSpace}. For all $s\in\spans$, for all $e\in\spane$, it holds that
\begin{equation}
\scal{\proj{\mainr}{s}}{\proj{\mainr}{e}}{\mainr}
=
\scal{f(s)}{g(e)}{\altr\mainr}.
\end{equation}
\end{lemma}
\begin{myproof}
It suffices to extend by linearity \cref{prop:reducedscalar} and \eqref{eq:AltScal} of \cref{def:AltReducedSpace}:
\begin{subequations}
\begin{align}
\scal{\proj{\mainr}{s}}{\proj{\mainr}{e}}{\mainr} &= \scal{s}{e}{\lh}, \\
\scal{f(s)}{g(e)}{\altr\mainr} &= \scal{s}{e}{\lh}. \qedhere
\end{align}
\end{subequations}
\end{myproof}

We now restate and prove the equality between the dimensions of the alternative reduced spaces.

\PropAltDimensions*

\begin{myproof}
We will first prove the existence of an invertible linear map between the two vector spaces $\altr\mainr$ and $\mainr$. Then, theorem~2.35 in~\cite{LinearAlgebra} allows to conclude that $\dim(\altr\mainr)=\dim(\mainr)$.

Let $d:= \dim(\mainr)$ and $\altr d := \dim(\altr\mainr)$ (which is finite by \cref{def:AltReducedSpace}). Let $\{T_i\in\altr\mainr\}_{i=1}^{\altr d}$ be an orthonormal basis of $\altr\mainr$. By equation \eqref{eq:spanfs}, it is possible to find for all $i=1,\dots,\altr d$ an element $s_i\in\spans$ such that $T_i = f(s_i)$.

Now let $\{R_j\in\mainr\}_{j=1}^d $ be an orthonormal basis of $\mainr$. Then, using \cref{prop:spans}, choose for all $j=1,\dots,d$ elements $u_j\in\spans$ such that $R_j = \proj{\mainr}{u_j}$.

Let $\Phi\funcrange{\mainr}{\altr\mainr}$ and $\phi\funcrange{\altr\mainr}{\mainr}$ be defined by: for all $r\in\mainr$, for all $t\in\altr\mainr$,
\begin{align}
\Phi(r) &= \sum_{i=1}^{\altr d} \scal{ \proj{\mainr}{s_i} }{r}{\mainr} T_i, \\
\phi(t) &= \sum_{j=1}^d \scal{ f(u_j) }{t}{\altr\mainr} R_j.
\end{align}
The main property of $\Phi$ is that, for any $e\in\spane$, $\Phi(\proj{\mainr}{e}) = g(e)$. Indeed, using \cref{corollary:Reflection},
\begin{align}
\nonumber \Phi(\proj{\mainr}{e}) &= \sum_{i=1}^{\altr d} 
\scal{\proj{\mainr}{s_i}}{\proj{\mainr}{e}}{\mainr} T_i \\
\nonumber  &= \sum_{i=1}^{\altr d} \scal{f(s_i)}{g(e)}{\altr\mainr} T_i \\
&= \sum_{i=1}^{\altr d}\scal{T_i}{g(e)}{\altr\mainr} T_i
= g(e). \label{eq:phiprege}
\end{align}
In the last line, we used the completeness relation of $\altr\mainr$. Similarly, for any $e\in\spane$, it holds that $\phi(g(e)) = \proj{\mainr}{e}$. Indeed:
\begin{align}
\nonumber \phi(g(e)) &= \sum_{j=1}^{d} 
\scal{f(u_j)}{g(e)}{\mainr} R_j \\
\nonumber  &= \sum_{j=1}^{d} \scal{\proj{\mainr}{u_j}}{\proj{\mainr}{e}}{\mainr} R_j \\
&= \sum_{j=1}^{d}\scal{R_j}{\proj{\mainr}{e}}{\mainr} R_j
= \proj{\mainr}{e}, \label{eq:phigepre}
\end{align}
where we used the completeness relation of $\mainr$.

Let us now compute $\phi(\Phi(r))$ for any $r\in\mainr$. By \cref{prop:spans}, there must exist $e\in\spane$ such that $r = \proj{\mainr}{e}$. Using \eqref{eq:phiprege} and \eqref{eq:phigepre}:
\begin{multline}
\phi(\Phi(r)) = \phi(\Phi(\proj{\mainr}{e})) = \phi(g(e)) = \proj{\mainr}{e} \\ = r.
\end{multline}
Similarly, for any $t\in\altr\mainr$, there exists by equation \eqref{eq:spange} $e'\in\spane$ such that $t=g(e')$. Then, using \eqref{eq:phiprege} and \eqref{eq:phigepre} again:
\begin{multline}
\Phi(\phi(t)) = \Phi(\phi(g(e')) = \Phi(\proj{\mainr}{e'}) = g(e') \\ = t.
\end{multline}
Thus we have proven that $\phi = \Phi^{-1}$, and hence $\Phi\funcrange{\mainr}{\altr\mainr}$ is an invertible linear map. The claim follows follows from the fact that two vector spaces are isomorphic if and only if they have the same dimension as proven, e.g., in theorem~2.35 of~\cite{LinearAlgebra}.
\end{myproof}

Let us now define the \vgrSC{classical}\vgr{operationally noncontextual ontological} model on any alternative reduced space $\altr\mainr$.

\begin{definition}
\label{def:altclassicalmodel}
The \vgrSC{classical}\vgr{operationally noncontextual ontological} model for \resources{} on a given alternative reduced space $\altr\mainr$ is specified as follows. $\altr\mainr$ and the mappings $f$ and $g$ are the primitives of \cref{def:AltReducedSpace}. Let $\altr\ospace$ be the ontic space.
Let $\altr\mu$ be the ontic state mapping that has domain
\begin{equation}
\altr\mu\funcrange{f(\sets)\times\altr\ospace}{\mathbb R}
\end{equation}
and that satisfies
\begin{subequations}
\begin{align}
\forall \red\rho\in f(\sets)\st& \myint{\altr\ospace}{\hv}{\altr\mu(\red\rho,\hv)} = 1, \\
\hspace{-0.3cm}\forall\hv\in\altr\ospace,\forall \red\rho\in f(\sets)\st& \altr\mu(\red\rho,\hv) \geq 0, 
\end{align}
\begin{multline*}
\forall\hv\in\altr\ospace,\forall p\in[0,1],\forall \red\rho_1,\red\rho_2\in f(\sets)\st \\
\end{multline*}
\vspace{-1.4cm}
\begin{multline}
\altr\mu(p\red\rho_1 + (1-p)\red\rho_2,\hv) \\
= p \altr\mu(\red\rho_1,\hv)
+ (1-p)\altr\mu(\red\rho_2,\hv). 
\end{multline}
\end{subequations}
Let $\altr\xi$ be the response function mapping that has domain
\begin{subequations}
\begin{equation}
\altr\xi\funcrange{ g(\sete)\times\ospace}{\mathbb R}
\end{equation}
and that satisfies
\begin{multline}
\forall \hv\in\altr\ospace,\forall K\in\mathbb N\cup\{+\infty\}, \\
\forall \left\{ E_k\in\sete\st\textstyle{\sum_{k=1}^K E_k} = \id\hil\right\}_{k=1}^K: \\
\textstyle \sum_{k=1}^K \altr\xi(g(E_k),\hv) = 1, 
\end{multline}
\vspace{-0.7cm}
\begin{align}
&\forall\hv\in\altr\ospace,\forall \red E\in g(\sete)\st&&\hspace{0.78cm} \altr\xi(\red E,\hv) \geq 0,\hspace{0.05cm}
\end{align}
\vspace{-0.75cm}
\begin{multline*}
\forall\hv\in\altr\ospace,\forall p\in[0,1],\forall \red E_1,\red E_2\in g(\sete)\st \\
\end{multline*}
\vspace{-1.4cm}
\begin{multline}
\altr\xi(p\red E_1 + (1-p)\red E_2,\hv) \\
= p \altr\xi(\red E_1,\hv)
+ (1-p)\altr\xi(\red E_2,\hv).
\end{multline}
\end{subequations}
The \vgrSC{classical}\vgr{operationally noncontextual ontological} model is required to reproduce the statistics that quantum mechanics predicts for the available states and measurements --- this is formulated using \eqref{eq:AltScal}:
\begin{multline}
\forall\red \rho\in f(\sets),\forall \red E\in g(\sete)\st
\\
\scal{\red\rho}{\red E}{\altr\mainr}
= \myint{\altr\ospace}{\hv}{\altr\mu(\red \rho,\hv)\altr\xi(\red E,\hv)}.
\end{multline}
\end{definition}

\begin{lemma}
\label{corollary:traceonealt}
For any reduced space $\altr\mainr$ (\cref{def:AltReducedSpace}), for all $\red\rho\in f(\sets)$,
\begin{equation}
\scal{\red\rho}{g(\id\hil)}{\altr\mainr} = 1.
\end{equation}
\end{lemma}
\begin{myproof}
Simply note that there must exist $\rho\in\sets$ such that $\red\rho = f(\rho)$, and then by virtue of \eqref{eq:AltScal},
\begin{gather}
\scal{\red\rho}{g(\id\hil)}{\altr\mainr} = \scal{\rho}{\id\hil}{\lh} = 1. \qedhere
\end{gather}
\end{myproof}

\begin{prop}
\label{prop:LinExtAlt}
Let $\hv\in\altr\ospace$ be arbitrary. Starting from the convex-linear mappings
\begin{subequations}
\begin{align}
\altr\mu(\cdot,\hv)&\funcrange{ f(\sets)}{\mathbb R}, \\
\altr\xi(\cdot,\hv)&\funcrange{ g(\sete)}{\mathbb R},
\end{align}
\end{subequations}
there exist unique linear extensions
\begin{subequations}
\begin{align}
\altr\mu(\cdot,\hv)\funcrange{\altr\mainr}{\mathbb R}, \\
\altr\xi(\cdot,\hv)\funcrange{\altr\mainr}{\mathbb R}.
\end{align}
\end{subequations}
\end{prop}

\begin{myproof}
The proof is the same as those of \cref{prop:extensionmu} and \cref{prop:extensionxi}: the same constructions apply in this case. The results which needed to be verified are the span assumptions \eqref{eq:spanfs} and \eqref{eq:spange} as well as \cref{corollary:traceonealt} for the case of $\altr\mu$: replace $\tr{\hil}{\red\rho}$ with $\scal{\red\rho}{g(\id\hil)}{\altr\mainr}$ in the proof of \cref{prop:extensionmu}.
\end{myproof}

\begin{theorem}[Basic criterion for the existence of \vgr{an operationally noncontextual ontological} model on $\altr\mainr$]
\label{th:BasicCriterionAlt}
Given \resources{} that lead to an alternative reduced space $\altr\mainr$ with associated mappings $f,g$ (\cref{def:AltReducedSpace}), there exists \vgrSC{a classical}\vgr{an operationally noncontextual ontological} model on $\altr\mainr$ with ontic space $\altr\ospace$ if and only if there exist mappings $\altr F$, $\altr\sigma$ with ranges
\begin{subequations}
\label{eq:SimpleRangesAlt}
\begin{align}
\altr F\funcrange{\altr\ospace}{\bigpolar{f(\sets)}{\altr\mainr}}, \\
\altr\sigma\funcrange{\altr\ospace}{\bigpolar{g(\sete)}{\altr\mainr}},
\end{align}
\end{subequations}
satisfying the normalization condition
\begin{equation}
\label{eq:NormSigmaAlt}
\forall\hv\in\altr\ospace\st
\scal{\altr\sigma(\hv)}{g(\id\hil)}{\altr\mainr} = 1
\end{equation}
as well as the consistency requirement: $\forall r,s\in\altr\mainr$,
\begin{equation}
\label{eq:ConsistencyAlt}
\scal{r}{s}{\altr\mainr} = \myint{\altr\ospace}{\hv}{\scal{r}{\altr F(\hv)}{\altr\mainr}
\scal{\altr\sigma(\hv)}{s}{\altr\mainr}}.
\end{equation}
\end{theorem}
\begin{myproof}
It suffices to apply Riesz' representation \cref{th:Riesz} to the linear extensions of \cref{prop:LinExtAlt} and to read off their properties from the \cref{def:altclassicalmodel} of the \vgrSC{classical}\vgr{operationally noncontextual ontological} model similarly to what was done in the proof of \cref{prop:basiccriterion}.
\end{myproof}

\subsection{Equivalence of reduced spaces}
\label{app:equivalence}

Let us restate and prove \cref{th:EquivalenceReducedSpaces}.

\ThEquivalenceReducedSpaces*

\begin{myproof}
Suppose first that \resources{} admits \vgrSC{a classical}\vgr{an operationally noncontextual ontological} model on $\mainr$ in the form of \cref{def:classicalmodel} with ontic space $\ospace$: we will show that this implies the existence of \vgrSC{a classical}\vgr{an operationally noncontextual ontological} model for \resources{} on $\altr\mainr$ equipped with the same ontic space $\ospace$.

The existence of \vgrSC{a classical}\vgr{an operationally noncontextual ontological} model for \resources{} of $\mainr$ is equivalent to the existence of $\ospace$, $F$ and $\sigma$ as in \cref{prop:basiccriterion}. Consider \cref{prop:spans}, which we restate in a slightly different form:
\begin{subequations}
\begin{align}
\proj{\mainr}{\spans} &= \mainr, \\
\proj{\mainr}{\spane} &= \mainr.
\end{align}
\end{subequations}
This implies that for any element $r\in\mainr$, there exists $s\in\spans$ such that $r=\proj{\mainr}{s}$; and there exists $e\in\spane$ such that $r=\proj{\mainr}{e}$. 
Applying this reasoning for every $\hv\in\ospace$ to $F(\hv)\in\polars\subseteq\mainr$, $\sigma(\hv)\in\polare\subseteq\mainr$, there must exist mappings
\begin{subequations}
\begin{align}
S&\funcrange{\ospace}{\spans}, \\
E&\funcrange{\ospace}{\spane},
\end{align}
\end{subequations}
such that for all $\hv\in\ospace$, the primitives from \cref{prop:basiccriterion} satisfy
\begin{subequations}
\label{eq:DefES}
\begin{align}
F(\hv) &= \proj{\mainr}{E(\hv)}, \\
\sigma(\hv) &= \proj{\mainr}{S(\hv)}. 
\end{align}
\end{subequations}
At this point, the existence of \vgrSC{a classical}\vgr{an operationally noncontextual ontological} model on $\mainr$ is equivalent to the existence of the mappings $S$ and $E$ as in \eqref{eq:DefES}. But the existence of these mappings implies the existence of a valid \vgrSC{classical}\vgr{operationally noncontextual ontological} model on $\altr\mainr$: indeed, define for all $\hv\in\ospace$,
\begin{subequations}
\begin{align}
\altr F(\hv) &:= g(E(\hv)),\\
\altr\sigma(\hv) &:= f(S(\hv)).
\end{align}
\end{subequations}
Let us verify that the primitives $\altr F$, $\altr \sigma$ satisfy all the requirements of \cref{th:BasicCriterionAlt},
starting with the verification of \eqref{eq:ConsistencyAlt}.
Let $r,s\in\altr\mainr$ be arbitrary. By equations \eqref{eq:spanfs} and \eqref{eq:spange}, there exist $r'\in\spans$, $s'\in\spane$ such that $r=f(r')$ and  $s=g(s')$. Using \cref{corollary:Reflection}:
\begin{multline}
\scal{r}{s}{\altr\mainr} = \scal{f(r')}{g(s')}{\altr\mainr} \\
= \scal{\proj{\mainr}{r'}}{\proj{\mainr}{s'}}{\mainr}.
\end{multline}
Using \eqref{eq:consistency} to expand the last term of the previous equation, the definition \eqref{eq:DefES} of the mappings $E$, $S$ and \cref{corollary:Reflection}:
\begin{align}
\nonumber &\scal{r}{s}{\altr\mainr}
= \inthv{\scal{\proj{\mainr}{r'}}{F(\hv)}{\mainr}
\scal{\sigma(\hv)}{\proj{\mainr}{s'}}{\mainr}} \\
\nonumber &= \inthv{\scal{\proj{\mainr}{r'}}{\proj{\mainr}{E(\hv)}}{\mainr}
\scal{\proj{\mainr}{S(\hv)}}{\proj{\mainr}{s'}}{\mainr}} \\
\nonumber &= \inthv{\scal{f(r')}{g(E(\hv))}{\mainr}
\scal{f(S(\hv))}{g(s')}{\mainr}} \\
&= \inthv{\scal{r}{\altr F(\hv)}{\altr\mainr}
\scal{\altr\sigma(\hv)}{s}{\altr\mainr}
}.
\end{align}
This derivation being valid for all $r,s\in\altr\mainr$ proves \eqref{eq:ConsistencyAlt}.
Equations \eqref{eq:SimpleRangesAlt} and \eqref{eq:NormSigmaAlt} are verified similarly. This proves that if there exists \vgrSC{a classical}\vgr{an operationally noncontextual ontological} model for \resources{} constructed on $\mainr$ with ontic space $\ospace$, then there exists \vgrSC{a classical}\vgr{an operationally noncontextual ontological} model for \resources{} constructed on $\altr\mainr$ with the same ontic space $\ospace$.

The other direction is proven analogously. Suppose that there exists \vgrSC{a classical}\vgr{an operationally noncontextual ontological} constructed on $\altr\mainr$ with ontic space $\altr\ospace$. Starting with the primitives $\altr F$, $\altr \sigma$ of \cref{th:BasicCriterionAlt}, choose for all $\hv\in\altr\ospace$ elements $\altr E(\hv)\in\spane$ such that $\proj{\altr\mainr}{\altr E(\hv)} = \altr F(\hv)$, and $\altr S(\hv)\in\spans$ such that 
$\proj{\altr\mainr}{\altr S(\hv)} = \altr\sigma(\hv)$. This is always possible thanks to equations \eqref{eq:spanfs} and \eqref{eq:spange}. Then, define for all $\hv\in\altr\ospace$:
\begin{subequations}
\begin{align}
F(\hv) &:= \proj{\mainr}{\altr E(\hv)}, \\
\sigma(\hv) &:= \proj{\mainr}{\altr S(\hv)}.
\end{align}
\end{subequations}
It is then easy to verify that these $F$ and $\sigma$ satisfy the requirements of \cref{prop:basiccriterion}, and thus that there exists \vgrSC{a classical}\vgr{an operationally noncontextual ontological} model on $\mainr$ with the same ontic space $\altr\ospace$. This concludes the proof.
\end{myproof}

\section{Algorithmic formulation, witnesses and certifiers}

\subsection{Resolution of convex cones}

\label{app:ConvexConeResolution}

Throughout this section, $\mathcal V$ is a finite dimensional real inner product space.

Let us recall the definition of a pointed cone, then prove \cref{prop:ResolutionPointedCone}.

\DefPointedCone*

\begin{lemma}[Adapted from theorem~8.1.3 in~\cite{ConvexAnalysis}]
\label{lem:ResolutionConvexSet}
For any nonempty compact convex set $X\subseteq\mathcal V$, it holds that $X$ is the convex hull of its extremal points (\cref{def:extreme points}):
\begin{equation}
X = \myconv{\convextr{X}}.
\end{equation}
\end{lemma}

The following lemma relates the extremal half-lines of a given pointed cone to the extremal points of a certain compact convex ``slice" of the cone.

\begin{lemma}
\label{lem:SlicedCone}
If $\mathcal C\subseteq \mathcal V$ is a pointed cone, then let $L$ be the linear functional of \cref{def:PointedCone} and let
\begin{equation}
\label{eq:TempDefH}
H := \{v\in\mathcal V\st L(v) = 1\}.
\end{equation}
It holds that $\mathcal C\cap H$ is a nonempty compact convex set such that:
\begin{equation}
\label{eq:ExtrCCapH}
\coneextr{\mathcal C} = \big\{\mycone{c}\st c\in\convextr{\mathcal C\cap H}\big\}.
\end{equation}
\end{lemma}
\begin{myproof}
$\mathcal C\cap H$ is a nonempty set: indeed, by the \cref{def:PointedCone} of pointed cones, $\mathcal C\setminus\{0\}$ is a nonempty set, and for any $c\in\mathcal C\setminus\{0\}$ which is a nonempty set by \cref{def:PointedCone}, it holds that $L(c) > 0$. But then, $c/L(c)\in\mathcal C$ satisfies $L(c/L(c)) = 1$, so that $c/L(c) \in \mathcal C\cap H$.
Clearly, $\mathcal C\cap H$ is a closed set: the closure follows from the closure of $\mathcal C$, by \cref{def:PointedCone} and the closure of $H$ in \eqref{eq:TempDefH}. The convexity of $\mathcal C\cap H$ follows from the convexity of $\mathcal C$ and that of $H$.

Let us prove that $\mathcal C\cap H$ is bounded.
We reason by contradiction: suppose that $\mathcal C\cap H$ was not bounded. Following the reasoning of section~8.1 of~\cite{ConvexAnalysis}, the unboundedness of the closed convex set $\mathcal C\cap H$ is equivalent to to the existence of $d\in\mathcal V$, $d\neq0$, $c\in\mathcal C$ such that for all $\lambda\in\realpos$, $(c+\lambda d) \in \mathcal C\cap H$. Clearly, this implies $L(d) = 0$. But by the convex structure of $\mathcal C$, it holds that for any $\lambda \geq 1$, the following convex combination lies in $\mathcal C$:
\begin{equation}
\big(1-\frac{1}{\lambda}\big) 0 + \frac{1}{\lambda}(c+\lambda d) \in\mathcal C.
\end{equation}
By the closure of $\mathcal C$, the limit point belongs to $\mathcal C$:
\begin{equation}
d = \lim_{\lambda\rightarrow\infty} \frac{1}{\lambda}(c+\lambda d) \in \mathcal C.
\end{equation}
Thus, $d$ is a point of $\mathcal C$ that satisfies $L(d) = 0$: this implies $d=0$ which is a contradiction. This proves that $\mathcal C\cap H$ is bounded.
This proves that $\mathcal C\cap H$ is a nonempty compact convex set.

Let us now prove \eqref{eq:ExtrCCapH}. For any $\halfl\in\coneextr{\mathcal C}$, there exists $c_0 \in \halfl$ be such that $L(c_0) = 1$. We will show that $c_0\in\convextr{\mathcal C\cap H}$. Consider any $\lambda\in\,]0,1[$ and $c_1,c_2\in\mathcal C\cap H$ such that
\begin{equation}
\label{eq:TempConvexC0}
c_0 = \lambda c_1 +(1-\lambda) c_2.
\end{equation}
Then, because $c_1,c_2\in\mathcal C\cap H\subset \mathcal C$, and because $c_0$ is part of an extremal direction (\cref{def:ExtremalHalfLines}) of $\mathcal C$, this implies without loss of generality (if $c_2 =0$ and $c_1\neq 0$, swap $c_1$ with $c_2$) that there exists $\alpha\in\mathbb R$ such that $c_1 = \alpha c_2$. Applying the linear map $L$ to this equation and recalling $c_1,c_2\in H$ shows that $\alpha = 1$, and thus $c_1 = c_2$. Together with \eqref{eq:TempConvexC0}, $c_1 = c_2 = c_0$. This proves that $c_0\in\convextr{\mathcal C\cap H}$. This shows that for any $\halfl\in\coneextr{\mathcal C}$, there exists $c_0\in\convextr{\mathcal C\cap H}$ such that
\begin{equation}
\halfl = \mycone{c_0},
\end{equation}
and thus it holds that
\begin{equation}
\label{eq:TempInclusion}
\textstyle \coneextr{\mathcal C} \subseteq \{\mycone{c}\st c\in\convextr{\mathcal C\cap H}\}.
\end{equation}
Now, let $c_0\in\convextr{\mathcal C\cap H}$, and we will show that $\mycone{c_0} \in\coneextr{\mathcal C}$.
It suffices to prove that $c_0$ is an extreme direction of $\mathcal C$ according to \cref{def:ExtremalHalfLines}.
Let $d_1,d_2\in\mathcal C$ be such that $c_0 = d_1 + d_2$.
Note that $1 = L(d_1) + L(d_2)$ due to $c_0\in H$, and since $d_1,d_2\in\mathcal C$: $L(d_1),L(d_2) \geq0$, which implies in particular $L(d_1),L(d_2)\in[0,1]$. We will show that $d_1$ and $d_2$ are linearly dependent. If either $d_1$ or $d_2$ is zero this is trivial, so assume that they are both nonzero. Thus we have $L(d_1),L(d_2) \in \,]0,1[$. Then, rewrite
\begin{equation}
\label{eq:TempConvC0}
c_0 = L(d_1)\frac{d_1}{L(d_1)} + (1-L(d_1))\frac{d_2}{L(d_2)}.
\end{equation}
Due to $d_i/L(d_i)\in\mathcal C\cap H$ for $i=1,2$ and the fact that $c_0$ is an extremal point of $\mathcal C\cap H$, the convex combination \eqref{eq:TempConvC0} implies 
\begin{equation}
\frac{d_1}{L(d_1)} = \frac{d_2}{L(d_2)},
\end{equation}
which implies the linear dependence of $d_1$ and $d_2$. Thus, $c_0$ is an extremal direction of $\mathcal C$ and $\mycone{c_0}\in\coneextr{\mathcal C}$. Since $c_0$ was arbitrary in $\convextr{\mathcal C\cap H}$, this proves the reverse inclusion to \eqref{eq:TempInclusion}. Thus, the equality
\begin{equation}
\label{eq:ReverseInclusion}
\coneextr{\mathcal C} = \{\mycone{c}\st c\in\convextr{\mathcal C\cap H}\}
\end{equation}
holds.
\end{myproof}

\PropResolutionPointedCone*

\begin{myproof}
By applying \cref{lem:ResolutionConvexSet} to the nonempty compact  convex set $\mathcal C\cap H$ (with $H$ defined in \cref{lem:SlicedCone}, and the nonempty compact convex set also established in \cref{lem:SlicedCone}):
\begin{equation}
\mathcal C\cap H = \myconv{\convextr{\mathcal C\cap H}}.
\end{equation}
It is clear that $\mathcal C = \mycone{\mathcal C\cap H}$. It is also easy to show that for any set $X\subseteq \mathcal V$, $\myconv{\mycone{X}} = \mycone{\myconv{X}}$. Thus,
\begin{multline}
\label{eq:TempCone}
\mathcal C = \mycone{\mathcal C\cap H} = \mycone{\myconv{\convextr{\mathcal C\cap H}}} \\
= \myconv{\mycone{\convextr{\mathcal C\cap H}}}.
\end{multline}
Note that $\mycone{\convextr{\mathcal C\cap H}} = \bigcup_{c\in\convextr{\mathcal C\cap H}}\mycone{c}$. By \eqref{eq:ExtrCCapH} of \cref{lem:SlicedCone}, it holds that
\begin{equation}
\textstyle \mycone{\convextr{\mathcal C\cap H}} = \bigcup_{\halfl\in\coneextr{\mathcal C}} \halfl.
\end{equation}
Together with \eqref{eq:TempCone} this proves
\begin{gather}
\textstyle \mathcal C = \myconv{\bigcup_{\halfl\in\coneextr{\mathcal C}} \halfl}. \qedhere
\end{gather}
\end{myproof}

Let us restate \cref{def:SpanningCone} and prove \cref{prop:ResolutionPolarCone}.

\DefSpanningCone*

We will need the following lemma.

\begin{lemma}
\label{lem:PolarPolar}
Let $\mathcal C\subseteq\mathcal V$ be a nonempty closed convex cone. Then,
\begin{equation}
\polar{[\polar{\mathcal C}{\mathcal V}]}{\mathcal V} = \mathcal C.\label{eq:polar polar identity}
\end{equation}
\end{lemma}
\begin{proof}
One direction is easy to verify directly:
\begin{equation}
\mathcal C\subseteq\polar{[\polar{\mathcal C}{\mathcal V}]}{\mathcal V}.
\end{equation}
Let us consider the other direction. Suppose there existed $v_0\in\polar{[\polar{\mathcal C}{\mathcal V}]}{\mathcal V}$ such that $v_0\notin\mathcal C$. Using that $\mathcal C$ is closed, by~theorem 3.2.2 in~\cite{ConvexAnalysis}, there exists $n\in\mathcal V$, $\alpha\in\mathbb R$ such that
\begin{subequations}
\begin{align}
\forall c\in\mathcal C\st 
\scal{n}{c}{\mathcal V} &\geq \alpha,\label{eq:TempNDVA} \\
\scal{n}{v_0}{\mathcal V} &< \alpha.
\end{align}
\end{subequations}
Because $0\in\mathcal C$ for any cone, \eqref{eq:TempNDVA} implies $\alpha \leq 0$. This proves that
\begin{equation}
\scal{n}{v_0}{\mathcal V} < 0.
\end{equation}
The conic structure of $\mathcal C$ and \eqref{eq:TempNDVA} imply, for all $c\in\mathcal C$,
\begin{equation}
\forall \lambda\in\realpos\st
\scal{n}{\lambda c}{\mathcal V} \geq \alpha,
\end{equation}
or, for all $\lambda > 0$, $\scal{n}{c}{\mathcal V} \geq (\alpha/\lambda)$.
By taking the limit $\lambda \rightarrow\infty$, this proves that $\scal{n}{c}{\mathcal V} \geq 0$, so that
\begin{subequations}
\begin{align}
\forall c\in\mathcal C\st \scal{n}{c}{\mathcal V} \geq 0, \label{eq:TempND0}\\
\scal{n}{v_0}{\mathcal V} < 0. \label{eq:TempContra}
\end{align}
\end{subequations}
Equation \eqref{eq:TempND0} proves that $n\in\polar{\mathcal C}{\mathcal V}$. Thus, it must holds that since $v_0\in\polar{[\polar{\mathcal C}{\mathcal V}]}{\mathcal V}$, $\scal{n}{v_0}{\mathcal V} \geq 0$, which contradicts equation \eqref{eq:TempContra}. This contradiction proves that there cannot exists $v_0\in\polar{[\polar{\mathcal C}{\mathcal V}]}{\mathcal V}$ and $v_0\notin\mathcal C$, which proves that $\polar{[\polar{\mathcal C}{\mathcal V}]}{\mathcal V} \subseteq \mathcal C$. Thus both inclusions are proven and the claim follows.
\end{proof}

\PropResolutionPolarCone*

\begin{myproof}
Let us prove that $\polar{\mathcal C}{\mathcal V}$ is a pointed cone by verifying the assumptions of \cref{def:PointedCone}.
Clearly, from \cref{def:Polar} of the polar cone, $\polar{\mathcal C}{\mathcal V}$ is closed.
Now, let us prove that $\polar{\mathcal C}{\mathcal V} \neq \{0\}$.
We reason by contradiction: if $\polar{\mathcal C}{\mathcal V} = \{0\}$, this means that $\polar{[\polar{\mathcal C}{\mathcal V}]}{\mathcal V} = \mathcal V$.
But since $\mathcal C$ is closed by virtue of~\ref{item:SpanningCone(i)} in \cref{def:SpanningCone},  by \cref{lem:PolarPolar}, it holds that $\polar{[\polar{\mathcal C}{\mathcal V}]}{\mathcal V} = \mathcal C$.
Thus, if $\polar{\mathcal C}{\mathcal V} = \{0\}$, then $\mathcal C = \mathcal V$, which contradicts assumption~\ref{item:ClosedSpanningCone} of \cref{def:SpanningCone}.

It remains to verify the property~\ref{item:LinPointedCone} of pointed cones in \cref{def:PointedCone}. By property~\ref{item:SpanningSpanningCone} of spanning cones in \cref{def:SpanningCone}, $\myspan{\mathcal C} = \mathcal V$, so that we may choose a basis of $\mathcal V$ of the form $\{c_i\in\mathcal C\}_{i=1}^{\dim(\mathcal V)}$.
Consider the linear map $L\funcrange{\mathcal V}{\mathbb R}$ defined by
\begin{equation}
L(\cdot) := \sum_{i=1}^{\dim(\mathcal V)} \scal{c_i}{\cdot}{\mathcal V}.
\end{equation}
Clearly, for all $d\in\polar{\mathcal C}{\mathcal V}$, $L(d) \geq 0$. If $L(d) = 0$ for some $d\in\polar{\mathcal C}{\mathcal V}$, then by the nonnegativity of each term in the sum we must have for all $i=1,\dots,\dim(\mathcal V)$: $\scal{c_i}{d}{\mathcal V} = 0$. Because the $c_i$'s span $\mathcal V$, by linearity of the scalar product it holds that $\scal{v}{d}{\mathcal V} = 0$ for all $v\in\mathcal V$. 
The non-degeneracy of the inner product proves $d=0$, so that $L$ satisfies indeed assumption~\ref{item:LinPointedCone} of \cref{def:PointedCone}.

We have proven that spanning cones defined in \cref{def:SpanningCone} have polar cones which are pointed cone as defined in \cref{def:PointedCone}.
\end{myproof}

\PropHyperspaceRepresentation*

\begin{myproof}
Since $\mathcal C$ is a spanning pointed cone, by \cref{def:PointedCone} it is closed, so that by \cref{lem:PolarPolar},
\begin{multline}
\label{eq:TempCpp}
\mathcal C = \polar{[\polar{\mathcal C}{\mathcal V}]}{\mathcal V}
= \{v\in\mathcal V\st
\scal{v}{d}{\mathcal V} \geq 0 \ \forall d\in\polar{\mathcal C}{\mathcal V}\}.
\end{multline}
In particular, for all $v\in\mathcal C$, it holds that $\scal{v}{d}{\mathcal V} \geq 0$ for all $d\in \halfl\in\coneextr{\polar{\mathcal C}{\mathcal V}}$. Thus,
\begin{equation}
\textstyle \mathcal C \subseteq \bigcap_{\halfl\in\coneextr{\polar{\mathcal C}{\mathcal V}}} \polar{\halfl}{\mathcal V}.
\end{equation}
Now, let $v\in\bigcap_{\halfl\in\coneextr{\polar{\mathcal C}{\mathcal V}}} \polar{\halfl}{\mathcal V}$. We will show that also $v\in\mathcal C$. By equation \eqref{eq:TempCpp}, it suffices to verify that for an arbitrary $d\in\polar{\mathcal C}{\mathcal V}$, $\scal{v}{d}{\mathcal V}\geq 0$. Now, since $\mathcal C$ is a spanning cone, by \cref{prop:ResolutionPolarCone}, it holds that there must exist a finite number $N\in\mathbb N$ of elements $d_i$ each belonging to an extremal line $\halfl_i$ of $\polar{\mathcal C}{\mathcal V}$ such that
\begin{equation}
d = \sum_{i=1}^N d_i.
\end{equation}
Then, $\scal{v}{d}{\mathcal V} = \sum_{i=1}^N \scal{v}{d_i}{\mathcal V}$. Due to $v\in \bigcap_{\halfl\in\coneextr{\polar{\mathcal C}{\mathcal V}}} \polar{\halfl}{\mathcal V}$, and $d_i\in \halfl_i\in\coneextr{\polar{\mathcal C}{\mathcal V}}$, it is clear that $\scal{v}{d}{\mathcal V} \geq 0$. Thus, it holds that
\begin{equation}
\textstyle \mathcal C = \bigcap_{\halfl\in\coneextr{\polar{\mathcal C}{\mathcal V}}} \polar{\halfl}{\mathcal V}.
\end{equation}
Because $\mathcal C$ is a spanning pointed cone, the vertex enumeration problem of \cref{def:venum} is well-defined and it holds that $\coneextr{\polar{\mathcal C}{\mathcal V}} = \venum{\mathcal V}{\coneextr{\mathcal C}}$, which concludes the proof.
\end{myproof}

\subsection{General aspects of the algorithm}

\label{app:algo}

\begin{lemma}
\label{lem:ClosureProjection}
It holds that
\begin{subequations} 
\begin{align}
\proj{\mainr}{\closure\sets} &= \closure{\proj{\mainr}{\sets}}, \label{eq:ProjClosureSets} \\
\proj{\mainr}{\closure\sete} &= \closure{\proj{\mainr}{\sete}}. \label{eq:ProjClosureSete}
\end{align}
\end{subequations} 
\end{lemma}
\begin{myproof}
Let us prove \eqref{eq:ProjClosureSets}. Clearly, $\proj{\mainr}{\closure\sets} \subseteq \closure{\proj{\mainr}{\sets}}$. Then, let $\red\rho^*\in\closure{\proj{\mainr}{\sets}}$ be arbitrary. By \cref{lem:ClosedEquiv}, there exists a sequence $(\red\rho_n\in\reds)_{n\in\mathbb N}$ such that $\red\rho^* = \lim_{n\rightarrow\infty} \red\rho_n$. For all $n$, choose any $\rho_n\in\sets$ such that $\proj{\mainr}{\rho_n} = \red\rho_n$. Because the set $\sets$ is bounded, the sequence $(\rho_n\in\sets)_{n\in\mathbb N}$ is also bounded. By the Bolzano-Weierstrass \cref{th:bolzano}, there exists a set of indices $\{n_k\}_k\subseteq\mathbb N$ and $\rho^*\in\closure\sets$ such that
\begin{equation}
\lim_{k\rightarrow\infty} \rho_{n_k} = \rho^*.
\end{equation}
Then, by \cref{lem:SubsquenceLimit}, the subsequence $(\red\rho_{n_k})_k$ must converge to $\red\rho^*$, so that
\begin{equation}
\red\rho^* = \lim_{k\rightarrow\infty} \red\rho_{n_k} = \lim_{k\rightarrow\infty} \proj{\mainr}{\rho_{n_k}} = \proj{\mainr}{\rho^*} \in \proj{\mainr}{\closure\sets}.
\end{equation}
Note that we used the continuity of $\proj{\mainr}{\cdot}$ to conclude. The proof of \eqref{eq:ProjClosureSete} is analogous and relies on the boundedness of $\sete$.
\end{myproof}

\PropPointedConeExpression*

\begin{myproof}
Let us prove equation \eqref{eq:PolarPRS} explicitly.
Strictly speaking, we could simply use $\closure{\proj{\mainr}{\sets}}$ rather than $\proj{\mainr}{\closure\sets}$ in every instance it appears, but since the two sets are equal (\cref{lem:ClosureProjection}), for notational purposes we prefer the use of $\proj{\mainr}{\closure\sets}$.
First, note that the polar of a set is equal to the polar of its closure. Indeed, consider $\polars$.
One direction is clear thanks to \cref{lem:SubPolar}: $\polar{\proj{\mainr}{\closure\sets}}{\mainr} \subseteq \polars$.
Now, consider any $s\in\polars$, and let us show that also $s\in\polar{\proj{\mainr}{\closure\sets}}{\mainr}$. 
It suffices to show that $\scal{s}{\red\rho}{\mainr} \geq 0$ for any element $\red \rho\in\proj{\mainr}{\closure\sets}$.
Such elements $\red\rho$ can be written as the limit of a converging sequence $(\red\rho_n\in\proj{\mainr}{\sets})_{n\in\mathbb N}$, thanks to \cref{lem:ClosedEquiv}.
Then, by \cref{lem:NormScalContinuous} which states the continuity of the scalar product, and by the closure of the set $\realpos$, it holds that
\begin{equation}
\scal{s}{\red\rho}{\mainr} = \lim_{n\rightarrow\infty} \scal{s}{\red\rho_i}{\mainr} \in\realpos.
\end{equation}
Thus we have
$
\polars = \polar{\proj{\mainr}{\closure\sets}}{\mainr}$.
Clearly, the polar cone of a set $X$ and the polar cone to $\mycone{X}$ are equal, so that
$
\polars = \polar{\mycone{\proj{\mainr}{\closure\sets}}}{\mainr}$.
This proves equation \eqref{eq:PolarPRS}, and the proof of equation \eqref{eq:PolarPRE} is completely analogous.
\end{myproof}

The following lemmas is an intermediate step towards proving \cref{prop:PointedSpanningPRSE}.

\begin{lemma}
\label{lem:LinMapS}
There exists a linear map $L\funcrange{\mainr}{\mathbb R}$ such that for all $s\in\mycone{\proj{\mainr}{\closure\sets}}$,
\begin{equation}
L(s) \geq 0
\end{equation}
with equality if and only if $s=0$.
\end{lemma}
\begin{myproof}
Choose $L(\cdot) = \otrace{\cdot}$. For all $s\in\mycone{\proj{\mainr}{\closure\sets}}$, there exists $\lambda\in\realpos$ and $\red\rho\in\proj{\mainr}{\closure\sets}$ such that $s = \lambda\red\rho$. Then, $L(s) = \lambda L(\red\rho) = \lambda \geq 0$ where we used \cref{lem:traceone} to assert $\otrace{\red\rho} = 1$ (strictly speaking, if $\red\rho\in\proj{\mainr}{\closure\sets}\setminus{\reds}$, one needs to consider a converging sequence of elements of $\reds$ to assert that the limit also has unit trace). Also, $L(s) = 0$ implies $\lambda= 0$ so that $s=\lambda\red\rho = 0$. Thus, for all $s\in\mycone{\proj{\mainr}{\closure\sets}}$, $L(s) \geq 0$ with equality if and only if $s=0$.
\end{myproof}

\begin{lemma}
\label{lem:LinMapE}
There exists a linear map $L\funcrange{\mainr}{\mathbb R}$ such that for all $e\in\mycone{\proj{\mainr}{\closure\sete}}$,
\begin{equation}
L(e) \geq 0
\end{equation}
with equality if and only if $e=0$.
\end{lemma}
\begin{myproof}
Choose a (non-orthonormal) basis of $\mainr$ in the form $\{\proj{\mainr}{\rho_i}\st\rho_i\in\sets\}_{i=1}^{\dim(\mainr)}$, which is always possible by \cref{prop:spans}. Then, define for all $r\in\mainr$:
\begin{equation}
L(r) = \sum_{i=1}^{\dim(\mainr)} \scal{\proj{\mainr}{\rho_i}}{r}{\mainr}.
\end{equation}
Then, consider an arbitrary element of $\mycone{\proj{\mainr}{\closure\sete}}$ written in the form $\lambda\proj{\mainr}{E}$ for some $\lambda\in\realpos$ and $E\in\closure\sete$.
Thanks to \cref{prop:reducedscalar},
\begin{equation}
L(\lambda\proj{\mainr}{E}) = \lambda \sum_{i=1}^{\dim(\mainr)} \scal{\rho_i}{E}{\lh} \geq 0.
\end{equation}
Equality implies for all $i=1,\dots,\dim(\mainr)$ that
\begin{equation}
0 = \lambda\scal{\rho_i}{E}{\lh} = \scal{\proj{\mainr}{\rho_i}}{\lambda\proj{\mainr}{E}}{\mainr},
\end{equation}
but since the set $\{\proj{\mainr}{\rho_i}\}_i$ spans $\mainr$, this implies $\scal{r}{\lambda\proj{\mainr}{E}}{\mainr} = 0$ for all $r\in\mainr$, which by the non-degeneracy of the inner product implies
\begin{gather}
\lambda\proj{\mainr}{E} = 0. \qedhere
\end{gather}
\end{myproof}

\PropPointedSpanningPRSE*

\begin{myproof}
Let us verify the \cref{def:PointedCone} of pointed cones. The closure of $\mycone{\proj{\mainr}{\closure\sets}}$ and $\mycone{\proj{\mainr}{\closure\sete}}$ is clear, so~\ref{item:ClosedPointedCone} of \cref{def:PointedCone} is verified.
By \cref{lem:emptysets}, $\reds$ and $\rede$ are strict supersets of $\{0\in\mainr\}$, which proves that also $\mycone{\proj{\mainr}{\closure\sets}}$ and $\mycone{\proj{\mainr}{\closure\sete}}$ are strict supersets of $\{0\}$, satisfying~\ref{item:NonTrivalPointedCone} of \cref{def:SpanningCone}.
The property~\ref{item:LinPointedCone} was proven separately in \cref{lem:LinMapS,lem:LinMapE}.
Thus, $\mycone{\proj{\mainr}{\closure\sets}}$ and $\mycone{\proj{\mainr}{\closure\sete}}$ are pointed cones in $\mainr$.

Now we verify the \cref{def:SpanningCone} of spanning cones. Consider the property~\ref{item:ClosedSpanningCone} (``$\mathcal C\neq \mathcal V$") of \cref{def:SpanningCone}: it is automatically verified thanks to the fact that~\ref{item:NonTrivalPointedCone} (``$\mathcal C\neq \emptyset,\mathcal C\neq \{0\}$") of \cref{def:PointedCone} holds.
Property~\ref{item:SpanningSpanningCone} follows directly from the subset inclusion $\reds\subset \mycone{\proj{\mainr}{\closure\sets}}$ and $\rede\subset\mycone{\proj{\mainr}{\closure\sete}}$ together with \cref{prop:spans}. This proves that $\mycone{\proj{\mainr}{\closure\sets}}$ and $\mycone{\proj{\mainr}{\closure\sete}}$ are also spanning cones.
\end{myproof}

\PropSepesSpanningPointed*

\begin{myproof}
The closure of $\sepes$ was proven in \cref{prop:sepclosed}. By \cref{lem:emptysets}, $\sepes\neq \{0\}\subset\mainr\otimes\mainr$. Then, define the linear map $L\funcrange{\mainr\otimes\mainr}{\mathbb R}$ by: for all $\Omega\in\mainr\otimes\mainr$,
\begin{equation}
L(\Omega) := \sum_{i,j=1}^{\dim(\mainr)} \scal{\red\rho_i\otimes\red E_j}{\Omega}{\mainr\otimes\mainr},
\end{equation}
where $\{\red \rho_i\in\reds \}_{i=1}^{\dim(\mainr)}$ and $\{\red E_j\in\rede\}_{j=1}^{\dim(\mainr)}$ are two bases of $\mainr$, which is possible thanks to \cref{prop:spans}. Now consider any $\Omega\in\sepes$: then, by \cref{def:sepes}, there exists $n\in\mathbb N$ and $\{F_k\in\polars\}_{k=1}^n$, $\{\sigma_k\in\polare\}_{k=1}^n$ such that $\Omega = \sum_{k=1}^n F_k\otimes\sigma_k$. Then,
\begin{equation}
L(\Omega) = \sum_{i,j,k} \scal{\red\rho_i}{F_k}{\mainr}\scal{\red E_j}{\sigma_k}{\mainr}.
\end{equation}
Thanks to the domains of the respective elements, each scalar product is a nonnegative number. Thus $L(\Omega) \geq0$, and equality implies for all $k=1,\dots,n$ that
\begin{multline}
\forall i,j=1,\dots,\dim(\mainr)\st \\
\scal{\red\rho_i\otimes \red E_j}{F_k\otimes \sigma_k}{\mainr\otimes\mainr} = 0.
\end{multline}
But $\{\red\rho_i\otimes\red E_j\}_{i,j=1}^{\dim(\mainr)}$ is a basis of $\mainr\otimes\mainr$, so that in fact, for all $k=1,\dots,n$:
\begin{equation}
\forall R\in\mainr\otimes\mainr\st 
\scal{R}{F_k\otimes \sigma_k }{\mainr\otimes\mainr}=0.
\end{equation}
By the non-degeneracy of the inner product, each term $F_k\otimes\sigma_k$ is zero so that $\Omega=0$. This proves that $\sepes$ is a pointed cone.

For the spanning cone aspect,  \ref{item:ClosedSpanningCone} of \cref{def:SpanningCone} holds thanks to~\ref{item:LinPointedCone} of \cref{def:PointedCone}. The spanning property~\ref{item:SpanningSpanningCone} of \cref{def:SpanningCone} follows from the fact that the following basis of $\mainr\otimes\mainr$:
\begin{equation}
\{\red E_i\otimes \red\rho_j\st \red E_i\in\rede,\red\rho_j\in\reds\}_{i,j=1}^{\dim(\mainr)}
\end{equation}
is a subset of $\sepes$. This concludes the proof.
\end{myproof}

In the course of the proof of \cref{prop:ResolutionSepes}, we will need the notion of partial scalar product over a tensor product space, which took its inspiration from the partial trace familiar from quantum mechanics, but also from the insightful proof techniques of~\cite{TensorProducts}.

\begin{definition}[Partial scalar product]
\label{def:PartialScal}
For any $a\otimes b\in\mainr\otimes\mainr$, for any $r\in\mainr$, define 
\begin{subequations}
\begin{align}
\scala{a\otimes b}{r} := \scal{a}{r}{\mainr} b, \\
\scalb{a\otimes b}{r} := \scal{b}{r}{\mainr} a,
\end{align}
\end{subequations}
and extend these definitions by linearity to obtain bilinear maps:
\begin{subequations}
\begin{align}
\scala\cdot\cdot\funcrange{(\mainr\otimes\mainr) \times \mainr}{\mainr}, \\
\scalb\cdot\cdot\funcrange{(\mainr\otimes\mainr)\times\mainr}{\mainr}.
\end{align}
\end{subequations}
\end{definition}

\begin{lemma}
\label{lem:PartialScal}
For any $\Omega\in\sepes$, for any $\red\rho\in\reds$, it holds that
\begin{equation}
\scala{\Omega}{\red\rho} \in\polare.
\end{equation}
For any $\Omega\in\sepes$, for any $\red E\in\rede$, it holds that
\begin{equation}
\scalb{\Omega}{\red E} \in\polars.
\end{equation}
\end{lemma}
\begin{myproof}
Consider any $\Omega\in\sepes$. By \cref{def:sepes}, there exist $n\in\mathbb N$ and $\{F_i\in\polars\}_{i=1}^n$, $\{\sigma_i\in\polare\}_{i=1}^n$ such that $\Omega = \sum_{i=1}^n F_i\otimes\sigma_i$. Then, by \cref{def:PartialScal}, for any $\red\rho\in\reds$,
\begin{equation}
\scala{\Omega}{\red\rho} = \sum_{i=1}^n \scal{F_i}{\red\rho}{\mainr} \sigma_i.
\end{equation}
Since $\polare$ is a convex cone, $\sigma_i\in\polare$ and $\scal{F_i}{\red\rho}{\mainr} \geq 0$, it holds that $\scala{\Omega}{\red\rho} \in\polare$. The proof that for any $\red E\in\rede$: $\scalb{\Omega}{\red E} \in\polars$ is analogous.
\end{myproof}

\PropResolutionSepes*

\begin{myproof}
It is easy to see from \cref{def:sepes} of $\sepes$ that
\begin{multline}
\sepes = \\
\textstyle \myconv{\bigcup_{\halfl_1\in\coneextr{\polars},\halfl_2\in\coneextr{\polare}} \halfl_1\oprod \halfl_2}.
\end{multline}
This shows that the set of extremal half-lines of $\sepes$ have to be a subset of or equal to the set
\begin{multline}
\Big\{
\halfl_1 \oprod \halfl_2\st 
\halfl_1 \in \coneextr{\polars}, \\
\halfl_2 \in \coneextr{\polare}
\label{eq:estremal hl for Sep}
\Big\}.
\end{multline}
It remains to show that the set \eqref{eq:estremal hl for Sep} contains no more than the extremal half-lines of $\sepes$, i.e., that each half-line in \eqref{eq:estremal hl for Sep} is indeed an extremal half-line of $\sepes$.

For any $\halfl_1 \in \coneextr{\polars}$, choose $F\in \halfl_1$, $F\neq 0$.
Then, for any
$\halfl_2 \in \coneextr{\polare}$, choose $\sigma\in \halfl_2$, $\sigma \neq 0$. Now consider any $\Omega_1,\Omega_2\in\sepes$ that satisfy
\begin{equation}
\label{eq:TempFSO}
F\otimes \sigma = \Omega_1 + \Omega_2.
\end{equation}
Choose now any $\red\rho\in\reds$, and take the partial scalar product defined in \cref{def:PartialScal}:
\begin{equation}
\label{eq:TempS1Omega}
\scal{F}{\red\rho}{\mainr}\sigma
= \scala{\Omega_1}{\red\rho} + \scala{\Omega_2}{\red\rho}.
\end{equation}
Note that $\scal{F}{\red\rho}\mainr\geq 0$ and thanks to \cref{lem:PartialScal}, 
$\scala{\Omega_1}{\red\rho}$ and  $\scala{\Omega_2}{\red\rho}$ belong to the cone $\polare$. There are two cases to consider. If $\scal{F}{\red\rho}\mainr > 0$, then $\scal{F}{\red\rho}\mainr\sigma$ is part of the extremal half-line $\halfl_2$ of $\polare$. Then, the \cref{def:ExtremalHalfLines} of extremal half-lines and equation \eqref{eq:TempS1Omega} imply that $\scala{\Omega_1}{\red\rho}$ and $\scala{\Omega_2}{\red\rho}$ are both scalar multiples of $\sigma$:
\begin{subequations}
\label{eq:TempOmegaF}
\begin{align}
\scala{\Omega_1}{\red\rho} = f_1(\red\rho)\sigma, \\
\scala{\Omega_2}{\red\rho} = f_2(\red\rho)\sigma,
\end{align}
\end{subequations}
where we allowed the scalar multiples $f_i(\red\rho)\in\mathbb R$ to depend on $\red\rho$.

This was when $\scal{F}{\red\rho}\mainr >0$. If instead, $\scal{F}{\red\rho}\mainr = 0$, then \eqref{eq:TempS1Omega} becomes $\scala{\Omega_1}{\red\rho} + \scala{\Omega_2}{\red\rho} = 0$. Both elements $\scala{\Omega_i}{\red\rho}$ belong to $\polare$, so that by taking the inner product with any $\red E\in\rede$, we must have
$
\scal{\scala{\Omega_i}{\red\rho}}{\red E}{\mainr} = 0
$ for $i=1,2$. Since the set $\rede$ spans $\mainr$ according to \cref{prop:spans}, it must be that $\scala{\Omega_i}{\red\rho} = 0\in\mainr$. In this case, extend the maps $f_i$ defined in \eqref{eq:TempOmegaF} to be $0$ for such $\red\rho$.

This proves that for any $\red\rho\in\reds$,
there exist maps $f_i\funcrange{\red\rho\in\reds}{\mathbb R}$ such that \eqref{eq:TempOmegaF} still holds. The mappings $f_i$ inherit properties from the left-hand sides of \eqref{eq:TempOmegaF}: in particular, the $f_i$ must be convex linear over the domain $\reds$. By repeating the argument of \cref{prop:extensionmu}, it holds that there exist unique linear extensions $f_i\funcrange{\mainr}{\mathbb R}$. By Riesz' representation \cref{th:Riesz}, there exist $g_1,g_2\in\mainr$ such that $f_i(r) = \scal{g_i}{r}{\mainr}$ for all $r\in\mainr$. Thus, for $i=1,2$, we must have: for all $r,r'\in\mainr$,
\begin{equation}
\scal{\scala{\Omega_i}{r}}{r'}{\mainr}
=
\scal{\scal{g_i}{r}{\mainr}\sigma}{r'}{\mainr},
\end{equation}
which may be rewritten as
\begin{equation}
\scal{\Omega_i}{r\otimes r'}{\mainr\otimes\mainr}
=
\scal{g_i\otimes \sigma}{r\otimes r'}{\mainr\otimes\mainr}.
\end{equation}
The non-degeneracy of the inner product implies $\Omega_i = g_i\otimes \sigma$. Now, choose any $\red E\in\rede$ such that $\scal{\sigma}{\red E}{\mainr} \geq 0$. This is always possible, otherwise it implies that $\sigma = 0$. By \cref{lem:PartialScal}, it must be that $\scalb{\Omega_i}{\red E}\in\polars$, which implies that $g_i\in\polars$.

We return to equation \eqref{eq:TempFSO} which now reads:
\begin{equation}
F\otimes \sigma = g_1\otimes \sigma + g_2 \otimes \sigma.
\end{equation}
Since we assumed $\sigma \neq 0$, this implies
\begin{equation}
F = g_1 + g_2.
\end{equation}
This is a decomposition of the extremal direction $F\in\polars$ over two other directions $g_i\in\polars$: is must be that the $g_i$'s are linearly dependent, which implies the linear dependence of $\Omega_1 = g_1\otimes\sigma$ and $\Omega_2 = g_2\otimes \sigma$.

This proves that the direction $F\otimes\sigma$ is extremal, which proves that any half-lines in the set of equation \eqref{eq:estremal hl for Sep}
is an extremal half-line of $\sepes$. This concludes the proof.
\end{myproof}

\subsection{Computational equivalence: varying reduced spaces and quantum descriptions}
\label{app:AlgorithmicEquivalence}

\PropIsoCones*

\begin{myproof}
Since $\mathcal C\sim\mathcal D$, let $\Phi\funcrange{\mathcal U}{\mathcal V}$ be the invertible linear map of \cref{def:IsoCone} that satisfies $\Phi(\mathcal C) = \mathcal D$.

(i) It suffices to prove that given an extremal direction $c\in\mathcal C$, $\Phi(c)$ is an extremal direction of $\mathcal D$. Take $d_1,d_2\in \mathcal D$ such that
\begin{equation}
\Phi(c) = d_1 + d_2.
\end{equation}
Then, apply the inverse linear map $\Phi^{-1}$:
\begin{equation}
c = \Phi^{-1}(d_1) + \Phi^{-1}(d_2).
\end{equation}
The fact that $c\in \mathcal C$ is an extremal direction of $\mathcal C$ and the fact that $\Phi^{-1}(d_1),\Phi^{-1}(d_2) \in \mathcal C$ imply that there exist $\lambda_1,\lambda_2\in\realpos$ such that $\Phi^{-1}(d_i) = \lambda_i c$, $i=1,2$; but by applying the linear map $\Phi$ this proves that $d_i = \lambda_i \Phi(c)$,
and thus $\Phi(c)$ is an extremal direction of $\mathcal D$.
Thus, $\Phi$ is a linear isomorphism that  makes a one-to-one correspondence between extremal lines of $\mathcal C$ and of $\mathcal D$.

(ii) Let us prove that $\polar{\mathcal C}{\mathcal U}\sim\polar{\mathcal D}{\mathcal V}$. It suffices  to exhibit an invertible isomorphism $\Psi\funcrange{\mathcal U}{\mathcal V}$ such that
\begin{equation}
\label{eq:VerifPsiCD}
\Psi(\polar{\mathcal C}{\mathcal U}) = \polar{\mathcal D}{\mathcal V}.
\end{equation}
Let $\Psi$ be the dual map to the inverse of $\Phi$: $\Psi := (\Phi^{-1})^*$. This map is invertible and linear: its inverse is simply $\Phi^*$. Let us verify \eqref{eq:VerifPsiCD}: let $u\in\polar{\mathcal C}{\mathcal U}$, and we will first prove that $(\Phi^{-1})^*(u)\in\polar{\mathcal D}{\mathcal V}$. Indeed, for all $d\in\mathcal D$, the dual property reads
\begin{equation}
\label{eq:TempScalUD}
\scal{(\Phi^{-1})^*(u)}{d}{\mathcal V}
= \scal{u}{\Phi^{-1}(d)}{\mathcal U}.
\end{equation}
The properties of $\Phi$ inherited from \cref{def:IsoCone} prove that $\Phi^{-1}(d)\in\mathcal C$.
Due to $u\in\polar{\mathcal C}{\mathcal U}$ and $\Phi^{-1}(d) \in \mathcal C$, the right-hand side of \eqref{eq:TempScalUD} is nonnegative. $d\in\mathcal D$ was arbitrary so that indeed, for all $u\in\polar{\mathcal C}{\mathcal U}$,
\begin{equation}
\Psi(u) = (\Phi^{-1})^*(u) \in\polar{\mathcal D}{\mathcal V},
\end{equation}
thus proving $\Psi(\polar{\mathcal C}{\mathcal U}) \subseteq \polar{\mathcal D}{\mathcal V}$.
The other direction can be proven as follows. First, it is easy to verify that for all $t\in\polar{\mathcal D}{\mathcal V}$, $\Phi^*(t) \in \polar{\mathcal C}{\mathcal U}$ due to $\Phi(\mathcal C) = \mathcal D$. Using the invertibility of $\Phi$, we obtain that $t = (\Phi^{-1})^*\circ \Phi^*(t) \in (\Phi^{-1})^*(\polar{\mathcal C}{\mathcal U}) \equiv \Psi(\polar{\mathcal C}{\mathcal U})$, which proves that also $\polar{\mathcal D}{\mathcal V} \subseteq \Psi(\polar{\mathcal C}{\mathcal U})$.
\end{myproof}

\begin{lemma}
\label{lem:IsoPrsPre}
For any $\altr\mainr$ with associated mappings $f$,$g$ (\cref{def:AltReducedSpace}), there exist invertible linear maps $\Phi_\sets,\Phi_\sete\funcrange{\mainr}{\altr\mainr}$ such that for all $s\in\spans$, for all $e\in\spane$,
\begin{subequations}
\begin{align}
\Phi_\sets(\proj{\mainr}{s}) &= f(s), \label{eq:PhiSets} \\
\Phi_\sete(\proj{\mainr}{e}) &= g(e). \label{eq:PhiSete}
\end{align}
\end{subequations}
\end{lemma}
\begin{myproof}
A suitable choice for $\Phi_\sete$ was given in the proof of \cref{prop:AltDimensions}. We will slightly tune this construction to obtain a valid choice for $\Phi_\sets$.

Let $d:=\dim(\mainr) = \dim(\altr\mainr)$ (by \cref{prop:AltDimensions}). Let $\{T_i\in\altr\mainr\}_{i=1}^d$ be an orthonormal basis of $\altr\mainr$. By equation \eqref{eq:spange}, there exist elements $e_i\in\spane$ such that $T_i = g(e_i)$ for all $i=1,\dots,d$.

Now let $\{R_j\in\mainr\}_{j=1}^d$ be an orthonormal basis of $\mainr$. By \cref{prop:spans}, there exist elements $f_j\in\spans$ such that $R_j = \proj{\mainr}{f_j}$ for all $j=1,\dots,d$. 

Then, define the linear maps $\Phi_\sets\funcrange{\mainr}{\altr\mainr}$ and $\phi_\sets\funcrange{\altr\mainr}{\mainr}$ by: for all $r\in\mainr$, for all $t\in\altr\mainr$,
\begin{subequations}
\begin{align}
\Phi_\sets(r) &= \sum_{i=1}^d \scal{\proj{\mainr}{e_i}}{r}{\mainr} T_i, \\
\phi_\sets(t) &= \sum_{j=1}^d
\scal{g(f_j)}{t}{\altr\mainr} R_j.
\end{align}
\end{subequations}
It holds that $\Phi_\sets(\proj{\mainr}{s}) = f(s)$ for all $s\in\spans$. Indeed, using \cref{corollary:Reflection} but swapping the arguments of the symmetric scalar product: for all $ s\in\myspan{\sets}$,
\begin{align}
\label{eq:ReflectionR}
\nonumber \Phi_\sets(\proj{\mainr}{s}) &= \sum_{i=1}^d \scal{\proj{\mainr}{e_i}}{\proj{\mainr}{s}}{\mainr} T_i
\\
\nonumber &= \sum_{i=1}^d \scal{g(e_i)}{f(s)}{\altr\mainr} T_i \\
&= \sum_{i=1}^d \scal{T_i}{f(s)}{\altr\mainr} T_i 
= f(s).
\end{align}
In the last line, we used the resolution of the identity for $\altr\mainr$ in the basis $\{T_i\}_i$. Similarly, for all $s\in\spans$ it holds that $\phi_\sets(f(s)) = \proj{\mainr}{s}$. Indeed, using again \cref{corollary:Reflection}: for all
$s\in\myspan{\sets}$,
\begin{align}
\label{eq:ReflectionRalt}
\nonumber \phi_\sets(f(s)) 
&= \sum_{j=1}^d
\scal{g(f_j)}{f(s)}{\altr\mainr} R_j \\
\nonumber &= \sum_{j=1}^d \scal{\proj{\mainr}{f_j}}{\proj{\mainr}{s}}{\mainr} R_j \\
&= \sum_{j=1}^d
\scal{R_j}{\proj{\mainr}{s}}{\mainr} R_j
= \proj{\mainr}{s}.
\end{align}
We used the completeness relation of $\mainr$ in the basis $\{R_i\}_i$.
Let us now verify that $\phi_\sets = \Phi_\sets^{-1}$. For all $r\in\mainr$, let $s\in\myspan{\sets}$ be such that $r= \proj{\mainr}{s}$.
\begin{equation}
\phi_\sets\circ\Phi_\sets(r) = \phi_\sets(\Phi_\sets(\proj{\mainr}{s})) = \phi_\sets(f(s)) = \proj{\mainr}{s} = r,
\end{equation}
where we used the properties \eqref{eq:ReflectionR}  and \eqref{eq:ReflectionRalt} to conclude.
This shows $\phi_\sets\circ\Phi_\sets = \id\mainr$. The proof that $\Phi_\sets\circ\phi_\sets = \id{\altr\mainr}$ is analogous. Thus $\phi_\sets = \Phi_\sets^{-1}$.
\end{myproof}

\begin{lemma}
\label{lem:isosepes}
Let $\mathcal U, \mathcal V$ be two finite dimensional inner product spaces such that $\dim(\mathcal U) = \dim(\mathcal V)$, and let $\mathcal C_1,\mathcal C_2\subset\mathcal U$, $\mathcal D_1,\mathcal D_2\subset \mathcal V$ be convex cones. If it holds that
\begin{align}
\mathcal C_1 &\sim \mathcal D_1, \\
\mathcal C_2 &\sim \mathcal D_2,
\end{align}
then it also holds that
\begin{equation}
\myconv{\polar{\mathcal C_1}{\mathcal U} \oprod \polar{\mathcal C_2}{\mathcal U}} \sim \myconv{\polar{\mathcal D_1}{\mathcal V} \oprod\polar{\mathcal D_2}{\mathcal V}}.
\end{equation}
\end{lemma}

\begin{myproof}
\Cref{prop:IsoCones} proved that if two cones are isomorphic in the sense of \cref{def:IsoCone}, their polar cones are also isomorphic. For $i=1,2$, let $\Psi_i\funcrange{\mathcal U}{\mathcal V}$ be the invertible linear map such that
\begin{equation}
\Psi_i(\polar{\mathcal C_i}{\mathcal U}) = \polar{\mathcal D_i}{\mathcal V}.
\end{equation}
The linear map that establish \eqref{eq:IsoSepes} is then simply the tensor product map $\Psi_1\otimes\Psi_2\funcrange{\mathcal U\otimes\mathcal U}{\mathcal V\otimes\mathcal V}$ that acts as follows: for any $u,u'\in\mathcal U$,
\begin{equation}
\Psi_1\otimes\Psi_2(u\otimes u') = \Psi_1(u)\otimes \Psi_2(u'),
\end{equation}
and extend this definition by linearity. This linear map is invertible, and allows one to easily verify that
\begin{multline}
\Psi_1\otimes\Psi_2(\myconv{\polar{\mathcal C_1}{\mathcal U} \oprod \polar{\mathcal C_2}{\mathcal U}}) \\
 = \myconv{\polar{\mathcal D_1}{\mathcal V} \oprod\polar{\mathcal D_2}{\mathcal V}}.
\end{multline}
which proves \eqref{eq:IsoSepes}.
\end{myproof}

\PropAltrAlgoEquiv*

\begin{myproof}
Due to $\mycone{\closure\sets}\subseteq \spans$, the invertible linear map $\Phi_\sets$ from \cref{lem:IsoPrsPre} satisfies, thanks to equation \eqref{eq:PhiSets}:
\begin{equation}
\Phi_\sets(\proj{\mainr}{\mycone{\closure\sets}}) = f(\mycone{\closure\sets}).
\end{equation}
The fact that linear operations and conical hulls commute implies \eqref{eq:IsoPrs}, and the proof of \eqref{eq:IsoPre} is analogous. \Cref{lem:isosepes} together with \cref{prop:PointedConeExpression} then allows to prove equation \eqref{eq:IsoSepes}.
\end{myproof}

\PropAncillaComp*

\begin{myproof}
Let $d= \dim(\mainr) = \dim(\tilde\mainr)$ (\cref{prop:ancilla}). We first construct a linear map $\Phi_\sets\funcrange{\mainr}{\tilde\mainr}$ such that for all $k\in I$,
\begin{equation}
\label{eq:ancillaphis}
\Phi_\sets(\proj{\mainr}{\rho_k}) = \proj{\tilde\mainr}{\tilde\rho_k}.
\end{equation}
To do so, let $\{\tilde R_i\in\tilde\mainr\}_{i=1}^d$ be an orthonormal basis of $\tilde\mainr$. Let $K\subseteq I$ be such that $\{P_{\tilde\mainr}(\tilde E_j)\}_{j\in K}$ is a basis of $\tilde\mainr$ (\cref{prop:spans}). For each $i$, there exist $\{\alpha_{ij}\in\mathbb R\}_{1\leq i \leq d;j\in K}$  such that
\begin{equation}
\label{eq:ancillari}
\tilde R_i = \sum_{1\leq i\leq d; j\in K} \alpha_{ij} P_{\tilde\mainr}(\tilde E_j).
\end{equation}
Then define for all $r\in\mainr$
\begin{equation}
\Phi_\sets(r) = \sum_{1\leq i\leq d; j\in K} \alpha_{ij} \scal{\proj{\mainr}{E_j}}{r}{\mainr} \tilde R_i.
\end{equation}
This map is linear and satisfies equation \eqref{eq:ancillaphis} (here one uses equations \eqref{eq:reducedancillastats}, \eqref{eq:ancillari} and the completeness relation of the orthonormal basis $\{\tilde R_i\}_i$). It is easy to mimic this construction to obtain a linear map that maps $P_{\tilde\mainr}(\tilde\rho_k)$ to $\proj{\mainr}{\rho_k}$ for all $k\in I$: this map is exactly the inverse of $\Phi_\sets$ since the $\{\proj{\mainr}{\rho_k}\}_{k\in I}$ span $\mainr$ (\cref{prop:spans}), which proves the invertibility of $\Phi_\sets$ (alternatively, it is easy to prove its injectivity and surjectivity). Having this map at hand, it follows easily (using that linear maps commute with the conical and convex hull) that
\begin{equation}
\Phi_\sets\big(\mycone{\reds}
\big) = 
\big(
\mycone{P_{\tilde\mainr}(\tilde\sets)}\big),
\end{equation}
which proves \eqref{eq:ancillacones}.

The proof of equation \eqref{eq:ancillaconee} is completely analogous: one starts by building a linear map $\Phi_\sete\funcrange{\mainr}{\tilde\mainr}$ such that for all $k\in I$,
\begin{equation}
\label{eq:ancillaphie}
\Phi_\sete(\proj{\mainr}{E_j}) = P_{\tilde\mainr}(\tilde E_j).
\end{equation}
Then, one uses a decomposition of the form
\begin{equation}
\tilde R_i = \sum_{1\leq i \leq d; j \in J} \beta_{ij} P_{\tilde\mainr}(\tilde \rho_j)
\end{equation}
instead of \eqref{eq:ancillari}, with $J$ such that $\{P_{\tilde\mainr}(\tilde\rho_j)\}_{j\in J}$ is a basis of $\tilde\mainr$ and $\beta_{ij} \in \mathbb R$. The definition of $\Phi_\sete$ is then for all $r\in\mainr$,
\begin{equation}
\Phi_\sete(r) = \sum_{1\leq i \leq d; j\in J} \beta_{ij}\scal{\proj{\mainr}{\rho_j}}{r}{\mainr} \tilde R_i.
\end{equation}
This map is invertible, satisfies \eqref{eq:ancillaphie}, and allows to prove \eqref{eq:ancillaconee}.

Equation \eqref{eq:ancillaconesepes} then follows directly from \cref{lem:isosepes} and from the simplification of  \cref{prop:PointedConeExpression}.
\end{myproof}

\section{Connections with generalized probabilistic theories}

\label{app:connection}

Let us restate the definition of a simplex-embeddable generalized probabilistic theory. We build on top of the notation of \cref{sec:Connection}, and additional notation comes from~\cite{Spekkens19}.

\begin{definition}[Adapted from~\cite{Spekkens19}]
\label{def:Simplex}
A tomographically complete  generalized probabilistic theory $(\mathcal V,\Omega,\mathcal E)$ is simplex embeddable
in $d$ dimensions if and only if there exist:
\begin{myitem}
\item a $d$-dimensional real inner product space $\mathcal W$;
\item a simplex $\Delta_d\subset\mathcal W$ with $d$ linearly independent
vertices denoted $\{\delta_i\in\mathcal W\}_{i=1}^d$; \label{item:Simplex}
\item a linear map $\iota\funcrange{\mathcal V}{\mathcal W}$ such that $\iota(\Omega) \subseteq \Delta_d$; \label{item:SimplexIota}
\item a linear map $\kappa\funcrange{\mathcal V}{\mathcal W}$ such that for all $E\in\mathcal E$, for all $i=1,\dots,d$: $\scal{\kappa(E)}{\delta_i}{\mathcal W}\in[0,1]$; \label{item:SimplexKappa}
\end{myitem}
where the maps $\iota,\kappa$ must satisfy the consistency requirement: for all $\rho\in\Omega$, $E\in\mathcal E$:
\begin{equation}
\label{eq:SimplexConsist}
\scal{\rho}{E}{\mathcal V} = \scal{\iota(\rho)}{\kappa(E)}{\mathcal W}.
\end{equation}
\end{definition}

The fact that the vertices of the simplex $\Delta_d$ are linearly independent in~\ref{item:Simplex} of \cref{def:Simplex} is equivalent to the fact that their affine span does not contain the origin $0\in\mathcal W$, which was the condition stated in definition~1 of~\cite{Spekkens19}. Let us now prove \cref{prop:SimplexImpliesClassical}.

\PropSimplexImpliesClassical*

\begin{myproof}
First, note that the existence of \vgrSC{a classical}\vgr{an operationally noncontextual ontological} model as in \cref{def:classicalmodel} under the substitution \eqref{eq:GPTsubst} is equivalent to the criterion given in \cref{prop:basiccriterion} under the substitution \eqref{eq:GPTsubst}. We will prove the claim using the latter rather than the former.

Suppose that the generalized probabilistic theory $(\mathcal V,\Omega,\mathcal E)$ is simplex-embeddable in $d$ dimensions as in \cref{def:Simplex}:
we will first prove that there exists \vgrSC{a classical}\vgr{an operationally noncontextual ontological} model as in \cref{prop:basiccriterion} under the substitution \eqref{eq:GPTsubst} with a discrete, finite ontic space of cardinality $d$.
The set $\{\delta_i\in\mathcal W\}_{i=1}^d$ forms a basis of $\mathcal W$, so that there must exist coordinate functions $\lambda_i\funcrange{\mathcal V}{\mathbb R}$ for all $i=1,\dots,d$ such that
\begin{equation}
\label{eq:DefLambdas}
\forall v\in\mathcal V\st \iota(v) = \sum_{i=1}^d \lambda_i(v) \delta_i.
\end{equation}
By Riesz' representation \cref{th:Riesz}, there must exist $\{F_i\in\mathcal V\}_{i=1}^d$ such that: for all $v\in\mathcal V$, for all $i=1,\dots,d$,
\begin{equation}
\label{eq:DefFi}
\lambda_i(v) = \scal{v}{F_i}{\mathcal V}
\end{equation}
Consider now for any $i=1,\dots,d$ the linear maps
\begin{equation}
\label{eq:kappa}
\scal{\kappa(\cdot)}{\delta_i}{\mathcal W}\funcrange{\mathcal V}{\mathbb R}.
\end{equation}
Applying Riesz' representation \cref{th:Riesz} again, there must exist $\{\sigma_i\in\mathcal V\}_{i=1}^d$ such that: for all $v\in\mathcal V$, for all $i=1,\dots,d$,
\begin{equation}
\label{eq:ScalKappa}
\scal{\kappa(v)}{\delta_i}{\mathcal W} = \scal{\sigma_i}{v}{\mathcal V}.
\end{equation}
The consistency requirement \eqref{eq:SimplexConsist} of \cref{def:Simplex} reads: for all $\rho\in\Omega$, for all $E\in\mathcal E$,
\begin{multline}
\label{eq:ExpansionSimplex}
\scal{ \rho}{ E}{\mathcal V}
= \scal{\iota(\rho)}{\kappa(E)}{\mathcal W}
= \sum_{i=1}^d \lambda_i(\rho) \scal{\delta_i}{\kappa( E)}{\mathcal W} \\
= \sum_{i=1}^d  \scal{\rho}{F_i}{\mathcal V}\scal{\sigma_i}{ E}{\mathcal V}.
\end{multline}
Consider now \cref{prop:basiccriterion} under the substitution \eqref{eq:GPTsubst}. Recall that for a tomographically complete generalized probabilistic theory, the reduced space is simply given by the whole vector space $\mathcal V$ as was illustrated in \eqref{eq:TomographicallyCompleteReducedSpace}.
We will show that the primitives $\ospace = \{1,\dots,d\}$, $\{F_i\in\mathcal V\}_{i=1}^d$ and $\{\sigma_i\in\mathcal V\}_{i=1}^d$ match the requirements of \cref{prop:basiccriterion}.
First off, the generalized version of the consistency requirement \eqref{eq:consistency} of \cref{prop:basiccriterion} is equivalent to \eqref{eq:ExpansionSimplex} together with the fact that $\myspan\Omega = \myspan{\mathcal E}  = \mathcal V$
which is the tomographic completeness assumption.

Let us now verify the positivity relations of equations \eqref{eq:simpleranges}. Let $\rho\in\Omega$. By \cref{def:Simplex}, it holds that
\begin{equation}
\iota(\rho) \in\Delta_d.
\end{equation}
The main property of a simplex such as $\Delta_d$ is that it is the convex hull of its $d$ extremal points which are linearly independent: thus, any point $\iota(\rho)\in\Delta_d$ may be written as a convex combination
\begin{equation}
\iota(\rho) = \sum_{i=1}^d \lambda_i(\rho) \delta_i,
\end{equation}
with
\begin{subequations}
\label{eq:PosLambdas}
\begin{align}
\forall i=1,\dots,d\st \lambda_i(\rho) &\geq 0, \\
\sum_{i=1}^d\lambda_i(\rho) &= 1.
\end{align}
\end{subequations}
Because the set $\{\delta_i\}_{i=1}^d$ forms a basis of $\mathcal W$, the $\lambda_i(\rho)$'s are unique and are thus the same as in \eqref{eq:DefLambdas}. Rewriting \eqref{eq:PosLambdas} with the operators $F_i$ defined in \eqref{eq:DefFi}: for all $\rho\in\Omega$, 
\begin{subequations}
\begin{align}
\forall i=1,\dots,d\st \scal{F_i}{\rho}{\mathcal V} \geq0, \label{eq:PosFi}\\
\sum_{i=1}^d \scal{F_i}{\rho}{\mathcal V} =  1. \label{eq:SumFi}
\end{align}
\end{subequations}
Equation \eqref{eq:PosFi} proves that
\begin{equation}
\label{eq:FiPolarOmega}
F_i \in \polar{\Omega}{\mathcal V},
\end{equation}
which corresponds to the nonnegativity requirement \eqref{eq:RangeF} under the substitution \eqref{eq:GPTsubst} (again recall $\mainr=\mathcal V$ in this case).
We now recall the property~\ref{item:SimplexKappa} of the \cref{def:Simplex} of simplex-embeddability, and rewrite it using equation \eqref{eq:ScalKappa}: for all $E\in\mathcal E$, for all $i=1,\dots,d$,
\begin{equation}
\label{eq:SigmaiEin01}
\scal{\sigma_i}{E}{\mathcal V} = \scal{\kappa(E)}{\delta_i}{\mathcal V} \in [0,1].
\end{equation}
The fact that $\scal{\sigma_i}{E}{\mathcal V} \geq0$ proves \eqref{eq:RangeSigma} under the substitution \eqref{eq:GPTsubst}: indeed, it holds that for all $i=1,\dots,d$,
\begin{equation}
\label{eq:SigmaiPolarE}
\sigma_i \in \polar{\mathcal E}{\mathcal V}.
\end{equation}

Equations \eqref{eq:simpleranges} and \eqref{eq:consistency} are verified, so let us verify \eqref{eq:normsigma}.
First, recall the defining property of the unit element $u\in\mathcal E$: for all $\rho\in\Omega$, $\scal{\rho}{u}{\mathcal V} = 1$.
Using \eqref{eq:ExpansionSimplex}, for any $\rho\in\Omega$,
\begin{equation}
1 = \scal{\rho}{u}{\mathcal V} = \sum_{i=1}^d \scal{\rho}{F_i}{\mathcal V}
\scal{\sigma_i}{u}{\mathcal V}.
\end{equation}
Since for all $i=1,\dots,d$, it holds that $\scal{\rho}{F_i}{\mathcal V} \geq 0$ according to \eqref{eq:FiPolarOmega} and $\scal{\sigma_i}{u}{\mathcal V} \in[0,1]$ according to \eqref{eq:SigmaiEin01}, if there existed $j$ such that $\scal{\sigma_j}{u}{\mathcal V} < 1$, then,
\begin{multline}
1 = \sum_{i=1}^d \scal{\rho}{F_i}{\mathcal V}\scal{\sigma_i}{u}{\mathcal V}
< 
\sum_{i=1}^d \scal{\rho}{F_i}{\mathcal V}
= 1.
\end{multline}
We used \eqref{eq:SumFi} to conclude. This yields a contradiction so that we conclude that $\scal{\sigma_i}{u}{\mathcal V} = 1$ for all $i=1,\dots,d$ which proves \eqref{eq:normsigma} under the substitution \eqref{eq:GPTsubst}.
Thus, we conclude that if the tomographically complete generalized probabilistic theory $(\mathcal V,\Omega,\mathcal E)$ is simplex-embeddable in $d$ dimensions, then there exists \vgrSC{a classical}\vgr{an operationally noncontextual ontological} model for the tomographically complete prepare-and-measure scenario $(\Omega,\mathcal E)$ as in \cref{prop:basiccriterion} under the substitution \eqref{eq:GPTsubst} with the ontic space $\ospace = \{1,\dots,d\}$.

Consider now the other direction:
suppose that the tomographically complete prepare-and-measure scenario $(\Omega,\mathcal E)$ admits \vgrSC{a classical}\vgr{an operationally noncontextual ontological} model with discrete ontic space of cardinality $d$. Again, the assumption of tomographically complete $(\Omega,\mathcal E)$ imply $\mainr = \mathcal V$ as in \eqref{eq:TomographicallyCompleteReducedSpace}.
Let the ontic primitives of \cref{prop:basiccriterion}, under the substitution \eqref{eq:GPTsubst}
be denoted $\{F_i\in\polar{\Omega}{\mathcal V}\}_{i=1}^{d}$ and $\{\sigma_i\in\polar{\mathcal E}{\mathcal V}\}_{i=1}^{d}$.
Now, we consider the euclidean space $\mathbb R^d$, equipped with an  orthonormal basis $\{\delta_i\in\mathbb R^d\}_{i=1}^d$. These define the simplex $\Delta_d \subset \mathbb R^d$:
\begin{equation}
\Delta_d := \myconv{\{\delta_i\}_{i=1}^d}.
\end{equation}
Now, define the linear maps $\iota,\kappa\funcrange{\mathcal V}{\mathbb R^d}$ by: for all $v\in\mathcal V$,
\begin{subequations}
\begin{align}
\iota(v) &= \sum_{i=1}^d \scal{v}{F_i}{\mathcal V} \delta_i, \label{eq:DefIota} \\
\kappa(v) &= \sum_{i=1}^d \scal{\sigma_i}{v}{\mathcal V} \delta_i. \label{eq:DefKappa}
\end{align}
\end{subequations}

We will now verify first the consistency requirement \eqref{eq:SimplexConsist}, then~\ref{item:SimplexIota} and~\ref{item:SimplexKappa} of \cref{def:Simplex}: this will prove the validity of the simplex-embedding under consideration.
Using the orthonormality of the basis $\{\delta_i\}_i$: for all $\rho\in\Omega$, for all $E\in\mathcal E$,
\begin{equation}
\scal{\iota(\rho)}{\kappa( E)}{\mathbb R^d}\! =
\sum_{i=1}^d \scal{\rho}{F_i}{\mathcal V}\scal{\sigma_i}{E}{\mathcal V}
= \scal{\rho}{ E}{\mathcal V},
\end{equation}
where we used the consistency requirement \eqref{eq:consistency} of the \cref{prop:basiccriterion} for the existence of the \vgrSC{classical}\vgr{operationally noncontextual ontological} model to conclude.
This proves the consistency requirement \eqref{eq:SimplexConsist} of \cref{def:Simplex}.
Let us now prove~\ref{item:SimplexIota}, i.e., that for all $\rho\in\Omega$: $\iota(\rho) \in\Delta_d$. 
Fix the argument $\rho\in\Omega$.
By the definition \eqref{eq:DefIota}, $\iota(\rho) = \sum_{i=1}^d \scal{\rho}{F_i}{\mathcal V}\delta_i$. The property \eqref{eq:RangeF} stating in this case that $F_i\in\polar{\Omega}{\mathcal V}$ proves that: for all $i=1,\dots,d$,
\begin{equation}
\label{eq:Lambdaspos}
\lambda_i :=\scal{\rho}{F_i}{\mathcal V} \geq 0.
\end{equation}
Furthermore, using  the normalization $\scal{\sigma_i}{u}{\mathcal V} = 1$ as in \eqref{eq:normsigma} first, and then the consistency requirement \eqref{eq:consistency}:
\begin{multline}
\label{eq:Lambdasnorm}
\sum_{i=1}^d \lambda_i =
\sum_{i=1}^d \scal{\rho}{F_i}{\mathcal V}
=\sum_{i=1}^d \scal{\rho}{F_i}{\mathcal V}
\scal{\sigma_i}{u}{\mathcal V}  \\
= \scal{\rho}{u}{\mathcal V} = 1.
\end{multline}
Equations \eqref{eq:Lambdaspos} and \eqref{eq:Lambdasnorm} prove that $\iota(\rho) = \sum_{i=1}^d \lambda_i \delta_i$ is a convex combination of the $\delta_i$'s, and hence for any $\rho\in\Omega$ it holds that $\iota(\rho)\in\Delta_d\subseteq\mathbb R^d$. Hence~\ref{item:SimplexIota} of \cref{def:Simplex} is verified.

Let us now verify~\ref{item:SimplexKappa}, i.e that for all $E\in\mathcal E$, for all $i=1,\dots,d$: $\scal{\kappa(E)}{\delta_i}{\mathbb R^d} \in [0,1]$. Fix $i\in\{1,\dots,d\}$ and $E\in\mathcal E$. 
We will make use of the fact that there exists $\{E_k\in\mathcal E\}_k$ such that $u = E + \sum_k E_k$.
By the definition \eqref{eq:DefKappa} of the mapping $\kappa$ and the orthonormality of the $\delta_i$'s:
$\scal{\kappa(E)}{\delta_i}{\mathbb R^d} = \scal{\sigma_i}{E}{\mathcal V}.$ Using the normalization \eqref{eq:normsigma}:
\begin{equation}
\label{eq:TempHypercube}
1 = \scal{\sigma_i}{u}{\mathcal V} = \scal{\sigma_i}{E}{\mathcal V} + \sum_k \scal{\sigma_i}{E_k}{\mathcal V}.
\end{equation}
Using that $E,E_k\in\mathcal E$ and using \eqref{eq:RangeSigma} that states here that $\sigma_i\in\polar{\mathcal E}{\mathcal V}$, it holds that $\scal{\sigma_i}{E}{\mathcal V},\scal{\sigma_i}{E_k}{\mathcal V}\geq 0$. Then, \eqref{eq:TempHypercube} implies also that $\scal{\sigma_i}{E}{\mathcal V} \leq 1$. 
This proves that
\begin{equation}
\scal{\kappa(E)}{\delta_i }{\mathcal V} = \scal{\sigma_i}{E}{\mathcal V} \in [0,1].
\end{equation}
Thus~\ref{item:SimplexKappa} of \cref{def:Simplex} is also verified.
This proves that indeed $(\mathcal V,\Omega,\mathcal E)$ is simplex-embeddable in $d$ dimensions.
\end{myproof}

\end{document}